\documentclass[aps,reprint,superscriptaddress,longbibliography,floatfix]{revtex4-2}

\usepackage[utf8]{inputenc}
\usepackage{hyperref}
\usepackage{graphicx}
\usepackage{amsmath}
\usepackage{bm}
\usepackage{amssymb}
\usepackage{soul}

\usepackage{color}

\begin{document}

\title{Capacitively-coupled and inductively-coupled excitons in bilayer MoS$_2$}

\author{Lukas Sponfeldner} 
\email{lukas.sponfeldner@unibas.ch}
\affiliation{Department of Physics, University of Basel, Klingelbergstrasse 82, CH-4056 Basel, Switzerland}

\author{Nadine Leisgang}
\affiliation{Department of Physics, University of Basel, Klingelbergstrasse 82, CH-4056 Basel, Switzerland}

\author{Shivangi Shree}
\affiliation{Universit\'e de Toulouse, INSA-CNRS-UPS, LPCNO, 135 Avenue Rangueil, 31077 Toulouse, France}

\author{Ioannis Paradisanos}
\affiliation{Universit\'e de Toulouse, INSA-CNRS-UPS, LPCNO, 135 Avenue Rangueil, 31077 Toulouse, France}

\author{Kenji Watanabe}
\affiliation{Research Center for Functional Materials, National Institute for Materials Science, 1-1 Namiki, Tsukuba 305-0044, Japan}

\author{Takashi Taniguchi}
\affiliation{International Center for Materials Nanoarchitectonics, National Institute for Materials Science, 1-1 Namiki, Tsukuba 305-0044, Japan}

\author{Cedric Robert}
\affiliation{Universit\'e de Toulouse, INSA-CNRS-UPS, LPCNO, 135 Avenue Rangueil, 31077 Toulouse, France}

\author{Delphine Lagarde}
\affiliation{Universit\'e de Toulouse, INSA-CNRS-UPS, LPCNO, 135 Avenue Rangueil, 31077 Toulouse, France}

\author{Andrea Balocchi}
\affiliation{Universit\'e de Toulouse, INSA-CNRS-UPS, LPCNO, 135 Avenue Rangueil, 31077 Toulouse, France}

\author{Xavier Marie}
\affiliation{Universit\'e de Toulouse, INSA-CNRS-UPS, LPCNO, 135 Avenue Rangueil, 31077 Toulouse, France}

\author{Iann C. Gerber}
\affiliation{Universit\'e de Toulouse, INSA-CNRS-UPS, LPCNO, 135 Avenue Rangueil, 31077 Toulouse, France}

\author{Bernhard Urbaszek}
\affiliation{Universit\'e de Toulouse, INSA-CNRS-UPS, LPCNO, 135 Avenue Rangueil, 31077 Toulouse, France}

\author{Richard J. Warburton}
\affiliation{Department of Physics, University of Basel, Klingelbergstrasse 82, CH-4056 Basel, Switzerland}

\begin{abstract}
The interaction of intralayer and interlayer excitons is studied in a two-dimensional semiconductor, homobilayer MoS$_2$. It is shown that the measured optical susceptibility reveals both the magnitude and the sign of the coupling constants. The interlayer exciton interacts capacitively with the intralayer B-exciton (positive coupling constant) consistent with hole tunnelling from one monolayer to the other. Conversely, the interlayer exciton interacts inductively with the intralayer A-exciton (negative coupling constant). First-principles many-body calculations show that this coupling arises via an intravalley exchange-interaction of A- and B-excitons.
\end{abstract}

\maketitle

The elementary excitation at energies close to the band gap of a semiconductor is the exciton, a bound electron-hole pair. The exciton is a prominent feature of the linear optical response of a two-dimensional (2D) semiconductor, for instance monolayer MoS$_2$, even at room temperature \cite{mak2010,splendiani2010}. This is a consequence of the giant exciton binding energies, hundreds of meV, in these materials \cite{chernikov2014}. Excitons interact with each other. These interactions lead to nonlinear optical effects \cite{wang2015,jakubczyk2016}, potentially useful in opto-electronics, and to condensation phenomena, for instance the creation of states with macroscopic quantum-correlations \cite{kasprzak2006}. At the few-exciton level, exciton-exciton repulsion in an appropriate trap is a potential way to engineer a single-photon emitter \cite{delteil2019,munoz-matutano2019}. 

Exciton-exciton interactions can arise via the charge of the constituent electrons and holes \cite{ciuti1998,shahnazaryan2017,erkensten2021}. Exciton-exciton interactions can also arise should one of the constituents undergo tunnelling. The canonical example is the double quantum well in which molecule-like electronic states form via tunnelling between the two quantum wells \cite{ferreira1990,fox1991,sivalertporn2012,andreakou2015}. Recently, molecule-like coupling has been discovered in bilayer 2D semiconductors \cite{alexeev2019,hsu2019,shimazaki2020,sung2020,merkl2020,zhang2020,tang2020,mcdonnell2021}. 

Here, we probe exciton-exciton interactions in gated-homobilayer MoS$_2$. This is a rich system for probing exciton-exciton interactions: an interlayer exciton (IE) interacts with both the B- and A-excitons of the constituent layers. We show that the optical susceptibility reveals not just the magnitude but also the sign of the exciton-exciton coupling. Remarkably, the IE-A and IE-B couplings have opposite signs. The IE-B coupling has a positive sign due to hole tunnelling from one monolayer to the other. Conversely, the IE-A coupling has a negative sign, showing that the coupling mechanism has a different microscopic origin. Beyond density functional theory (post-DFT) calculations enable us to ascribe the IE-A coupling to spin, specifically an exchange-based hybridisation of A- and B-excitons.

\begin{figure}[t!]
\centering
\includegraphics[width=86mm]{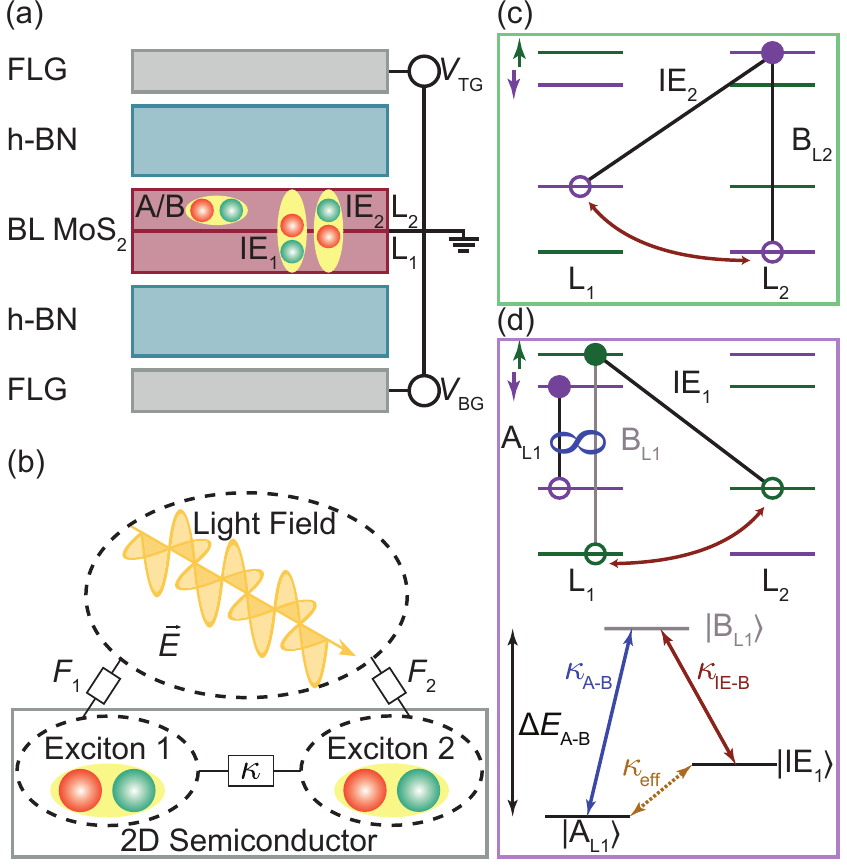}
\caption{(a) Schematic of the van der Waals heterostructure consisting of homobilayer MoS$_2$ embedded in h-BN. The MoS$_2$ bilayer is electrically grounded. The voltages $V_\text{TG}$ and $V_\text{BG}$ applied to the top and bottom few-layer graphene (FLG) layers are used to create a uniform electric field $F_z$ across the bilayer. The spatial extents of the intralayer A- and B-excitons and the interlayer excitons (IE) are sketched in the bilayer MoS$_2$ region. The electrons are depicted in green; the holes in red. MoS$_2$ bottom (top) layer is denoted as L$_1$ (L$_2$). (b) Sketch of two coupled excitons in a 2D semiconductor both driven by an external light field. The oscillator strength of each exciton is given by $F_{1,2}$; $\kappa$ is the coupling constant. (c,d) Band structure and schematics of excitons in homobilayer MoS$_2$ at the $K$-point. (c) Sketch of the microscopic origin of the IE-B interaction mediated by hole tunnelling at the $K$-point (red arrow). (d) Sketch of the IE-A interaction (top) in the bandstructure at the $K$-point and (bottom) in an excitonic transition energy diagram. IE and B are coupled ($\kappa_\text{IE-B}$, red arrow). A and B are coupled through the intralayer exchange interaction ($\kappa_\text{A-B}$, blue $\infty$-symbol). Through the common coupling to B, IE and A are effectively coupled ($\kappa_\text{eff}$).}
\label{fig1}
\end{figure}

\begin{figure*}[t!]
\centering
\includegraphics[width=172mm]{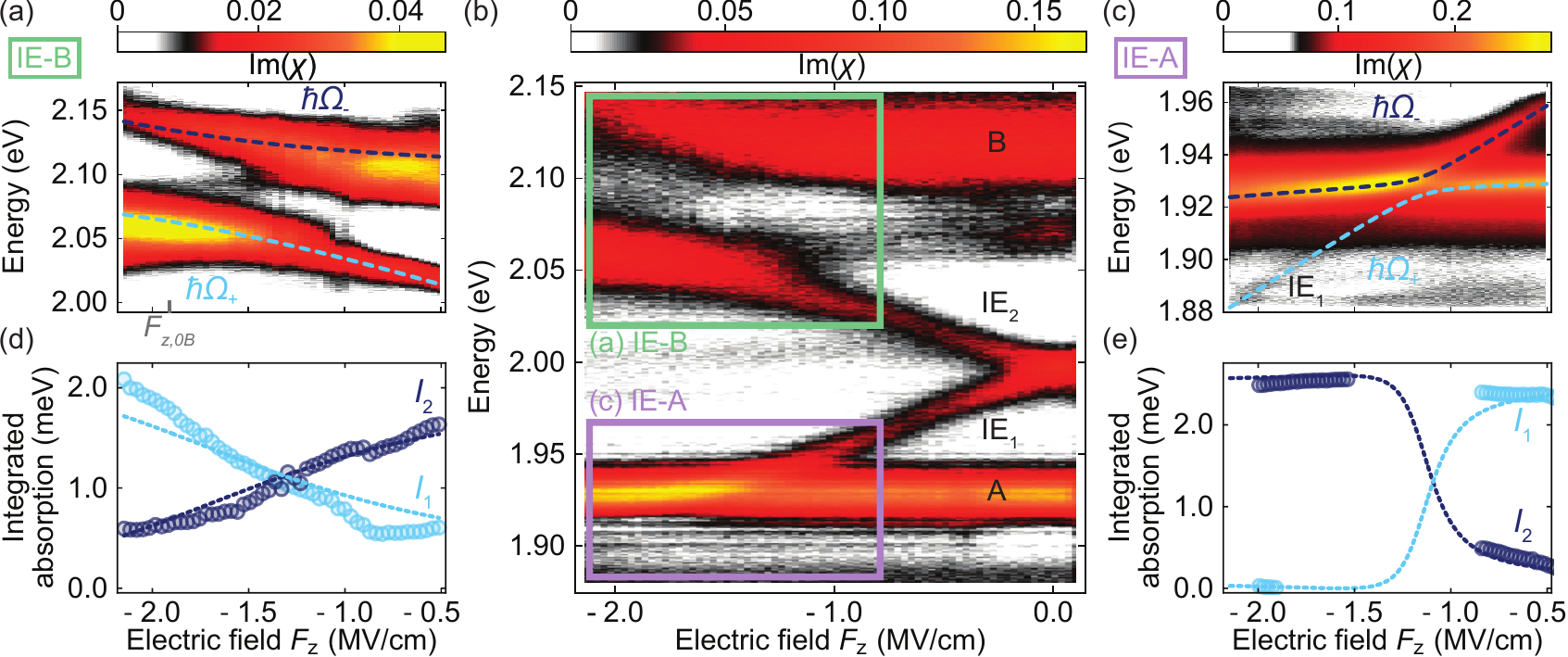}
\caption{Absorption ($\sigma^+$ polarisation) of homobilayer MoS$_2$ as a function of applied electric field $F_z$. (b) Absorption over the whole studied energy range. The green and purple rectangles correspond to the measurement regions centred around the B-exciton (a) and the A-exciton (c), respectively. The measurements were carried out at $T=4.2$~K and magnetic field $B_z=9$~T. For clarity, the non-interacting B$_\text{L1}$ exciton is removed from (a) and (b) and the non-interacting A$_\text{L2}$ exciton is removed from (b) and (c) in a data processing step (see Supplement). In (a), $F_{z,0B}$ marks the energy crossing for zero coupling. The coloured dashed lines in (a) and (c) are fits to the extracted peak energies according to equations S20 and S47 for the eigenenergies $\hbar\Omega_\pm$, respectively (see Supplement). (d) Integrated absorption extracted from the spectra shown in (a). The coloured dashed lines are fits to the absorption strength of the eigenmodes $I_{1,2}$ according to equation S26. (e) Integrated absorption extracted from the spectra shown in (c). As the coupling strength is much smaller than the linewdith of the A-exciton, only the spectra far from the crossing point can be fitted unambiguously. The coloured dashed lines are model fits with the same function as in (d) including the quadratic Stark shift of the A-exciton in the initial equations of motion (see supplement section IV). All parameters extracted from the measured data are summarised in SI~Table~I. Details on the fitting procedure can be found in supplement section V.}
\label{fig2}
\end{figure*}

The device is constructed using bilayer MoS$_2$ and sketched in Fig~\ref{fig1}a. The naturally 2H-stacked bilayer MoS$_2$ is embedded in h-BN; few-layer graphene layers provide a back-contact and a top-gate. In addition, the MoS$_2$ layer is contacted. Results are presented as a function of the vertical electric field ($F_z$) for the smallest possible electron density ($n$). The device is illuminated locally with a very weak broadband source. The reflectivity is measured, using the response at large $n$ as reference, and converted into the optical susceptibility using a Kramers-Kronig relationship (see supplement of Ref.\,\cite{roch2019} for details). At $F_z =0$, the imaginary part of $\chi$ (which determines the absorption) shows three peaks: at low energy the intralayer A-excitons, at high energy the intralayer B-excitons, and in between, the interlayer exciton (IE), see Fig.~\ref{fig2}b. These momentum-direct excitonic transitions are shown schematically in Fig.~\ref{fig1}a; the relevant band structure at the $K$-point is shown in Fig.~\ref{fig1}c and Fig.~\ref{fig1}d. The current understanding is that the IE consists of a hole state delocalised across the two layers bound to an electron localised in one of the two layers \cite{gerber2019,pisoni2019}. For $F_z \ne 0$, the IE splits into two lines \cite{leisgang2020,lorchat2020,peimyoo2021}. At high $F_z$, there is a clear avoided crossing between the IE upper branch and the B-exciton; and a weak avoided crossing between the IE lower branch and the A-exciton, as reported in Ref.\,\cite{leisgang2020}.

We focus here on the avoided crossings. At the IE-B avoided crossing, one of the B-excitons couples to the IE, the other does not. In Fig.~\ref{fig2}a and Fig.~\ref{fig2}b, we subtract the peak arising from the uncoupled B-exciton. A simple two-level model describes the peak energies convincingly but not the relative intensities. Strikingly, at the $F_z$ for which the two transitions are closest together in energy ($F_{z,0B}$), the intensities are quite different: the lower-energy transition is considerably stronger than the higher-energy transition. The IE-A coupling shown in Fig.~\ref{fig2}c is weaker than the IE-B coupling. One of the A-excitons couples to the IE, the other does not. The absorption from the uncoupled A-exciton is substracted in Fig.~\ref{fig2}c. The minimum energy separation of the peaks is comparable to the peak broadening. Nevertheless, the avoided crossing has a strong effect on the intensities: as $F_z$ increases, the IE-like branch enters the avoided crossing with a relatively large intensity, but emerges surprisingly with a much lower intensity. Building on recent studies \cite{leisgang2020,lorchat2020,peimyoo2021}, we show here that the IE can be tuned energetically below the A-exciton. The IE-A intensity behaviour mimics the behaviour at the IE-B avoided crossing but with one crucial difference. The upper-IE is strong on the low-energy side of the B-exciton; the lower-IE is strong on the high-energy side of the A-exciton. Our target is to understand the origin of the different coupling behaviours.

The optical susceptibility of a quantum well can be calculated from the semiconductor Bloch equations \cite{lindberg1988,koch2006,haug2009,klingshirn2012}. In this approach, the quantum well is treated quantum mechanically. The final result is identical to a completely classical approach in which the quantum well is treated as a 2D array of optical dipoles \cite{karrai2003}. Inspired by the success of the purely classical approach, we set up a heuristic description of the IE-B avoided crossing as sketched in Fig.~\ref{fig1}b. The IE and B-excitons are treated as optical dipoles, each driven by the same driving field $\vec{E}$, but with different oscillator strengths $F_{1,2}$ \cite{deilmann2018,gerber2019}. A linear coupling term $\kappa$ is introduced: an IE-dipole induces a B-dipole, and vice versa. Solving the equations of motion of this system yields an analytic equation for both the eigenenergies $\hbar\Omega_{\pm}$ and the absorption strengths of the two eigenmodes (see Supplement). The calculated absorption strength $I_{1,2}$ of each eigenmode depends on the energy detuning, the coupling $\kappa$, and the oscillator strengths $F_{1,2}$. 

\begin{figure}[b!]
\centering
\includegraphics[width=86mm]{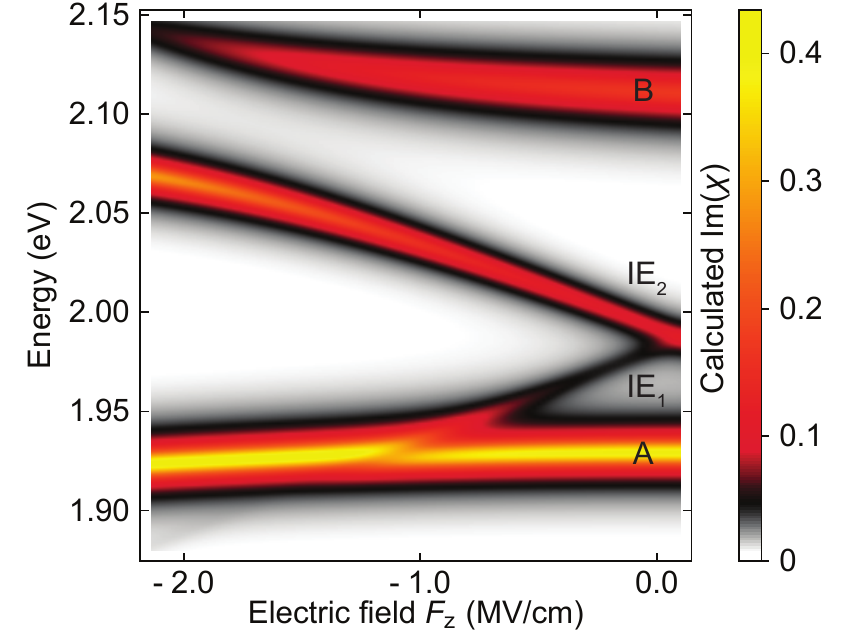}
\caption{Colour-map of the calculated absorption Im($\chi$) as a function of electric field $F_z$ with parameters extracted from model fits to the measured data in Fig.~\ref{fig2}. The parameters are listed in SI~Table~I. Separate models are used to describe the IE-A and IE-B couplings. The Im($\chi$) is a sum of the two. The colour-map is interpolated to first order.}
\label{fig3}
\end{figure}

The classical model reproduces the experimental results extraordinarily well provided the ratio of the oscillator strengths and the coupling constant are well chosen. The measured peak energies (Fig.~\ref{fig2}a) and integrated absorption (Fig.~\ref{fig2}d) of the IE-B avoided crossing are fitted by the coupled dipole model (light- and dark-blue dashed lines). The fits yield an oscillator strength ratio $F_1/F_2 \approx 5$ and an IE-B coupling strength of $\kappa=+35.8 \pm 3.6$~meV. The IE-A avoided crossing is described with the same model but with the energies appropriate to the lower IE-branch and the A-exciton along with a different choice of coupling constant. For the A-exciton energy, a quadratic Stark shift is included \cite{miller1984}. Our model reproduces also the IE-A avoided crossing very convincingly, capturing the $F_z$-dependence of the intensities (see Fig.~\ref{fig2}c and Fig.~\ref{fig2}e). The fits yield an oscillator strength ratio $F_1/F_2 \approx 5$ and an IE-A coupling strength of $\kappa=-3.5 \pm 0.4$~meV. The absorption spectra for the IE-A and IE-B couplings are calculated separately and added together to describe the full $F_z$-dependence, as shown in Fig.~\ref{fig3}. The calculated absorption reproduces very closely the experiments (Fig.~\ref{fig2}b).

This model unearths a crucial result: the IE-A coupling constant is of opposite sign to the IE-B coupling constant. The sign of the coupling constant is particularly important in determining the relative strengths of the absorption peaks (see Supplement). We interpret the sign of the coupling by making an analogy to driven, coupled RLC-circuits (see Supplement). Two such circuits can be coupled via an impedance. The equations of motion are analogues of those describing the driven optical dipole. The nature of the coupling impedance determines the sign of the coupling constant: coupling via a capacitance results in a positive coupling constant; coupling via an inductance results in a negative coupling constant. In this analogy, the IE-B coupling corresponds to a capacitive coupling. This suggests that, at a microscopic level, the IE-B coupling involves the movement of charge -- it is consistent with hole tunnelling from one layer to the other \cite{gerber2019}. Conversely, the IE-A coupling corresponds to an inductive coupling. This points to a completely different coupling mechanism, as discussed below.

The model provides an explanation for the $F_z$-dependent absorption strengths. We take the IE-A coupling as an example. Without a coupling, both IE and A respond directly to the driving field. A has the stronger response: the induced dipole moment is in-phase with the drive for energies well below the bare A-energy and out-of-phase for energies well above the bare A-energy (the standard behaviour for a driven harmonic oscillator). With a coupling, each eigenmode is a dressed state of IE and A. At detunings far from the avoided crossing, there is an A-like and an IE-like eigenmode. The IE-like mode is driven by two sources: the field acting directly on the IE, and via its coupling to A. Now the sign of the coupling plays a crucial role. If the coupling has a negative sign, then for energies above (below) the bare A-energy, these two terms have the same (opposite) sign and interfere constructively (destructively). At a particular detuning, the destructive interference is complete and the absorption of one mode disappears. Even at energies far from the avoided crossing, this interference has a strong effect on the IE-absorption. The picture inverts for a positive coupling, the IE-B interaction. In this case, the IE-like mode is boosted (suppressed) when it lies below (above) the bare B-energy. At $F_z=0$, the IE is far from the avoided crossing with both A- and B-excitons but its absorption strength is boosted by its coupling to both A and B. In simple terms, the dielectric constant at the IE-resonance is strongly influenced by the strong A- and B-resonances.

\begin{figure}[ht!]
\centering
\includegraphics[width=86mm]{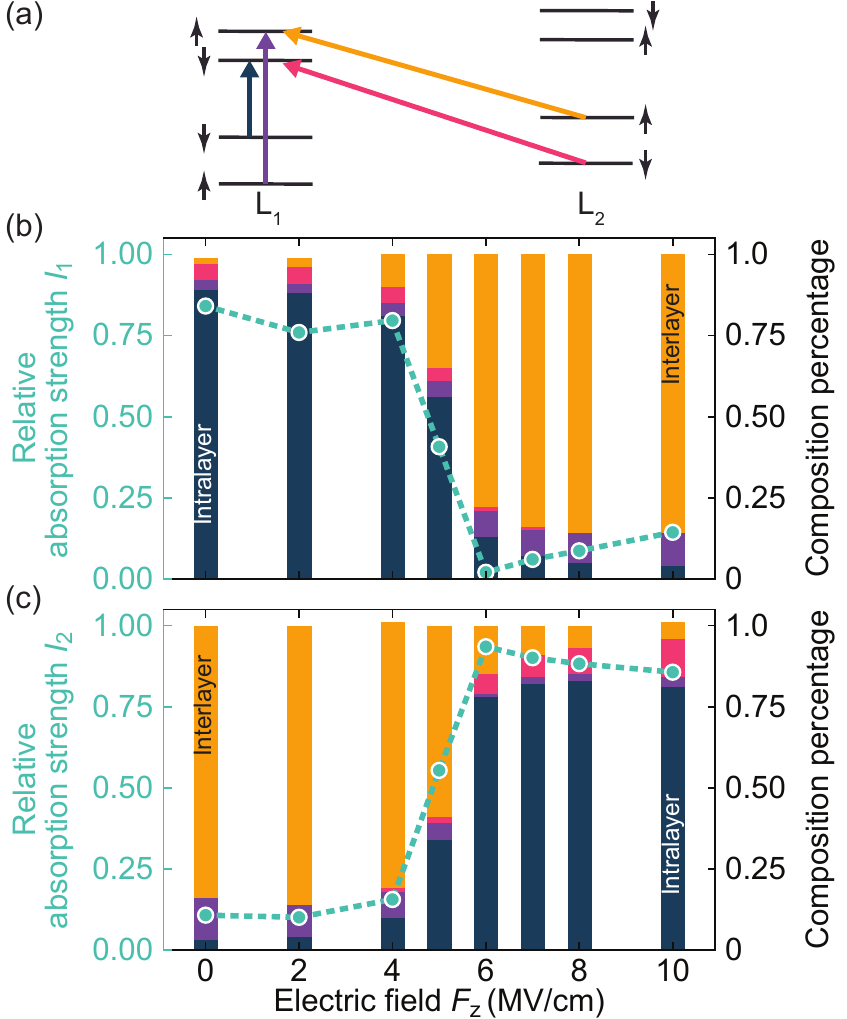}
\caption{(a) Main excitonic transitions contributing to the IE-A dressed state. They are sketched as the coloured arrows in the schematic band structure at the $K$-point of bilayer MoS$_2$ at a finite positive electric field $F_z$. (b,c) Relative absorption strength of eigenmode 1 $I_1$ (b) and eigenmode 2 $I_2$ (c) of the IE-A interaction as a function of electric field determined by GW+BSE calculations. The cyan dashed-line is a guide to the eye. The coloured bars indicate the composition percentage at each electric field step; the colours match those used to denote the excitonic transition in (a).}
\label{fig4}
\end{figure}

We look for a microscopic explanation for the different IE-A and IE-B couplings. To do this, we describe the band structure of bilayer MoS$_2$ at the GW-level (one-particle Green’s function G, dynamically screened Coulomb interaction W), and use these states to construct excitons by solving the Bethe-Salpeter equation BSE (see Fig.~\ref{fig4} and Supplement). The results describe the general behaviour of the experiments very well as revealed by a comparison of the calculated relative absorption strengths in Fig.~\ref{fig4}b and Fig.~\ref{fig4}c with the measured integrated absorption in Fig.~\ref{fig2}e. (Exact quantitative agreement is not expected as the model assumes that the MoS$_2$ bilayers are located in vacuum -- it does not take into account the full dielectric environment of the sample.) As in the experiment, the IE are relatively strong when they lie energetically between the bare A- and the bare B-resonances but weaker when they lie out of this energy window (Fig.~\ref{fig4}b and Fig.~\ref{fig4}c). The post-DFT results can be parameterised by the coupled-dipole model (see Supplement).

An analysis of the spin and orbital composition of the excitons provides an explanation for both the IE-B and IE-A couplings. The IE-B coupling arises via hole tunnelling (see Fig.~\ref{fig1}c). In the bare B-state, the electron and hole are localised in one monolayer; in the bare IE-state, the electron is localised in the same monolayer but the hole is localised in the other layer. Hole tunnelling couples these two states (Fig.~\ref{fig1}c). Consequently, the excitons observed in the experiment are mixed excitonic states rather than states consisting of a single excitonic transition. The large IE-B coupling constant, $+35.8$~meV, reflects the efficient hole tunnelling. Conversely, the IE-A coupling arises via a weak admixture between the A- and B-excitons in the same valley and layer, as shown schematically in Fig.~\ref{fig1}d. This exciton admixture was proposed in Ref.\,\cite{guo2019} for MoS$_2$ monolayers and is found also in our bilayer calculations (Fig.~\ref{fig4}). This means that both IE and A couple to B. IE and A then acquire an effective coupling, a second-order effect.

In this analysis, depicted in Fig.~\ref{fig1}d, the IE-A coupling is determined by $\kappa_\text{eff}=-\dfrac{\kappa_\text{A-B} \kappa_\text{IE-B}}{\Delta E_\text{A-B}}$ (with the sign convention employed here) where $\kappa_\text{IE-B}$ is the IE-B coupling, $\kappa_\text{A-B}$ the A-B coupling and $\Delta E_\text{A-B}$ is the energy splitting between A and B \cite{lowdin1951}. The experiment measures $\kappa_\text{eff}=-3.5$~meV, $\kappa_\text{IE-B}=35.8$~meV and $\Delta E_\text{A-B}=-170$~meV. In turn, this determines $\kappa_\text{A-B}$. We find $\kappa_\text{A-B}=-16.6 \pm 2.5$~meV. In other words, by using the IE as a tunable probe, we are able to determine the A-B coupling energy. This is an important quantity -- it arises even in a MoS$_2$ monolayer and determines to what extent spin is a good quantum number in the fundamental exciton \cite{guo2019,wang_2017}.

We state two main conclusions. First, in homobilayer MoS$_2$, the IE and B-excitons couple via hole tunnelling; the IE and A-excitons couple via an exchange-induced A-B admixture. The experiment enables the A-B coupling to be determined quantitatively. We find $-16.6$~meV. Its existence implies that spin is an imperfect quantum number for the A- and B-excitons in the same valley. Second, a measurement of the optical susceptibility enables not just the magnitude but also the sign of the exciton-exciton couplings to be determined.

As a final comment, we point out that the driven coupled-oscillator model shows that a weak resonance can be made visible by bringing it into near-degeneracy with a strong resonance. All that is required is a coupling between the two resonances. This point is not restricted to driven dipoles at optical frequencies: it applies quite generally.

We thank Christoph Bruder for fruitful discussions and Jonas Roch for expert experimental assistance. Basel acknowledges funding from the Swiss Nanoscience Institute (SNI), PhD School Quantum Computing and Quantum Technology (QCQT), SNF (project 200020\_175748), and NCCR QSIT. Toulouse acknowledges funding from ANR MagicValley, ANR IXTASE, ANR HiLight, 
ITN 4PHOTON Marie Sklodowska Curie Grant Agreement No.\ 721394 and the Institut Universitaire de France. K.W. and T.T. acknowledge support from the Elemental Strategy Initiative
conducted by the MEXT, Japan (Grant Number JPMXP0112101001) and JSPS KAKENHI (Grant Numbers 19H05790 and JP20H00354). I.C.G. thanks the CALMIP initiative for the generous allocation of computational time, through Project No.\ p0812, as well as GENCI-CINES and GENCI-IDRIS, Grant No.\ A010096649.

\bibliographystyle{apsrev4-2}
\bibliography{coupledExcitons_bib}

\begin{thebibliography}{39}%
\makeatletter
\providecommand \@ifxundefined [1]{%
 \@ifx{#1\undefined}
}%
\providecommand \@ifnum [1]{%
 \ifnum #1\expandafter \@firstoftwo
 \else \expandafter \@secondoftwo
 \fi
}%
\providecommand \@ifx [1]{%
 \ifx #1\expandafter \@firstoftwo
 \else \expandafter \@secondoftwo
 \fi
}%
\providecommand \natexlab [1]{#1}%
\providecommand \enquote  [1]{``#1''}%
\providecommand \bibnamefont  [1]{#1}%
\providecommand \bibfnamefont [1]{#1}%
\providecommand \citenamefont [1]{#1}%
\providecommand \href@noop [0]{\@secondoftwo}%
\providecommand \href [0]{\begingroup \@sanitize@url \@href}%
\providecommand \@href[1]{\@@startlink{#1}\@@href}%
\providecommand \@@href[1]{\endgroup#1\@@endlink}%
\providecommand \@sanitize@url [0]{\catcode `\\12\catcode `\$12\catcode
  `\&12\catcode `\#12\catcode `\^12\catcode `\_12\catcode `\%12\relax}%
\providecommand \@@startlink[1]{}%
\providecommand \@@endlink[0]{}%
\providecommand \url  [0]{\begingroup\@sanitize@url \@url }%
\providecommand \@url [1]{\endgroup\@href {#1}{\urlprefix }}%
\providecommand \urlprefix  [0]{URL }%
\providecommand \Eprint [0]{\href }%
\providecommand \doibase [0]{https://doi.org/}%
\providecommand \selectlanguage [0]{\@gobble}%
\providecommand \bibinfo  [0]{\@secondoftwo}%
\providecommand \bibfield  [0]{\@secondoftwo}%
\providecommand \translation [1]{[#1]}%
\providecommand \BibitemOpen [0]{}%
\providecommand \bibitemStop [0]{}%
\providecommand \bibitemNoStop [0]{.\EOS\space}%
\providecommand \EOS [0]{\spacefactor3000\relax}%
\providecommand \BibitemShut  [1]{\csname bibitem#1\endcsname}%
\let\auto@bib@innerbib\@empty
\bibitem [{\citenamefont {Mak}\ \emph {et~al.}(2010)\citenamefont {Mak},
  \citenamefont {Lee}, \citenamefont {Hone}, \citenamefont {Shan},\ and\
  \citenamefont {Heinz}}]{mak2010}%
  \BibitemOpen
  \bibfield  {author} {\bibinfo {author} {\bibfnamefont {K.~F.}\ \bibnamefont
  {Mak}}, \bibinfo {author} {\bibfnamefont {C.}~\bibnamefont {Lee}}, \bibinfo
  {author} {\bibfnamefont {J.}~\bibnamefont {Hone}}, \bibinfo {author}
  {\bibfnamefont {J.}~\bibnamefont {Shan}},\ and\ \bibinfo {author}
  {\bibfnamefont {T.~F.}\ \bibnamefont {Heinz}},\ }\href
  {https://doi.org/10.1103/PhysRevLett.105.136805} {\bibfield  {journal}
  {\bibinfo  {journal} {Phys. Rev. Lett.}\ }\textbf {\bibinfo {volume} {105}},\
  \bibinfo {pages} {136805} (\bibinfo {year} {2010})}\BibitemShut {NoStop}%
\bibitem [{\citenamefont {Splendiani}\ \emph {et~al.}(2010)\citenamefont
  {Splendiani}, \citenamefont {Sun}, \citenamefont {Zhang}, \citenamefont {Li},
  \citenamefont {Kim}, \citenamefont {Chim}, \citenamefont {Galli},\ and\
  \citenamefont {Wang}}]{splendiani2010}%
  \BibitemOpen
  \bibfield  {author} {\bibinfo {author} {\bibfnamefont {A.}~\bibnamefont
  {Splendiani}}, \bibinfo {author} {\bibfnamefont {L.}~\bibnamefont {Sun}},
  \bibinfo {author} {\bibfnamefont {Y.}~\bibnamefont {Zhang}}, \bibinfo
  {author} {\bibfnamefont {T.}~\bibnamefont {Li}}, \bibinfo {author}
  {\bibfnamefont {J.}~\bibnamefont {Kim}}, \bibinfo {author} {\bibfnamefont
  {C.-Y.}\ \bibnamefont {Chim}}, \bibinfo {author} {\bibfnamefont
  {G.}~\bibnamefont {Galli}},\ and\ \bibinfo {author} {\bibfnamefont
  {F.}~\bibnamefont {Wang}},\ }\href {https://doi.org/10.1021/nl903868w}
  {\bibfield  {journal} {\bibinfo  {journal} {Nano Lett.}\ }\textbf {\bibinfo
  {volume} {10}},\ \bibinfo {pages} {1271} (\bibinfo {year}
  {2010})}\BibitemShut {NoStop}%
\bibitem [{\citenamefont {Chernikov}\ \emph {et~al.}(2014)\citenamefont
  {Chernikov}, \citenamefont {Berkelbach}, \citenamefont {Hill}, \citenamefont
  {Rigosi}, \citenamefont {Li}, \citenamefont {Aslan}, \citenamefont
  {Reichman}, \citenamefont {Hybertsen},\ and\ \citenamefont
  {Heinz}}]{chernikov2014}%
  \BibitemOpen
  \bibfield  {author} {\bibinfo {author} {\bibfnamefont {A.}~\bibnamefont
  {Chernikov}}, \bibinfo {author} {\bibfnamefont {T.~C.}\ \bibnamefont
  {Berkelbach}}, \bibinfo {author} {\bibfnamefont {H.~M.}\ \bibnamefont
  {Hill}}, \bibinfo {author} {\bibfnamefont {A.}~\bibnamefont {Rigosi}},
  \bibinfo {author} {\bibfnamefont {Y.}~\bibnamefont {Li}}, \bibinfo {author}
  {\bibfnamefont {O.~B.}\ \bibnamefont {Aslan}}, \bibinfo {author}
  {\bibfnamefont {D.~R.}\ \bibnamefont {Reichman}}, \bibinfo {author}
  {\bibfnamefont {M.~S.}\ \bibnamefont {Hybertsen}},\ and\ \bibinfo {author}
  {\bibfnamefont {T.~F.}\ \bibnamefont {Heinz}},\ }\href
  {https://doi.org/10.1103/PhysRevLett.113.076802} {\bibfield  {journal}
  {\bibinfo  {journal} {Phys. Rev. Lett.}\ }\textbf {\bibinfo {volume} {113}},\
  \bibinfo {pages} {076802} (\bibinfo {year} {2014})}\BibitemShut {NoStop}%
\bibitem [{\citenamefont {Wang}\ \emph {et~al.}(2015)\citenamefont {Wang},
  \citenamefont {Marie}, \citenamefont {Gerber}, \citenamefont {Amand},
  \citenamefont {Lagarde}, \citenamefont {Bouet}, \citenamefont {Vidal},
  \citenamefont {Balocchi},\ and\ \citenamefont {Urbaszek}}]{wang2015}%
  \BibitemOpen
  \bibfield  {author} {\bibinfo {author} {\bibfnamefont {G.}~\bibnamefont
  {Wang}}, \bibinfo {author} {\bibfnamefont {X.}~\bibnamefont {Marie}},
  \bibinfo {author} {\bibfnamefont {I.}~\bibnamefont {Gerber}}, \bibinfo
  {author} {\bibfnamefont {T.}~\bibnamefont {Amand}}, \bibinfo {author}
  {\bibfnamefont {D.}~\bibnamefont {Lagarde}}, \bibinfo {author} {\bibfnamefont
  {L.}~\bibnamefont {Bouet}}, \bibinfo {author} {\bibfnamefont
  {M.}~\bibnamefont {Vidal}}, \bibinfo {author} {\bibfnamefont
  {A.}~\bibnamefont {Balocchi}},\ and\ \bibinfo {author} {\bibfnamefont
  {B.}~\bibnamefont {Urbaszek}},\ }\href
  {https://doi.org/10.1103/PhysRevLett.114.097403} {\bibfield  {journal}
  {\bibinfo  {journal} {Phys. Rev. Lett.}\ }\textbf {\bibinfo {volume} {114}},\
  \bibinfo {pages} {097403} (\bibinfo {year} {2015})}\BibitemShut {NoStop}%
\bibitem [{\citenamefont {Jakubczyk}\ \emph {et~al.}(2016)\citenamefont
  {Jakubczyk}, \citenamefont {Delmonte}, \citenamefont {Koperski},
  \citenamefont {Nogajewski}, \citenamefont {Faugeras}, \citenamefont
  {Langbein}, \citenamefont {Potemski},\ and\ \citenamefont
  {Kasprzak}}]{jakubczyk2016}%
  \BibitemOpen
  \bibfield  {author} {\bibinfo {author} {\bibfnamefont {T.}~\bibnamefont
  {Jakubczyk}}, \bibinfo {author} {\bibfnamefont {V.}~\bibnamefont {Delmonte}},
  \bibinfo {author} {\bibfnamefont {M.}~\bibnamefont {Koperski}}, \bibinfo
  {author} {\bibfnamefont {K.}~\bibnamefont {Nogajewski}}, \bibinfo {author}
  {\bibfnamefont {C.}~\bibnamefont {Faugeras}}, \bibinfo {author}
  {\bibfnamefont {W.}~\bibnamefont {Langbein}}, \bibinfo {author}
  {\bibfnamefont {M.}~\bibnamefont {Potemski}},\ and\ \bibinfo {author}
  {\bibfnamefont {J.}~\bibnamefont {Kasprzak}},\ }\href
  {https://doi.org/10.1021/acs.nanolett.6b01060} {\bibfield  {journal}
  {\bibinfo  {journal} {Nano Lett.}\ }\textbf {\bibinfo {volume} {16}},\
  \bibinfo {pages} {5333} (\bibinfo {year} {2016})}\BibitemShut {NoStop}%
\bibitem [{\citenamefont {Kasprzak}\ \emph {et~al.}(2006)\citenamefont
  {Kasprzak}, \citenamefont {Richard}, \citenamefont {Kundermann},
  \citenamefont {Baas}, \citenamefont {Jeambrun}, \citenamefont {Keeling},
  \citenamefont {Marchetti}, \citenamefont {Szymańska}, \citenamefont
  {André}, \citenamefont {Staehli}, \citenamefont {Savona}, \citenamefont
  {Littlewood}, \citenamefont {Deveaud},\ and\ \citenamefont
  {Dang}}]{kasprzak2006}%
  \BibitemOpen
  \bibfield  {author} {\bibinfo {author} {\bibfnamefont {J.}~\bibnamefont
  {Kasprzak}}, \bibinfo {author} {\bibfnamefont {M.}~\bibnamefont {Richard}},
  \bibinfo {author} {\bibfnamefont {S.}~\bibnamefont {Kundermann}}, \bibinfo
  {author} {\bibfnamefont {A.}~\bibnamefont {Baas}}, \bibinfo {author}
  {\bibfnamefont {P.}~\bibnamefont {Jeambrun}}, \bibinfo {author}
  {\bibfnamefont {J.~M.~J.}\ \bibnamefont {Keeling}}, \bibinfo {author}
  {\bibfnamefont {F.~M.}\ \bibnamefont {Marchetti}}, \bibinfo {author}
  {\bibfnamefont {M.~H.}\ \bibnamefont {Szymańska}}, \bibinfo {author}
  {\bibfnamefont {R.}~\bibnamefont {André}}, \bibinfo {author} {\bibfnamefont
  {J.~L.}\ \bibnamefont {Staehli}}, \bibinfo {author} {\bibfnamefont
  {V.}~\bibnamefont {Savona}}, \bibinfo {author} {\bibfnamefont {P.~B.}\
  \bibnamefont {Littlewood}}, \bibinfo {author} {\bibfnamefont
  {B.}~\bibnamefont {Deveaud}},\ and\ \bibinfo {author} {\bibfnamefont {L.~S.}\
  \bibnamefont {Dang}},\ }\href {https://doi.org/10.1038/nature05131}
  {\bibfield  {journal} {\bibinfo  {journal} {Nature}\ }\textbf {\bibinfo
  {volume} {443}},\ \bibinfo {pages} {409} (\bibinfo {year}
  {2006})}\BibitemShut {NoStop}%
\bibitem [{\citenamefont {Delteil}\ \emph {et~al.}(2019)\citenamefont
  {Delteil}, \citenamefont {Fink}, \citenamefont {Schade}, \citenamefont
  {Höfling}, \citenamefont {Schneider},\ and\ \citenamefont
  {İmamoğlu}}]{delteil2019}%
  \BibitemOpen
  \bibfield  {author} {\bibinfo {author} {\bibfnamefont {A.}~\bibnamefont
  {Delteil}}, \bibinfo {author} {\bibfnamefont {T.}~\bibnamefont {Fink}},
  \bibinfo {author} {\bibfnamefont {A.}~\bibnamefont {Schade}}, \bibinfo
  {author} {\bibfnamefont {S.}~\bibnamefont {Höfling}}, \bibinfo {author}
  {\bibfnamefont {C.}~\bibnamefont {Schneider}},\ and\ \bibinfo {author}
  {\bibfnamefont {A.}~\bibnamefont {İmamoğlu}},\ }\href
  {https://doi.org/10.1038/s41563-019-0282-y} {\bibfield  {journal} {\bibinfo
  {journal} {Nat. Mater.}\ }\textbf {\bibinfo {volume} {18}},\ \bibinfo {pages}
  {219} (\bibinfo {year} {2019})}\BibitemShut {NoStop}%
\bibitem [{\citenamefont {Muñoz-Matutano}\ \emph {et~al.}(2019)\citenamefont
  {Muñoz-Matutano}, \citenamefont {Wood}, \citenamefont {Johnsson},
  \citenamefont {Vidal}, \citenamefont {Baragiola}, \citenamefont {Reinhard},
  \citenamefont {Lemaître}, \citenamefont {Bloch}, \citenamefont {Amo},
  \citenamefont {Nogues}, \citenamefont {Besga}, \citenamefont {Richard},\ and\
  \citenamefont {Volz}}]{munoz-matutano2019}%
  \BibitemOpen
  \bibfield  {author} {\bibinfo {author} {\bibfnamefont {G.}~\bibnamefont
  {Muñoz-Matutano}}, \bibinfo {author} {\bibfnamefont {A.}~\bibnamefont
  {Wood}}, \bibinfo {author} {\bibfnamefont {M.}~\bibnamefont {Johnsson}},
  \bibinfo {author} {\bibfnamefont {X.}~\bibnamefont {Vidal}}, \bibinfo
  {author} {\bibfnamefont {B.~Q.}\ \bibnamefont {Baragiola}}, \bibinfo {author}
  {\bibfnamefont {A.}~\bibnamefont {Reinhard}}, \bibinfo {author}
  {\bibfnamefont {A.}~\bibnamefont {Lemaître}}, \bibinfo {author}
  {\bibfnamefont {J.}~\bibnamefont {Bloch}}, \bibinfo {author} {\bibfnamefont
  {A.}~\bibnamefont {Amo}}, \bibinfo {author} {\bibfnamefont {G.}~\bibnamefont
  {Nogues}}, \bibinfo {author} {\bibfnamefont {B.}~\bibnamefont {Besga}},
  \bibinfo {author} {\bibfnamefont {M.}~\bibnamefont {Richard}},\ and\ \bibinfo
  {author} {\bibfnamefont {T.}~\bibnamefont {Volz}},\ }\href
  {https://doi.org/10.1038/s41563-019-0281-z} {\bibfield  {journal} {\bibinfo
  {journal} {Nat. Mater.}\ }\textbf {\bibinfo {volume} {18}},\ \bibinfo {pages}
  {213} (\bibinfo {year} {2019})}\BibitemShut {NoStop}%
\bibitem [{\citenamefont {Ciuti}\ \emph {et~al.}(1998)\citenamefont {Ciuti},
  \citenamefont {Savona}, \citenamefont {Piermarocchi}, \citenamefont
  {Quattropani},\ and\ \citenamefont {Schwendimann}}]{ciuti1998}%
  \BibitemOpen
  \bibfield  {author} {\bibinfo {author} {\bibfnamefont {C.}~\bibnamefont
  {Ciuti}}, \bibinfo {author} {\bibfnamefont {V.}~\bibnamefont {Savona}},
  \bibinfo {author} {\bibfnamefont {C.}~\bibnamefont {Piermarocchi}}, \bibinfo
  {author} {\bibfnamefont {A.}~\bibnamefont {Quattropani}},\ and\ \bibinfo
  {author} {\bibfnamefont {P.}~\bibnamefont {Schwendimann}},\ }\href
  {https://doi.org/10.1103/PhysRevB.58.7926} {\bibfield  {journal} {\bibinfo
  {journal} {Phys. Rev. B}\ }\textbf {\bibinfo {volume} {58}},\ \bibinfo
  {pages} {7926} (\bibinfo {year} {1998})}\BibitemShut {NoStop}%
\bibitem [{\citenamefont {Shahnazaryan}\ \emph {et~al.}(2017)\citenamefont
  {Shahnazaryan}, \citenamefont {Iorsh}, \citenamefont {Shelykh},\ and\
  \citenamefont {Kyriienko}}]{shahnazaryan2017}%
  \BibitemOpen
  \bibfield  {author} {\bibinfo {author} {\bibfnamefont {V.}~\bibnamefont
  {Shahnazaryan}}, \bibinfo {author} {\bibfnamefont {I.}~\bibnamefont {Iorsh}},
  \bibinfo {author} {\bibfnamefont {I.~A.}\ \bibnamefont {Shelykh}},\ and\
  \bibinfo {author} {\bibfnamefont {O.}~\bibnamefont {Kyriienko}},\ }\href
  {https://doi.org/10.1103/PhysRevB.96.115409} {\bibfield  {journal} {\bibinfo
  {journal} {Phys. Rev. B}\ }\textbf {\bibinfo {volume} {96}},\ \bibinfo
  {pages} {115409} (\bibinfo {year} {2017})}\BibitemShut {NoStop}%
\bibitem [{\citenamefont {Erkensten}\ \emph {et~al.}(2021)\citenamefont
  {Erkensten}, \citenamefont {Brem},\ and\ \citenamefont
  {Malic}}]{erkensten2021}%
  \BibitemOpen
  \bibfield  {author} {\bibinfo {author} {\bibfnamefont {D.}~\bibnamefont
  {Erkensten}}, \bibinfo {author} {\bibfnamefont {S.}~\bibnamefont {Brem}},\
  and\ \bibinfo {author} {\bibfnamefont {E.}~\bibnamefont {Malic}},\ }\href
  {https://doi.org/10.1103/PhysRevB.103.045426} {\bibfield  {journal} {\bibinfo
   {journal} {Phys. Rev. B}\ }\textbf {\bibinfo {volume} {103}},\ \bibinfo
  {pages} {045426} (\bibinfo {year} {2021})}\BibitemShut {NoStop}%
\bibitem [{\citenamefont {Ferreira}\ \emph {et~al.}(1990)\citenamefont
  {Ferreira}, \citenamefont {Delalande}, \citenamefont {Liu}, \citenamefont
  {Bastard}, \citenamefont {Etienne},\ and\ \citenamefont
  {Palmier}}]{ferreira1990}%
  \BibitemOpen
  \bibfield  {author} {\bibinfo {author} {\bibfnamefont {R.}~\bibnamefont
  {Ferreira}}, \bibinfo {author} {\bibfnamefont {C.}~\bibnamefont {Delalande}},
  \bibinfo {author} {\bibfnamefont {H.~W.}\ \bibnamefont {Liu}}, \bibinfo
  {author} {\bibfnamefont {G.}~\bibnamefont {Bastard}}, \bibinfo {author}
  {\bibfnamefont {B.}~\bibnamefont {Etienne}},\ and\ \bibinfo {author}
  {\bibfnamefont {J.~F.}\ \bibnamefont {Palmier}},\ }\href
  {https://doi.org/10.1103/PhysRevB.42.9170} {\bibfield  {journal} {\bibinfo
  {journal} {Phys. Rev. B}\ }\textbf {\bibinfo {volume} {42}},\ \bibinfo
  {pages} {9170} (\bibinfo {year} {1990})}\BibitemShut {NoStop}%
\bibitem [{\citenamefont {Fox}\ \emph {et~al.}(1991)\citenamefont {Fox},
  \citenamefont {Miller}, \citenamefont {Livescu}, \citenamefont {Cunningham},\
  and\ \citenamefont {Jan}}]{fox1991}%
  \BibitemOpen
  \bibfield  {author} {\bibinfo {author} {\bibfnamefont {A.~M.}\ \bibnamefont
  {Fox}}, \bibinfo {author} {\bibfnamefont {D.~A.~B.}\ \bibnamefont {Miller}},
  \bibinfo {author} {\bibfnamefont {G.}~\bibnamefont {Livescu}}, \bibinfo
  {author} {\bibfnamefont {J.~E.}\ \bibnamefont {Cunningham}},\ and\ \bibinfo
  {author} {\bibfnamefont {W.~Y.}\ \bibnamefont {Jan}},\ }\href
  {https://doi.org/10.1103/PhysRevB.44.6231} {\bibfield  {journal} {\bibinfo
  {journal} {Phys. Rev. B}\ }\textbf {\bibinfo {volume} {44}},\ \bibinfo
  {pages} {6231} (\bibinfo {year} {1991})}\BibitemShut {NoStop}%
\bibitem [{\citenamefont {Sivalertporn}\ \emph {et~al.}(2012)\citenamefont
  {Sivalertporn}, \citenamefont {Mouchliadis}, \citenamefont {Ivanov},
  \citenamefont {Philp},\ and\ \citenamefont {Muljarov}}]{sivalertporn2012}%
  \BibitemOpen
  \bibfield  {author} {\bibinfo {author} {\bibfnamefont {K.}~\bibnamefont
  {Sivalertporn}}, \bibinfo {author} {\bibfnamefont {L.}~\bibnamefont
  {Mouchliadis}}, \bibinfo {author} {\bibfnamefont {A.~L.}\ \bibnamefont
  {Ivanov}}, \bibinfo {author} {\bibfnamefont {R.}~\bibnamefont {Philp}},\ and\
  \bibinfo {author} {\bibfnamefont {E.~A.}\ \bibnamefont {Muljarov}},\ }\href
  {https://doi.org/10.1103/PhysRevB.85.045207} {\bibfield  {journal} {\bibinfo
  {journal} {Phys. Rev. B}\ }\textbf {\bibinfo {volume} {85}},\ \bibinfo
  {pages} {045207} (\bibinfo {year} {2012})}\BibitemShut {NoStop}%
\bibitem [{\citenamefont {Andreakou}\ \emph {et~al.}(2015)\citenamefont
  {Andreakou}, \citenamefont {Cronenberger}, \citenamefont {Scalbert},
  \citenamefont {Nalitov}, \citenamefont {Gippius}, \citenamefont {Kavokin},
  \citenamefont {Nawrocki}, \citenamefont {Leonard}, \citenamefont {Butov},
  \citenamefont {Campman}, \citenamefont {Gossard},\ and\ \citenamefont
  {Vladimirova}}]{andreakou2015}%
  \BibitemOpen
  \bibfield  {author} {\bibinfo {author} {\bibfnamefont {P.}~\bibnamefont
  {Andreakou}}, \bibinfo {author} {\bibfnamefont {S.}~\bibnamefont
  {Cronenberger}}, \bibinfo {author} {\bibfnamefont {D.}~\bibnamefont
  {Scalbert}}, \bibinfo {author} {\bibfnamefont {A.}~\bibnamefont {Nalitov}},
  \bibinfo {author} {\bibfnamefont {N.~A.}\ \bibnamefont {Gippius}}, \bibinfo
  {author} {\bibfnamefont {A.~V.}\ \bibnamefont {Kavokin}}, \bibinfo {author}
  {\bibfnamefont {M.}~\bibnamefont {Nawrocki}}, \bibinfo {author}
  {\bibfnamefont {J.~R.}\ \bibnamefont {Leonard}}, \bibinfo {author}
  {\bibfnamefont {L.~V.}\ \bibnamefont {Butov}}, \bibinfo {author}
  {\bibfnamefont {K.~L.}\ \bibnamefont {Campman}}, \bibinfo {author}
  {\bibfnamefont {A.~C.}\ \bibnamefont {Gossard}},\ and\ \bibinfo {author}
  {\bibfnamefont {M.}~\bibnamefont {Vladimirova}},\ }\href
  {https://doi.org/10.1103/PhysRevB.91.125437} {\bibfield  {journal} {\bibinfo
  {journal} {Phys. Rev. B}\ }\textbf {\bibinfo {volume} {91}},\ \bibinfo
  {pages} {125437} (\bibinfo {year} {2015})}\BibitemShut {NoStop}%
\bibitem [{\citenamefont {Alexeev}\ \emph {et~al.}(2019)\citenamefont
  {Alexeev}, \citenamefont {Ruiz-Tijerina}, \citenamefont {Danovich},
  \citenamefont {Hamer}, \citenamefont {Terry}, \citenamefont {Nayak},
  \citenamefont {Ahn}, \citenamefont {Pak}, \citenamefont {Lee}, \citenamefont
  {Sohn}, \citenamefont {Molas}, \citenamefont {Koperski}, \citenamefont
  {Watanabe}, \citenamefont {Taniguchi}, \citenamefont {Novoselov},
  \citenamefont {Gorbachev}, \citenamefont {Shin}, \citenamefont {Fal’ko},\
  and\ \citenamefont {Tartakovskii}}]{alexeev2019}%
  \BibitemOpen
  \bibfield  {author} {\bibinfo {author} {\bibfnamefont {E.~M.}\ \bibnamefont
  {Alexeev}}, \bibinfo {author} {\bibfnamefont {D.~A.}\ \bibnamefont
  {Ruiz-Tijerina}}, \bibinfo {author} {\bibfnamefont {M.}~\bibnamefont
  {Danovich}}, \bibinfo {author} {\bibfnamefont {M.~J.}\ \bibnamefont {Hamer}},
  \bibinfo {author} {\bibfnamefont {D.~J.}\ \bibnamefont {Terry}}, \bibinfo
  {author} {\bibfnamefont {P.~K.}\ \bibnamefont {Nayak}}, \bibinfo {author}
  {\bibfnamefont {S.}~\bibnamefont {Ahn}}, \bibinfo {author} {\bibfnamefont
  {S.}~\bibnamefont {Pak}}, \bibinfo {author} {\bibfnamefont {J.}~\bibnamefont
  {Lee}}, \bibinfo {author} {\bibfnamefont {J.~I.}\ \bibnamefont {Sohn}},
  \bibinfo {author} {\bibfnamefont {M.~R.}\ \bibnamefont {Molas}}, \bibinfo
  {author} {\bibfnamefont {M.}~\bibnamefont {Koperski}}, \bibinfo {author}
  {\bibfnamefont {K.}~\bibnamefont {Watanabe}}, \bibinfo {author}
  {\bibfnamefont {T.}~\bibnamefont {Taniguchi}}, \bibinfo {author}
  {\bibfnamefont {K.~S.}\ \bibnamefont {Novoselov}}, \bibinfo {author}
  {\bibfnamefont {R.~V.}\ \bibnamefont {Gorbachev}}, \bibinfo {author}
  {\bibfnamefont {H.~S.}\ \bibnamefont {Shin}}, \bibinfo {author}
  {\bibfnamefont {V.~I.}\ \bibnamefont {Fal’ko}},\ and\ \bibinfo {author}
  {\bibfnamefont {A.~I.}\ \bibnamefont {Tartakovskii}},\ }\href
  {https://doi.org/10.1038/s41586-019-0986-9} {\bibfield  {journal} {\bibinfo
  {journal} {Nature}\ }\textbf {\bibinfo {volume} {567}},\ \bibinfo {pages}
  {81} (\bibinfo {year} {2019})}\BibitemShut {NoStop}%
\bibitem [{\citenamefont {Hsu}\ \emph {et~al.}(2019)\citenamefont {Hsu},
  \citenamefont {Lin}, \citenamefont {Lu}, \citenamefont {Lee}, \citenamefont
  {Chu}, \citenamefont {Li}, \citenamefont {Yao}, \citenamefont {Chang},\ and\
  \citenamefont {Shih}}]{hsu2019}%
  \BibitemOpen
  \bibfield  {author} {\bibinfo {author} {\bibfnamefont {W.-T.}\ \bibnamefont
  {Hsu}}, \bibinfo {author} {\bibfnamefont {B.-H.}\ \bibnamefont {Lin}},
  \bibinfo {author} {\bibfnamefont {L.-S.}\ \bibnamefont {Lu}}, \bibinfo
  {author} {\bibfnamefont {M.-H.}\ \bibnamefont {Lee}}, \bibinfo {author}
  {\bibfnamefont {M.-W.}\ \bibnamefont {Chu}}, \bibinfo {author} {\bibfnamefont
  {L.-J.}\ \bibnamefont {Li}}, \bibinfo {author} {\bibfnamefont
  {W.}~\bibnamefont {Yao}}, \bibinfo {author} {\bibfnamefont {W.-H.}\
  \bibnamefont {Chang}},\ and\ \bibinfo {author} {\bibfnamefont {C.-K.}\
  \bibnamefont {Shih}},\ }\href {https://doi.org/10.1126/sciadv.aax7407}
  {\bibfield  {journal} {\bibinfo  {journal} {Sci. Adv.}\ }\textbf {\bibinfo
  {volume} {5}},\ \bibinfo {pages} {eaax7407} (\bibinfo {year}
  {2019})}\BibitemShut {NoStop}%
\bibitem [{\citenamefont {Shimazaki}\ \emph {et~al.}(2020)\citenamefont
  {Shimazaki}, \citenamefont {Schwartz}, \citenamefont {Watanabe},
  \citenamefont {Taniguchi}, \citenamefont {Kroner},\ and\ \citenamefont
  {İmamoğlu}}]{shimazaki2020}%
  \BibitemOpen
  \bibfield  {author} {\bibinfo {author} {\bibfnamefont {Y.}~\bibnamefont
  {Shimazaki}}, \bibinfo {author} {\bibfnamefont {I.}~\bibnamefont {Schwartz}},
  \bibinfo {author} {\bibfnamefont {K.}~\bibnamefont {Watanabe}}, \bibinfo
  {author} {\bibfnamefont {T.}~\bibnamefont {Taniguchi}}, \bibinfo {author}
  {\bibfnamefont {M.}~\bibnamefont {Kroner}},\ and\ \bibinfo {author}
  {\bibfnamefont {A.}~\bibnamefont {İmamoğlu}},\ }\href
  {https://doi.org/10.1038/s41586-020-2191-2} {\bibfield  {journal} {\bibinfo
  {journal} {Nature}\ }\textbf {\bibinfo {volume} {580}},\ \bibinfo {pages}
  {472} (\bibinfo {year} {2020})}\BibitemShut {NoStop}%
\bibitem [{\citenamefont {Sung}\ \emph {et~al.}(2020)\citenamefont {Sung},
  \citenamefont {Zhou}, \citenamefont {Scuri}, \citenamefont {Zólyomi},
  \citenamefont {Andersen}, \citenamefont {Yoo}, \citenamefont {Wild},
  \citenamefont {Joe}, \citenamefont {Gelly}, \citenamefont {Heo},
  \citenamefont {Magorrian}, \citenamefont {Bérubé}, \citenamefont
  {Valdivia}, \citenamefont {Taniguchi}, \citenamefont {Watanabe},
  \citenamefont {Lukin}, \citenamefont {Kim}, \citenamefont {Fal’ko},\ and\
  \citenamefont {Park}}]{sung2020}%
  \BibitemOpen
  \bibfield  {author} {\bibinfo {author} {\bibfnamefont {J.}~\bibnamefont
  {Sung}}, \bibinfo {author} {\bibfnamefont {Y.}~\bibnamefont {Zhou}}, \bibinfo
  {author} {\bibfnamefont {G.}~\bibnamefont {Scuri}}, \bibinfo {author}
  {\bibfnamefont {V.}~\bibnamefont {Zólyomi}}, \bibinfo {author}
  {\bibfnamefont {T.~I.}\ \bibnamefont {Andersen}}, \bibinfo {author}
  {\bibfnamefont {H.}~\bibnamefont {Yoo}}, \bibinfo {author} {\bibfnamefont
  {D.~S.}\ \bibnamefont {Wild}}, \bibinfo {author} {\bibfnamefont {A.~Y.}\
  \bibnamefont {Joe}}, \bibinfo {author} {\bibfnamefont {R.~J.}\ \bibnamefont
  {Gelly}}, \bibinfo {author} {\bibfnamefont {H.}~\bibnamefont {Heo}}, \bibinfo
  {author} {\bibfnamefont {S.~J.}\ \bibnamefont {Magorrian}}, \bibinfo {author}
  {\bibfnamefont {D.}~\bibnamefont {Bérubé}}, \bibinfo {author}
  {\bibfnamefont {A.~M.~M.}\ \bibnamefont {Valdivia}}, \bibinfo {author}
  {\bibfnamefont {T.}~\bibnamefont {Taniguchi}}, \bibinfo {author}
  {\bibfnamefont {K.}~\bibnamefont {Watanabe}}, \bibinfo {author}
  {\bibfnamefont {M.~D.}\ \bibnamefont {Lukin}}, \bibinfo {author}
  {\bibfnamefont {P.}~\bibnamefont {Kim}}, \bibinfo {author} {\bibfnamefont
  {V.~I.}\ \bibnamefont {Fal’ko}},\ and\ \bibinfo {author} {\bibfnamefont
  {H.}~\bibnamefont {Park}},\ }\href
  {https://doi.org/10.1038/s41565-020-0728-z} {\bibfield  {journal} {\bibinfo
  {journal} {Nat. Nanotechnol.}\ }\textbf {\bibinfo {volume} {15}},\ \bibinfo
  {pages} {750} (\bibinfo {year} {2020})}\BibitemShut {NoStop}%
\bibitem [{\citenamefont {Merkl}\ \emph {et~al.}(2020)\citenamefont {Merkl},
  \citenamefont {Mooshammer}, \citenamefont {Brem}, \citenamefont {Girnghuber},
  \citenamefont {Lin}, \citenamefont {Weigl}, \citenamefont {Liebich},
  \citenamefont {Yong}, \citenamefont {Gillen}, \citenamefont {Maultzsch},
  \citenamefont {Lupton}, \citenamefont {Malic},\ and\ \citenamefont
  {Huber}}]{merkl2020}%
  \BibitemOpen
  \bibfield  {author} {\bibinfo {author} {\bibfnamefont {P.}~\bibnamefont
  {Merkl}}, \bibinfo {author} {\bibfnamefont {F.}~\bibnamefont {Mooshammer}},
  \bibinfo {author} {\bibfnamefont {S.}~\bibnamefont {Brem}}, \bibinfo {author}
  {\bibfnamefont {A.}~\bibnamefont {Girnghuber}}, \bibinfo {author}
  {\bibfnamefont {K.-Q.}\ \bibnamefont {Lin}}, \bibinfo {author} {\bibfnamefont
  {L.}~\bibnamefont {Weigl}}, \bibinfo {author} {\bibfnamefont
  {M.}~\bibnamefont {Liebich}}, \bibinfo {author} {\bibfnamefont {C.-K.}\
  \bibnamefont {Yong}}, \bibinfo {author} {\bibfnamefont {R.}~\bibnamefont
  {Gillen}}, \bibinfo {author} {\bibfnamefont {J.}~\bibnamefont {Maultzsch}},
  \bibinfo {author} {\bibfnamefont {J.~M.}\ \bibnamefont {Lupton}}, \bibinfo
  {author} {\bibfnamefont {E.}~\bibnamefont {Malic}},\ and\ \bibinfo {author}
  {\bibfnamefont {R.}~\bibnamefont {Huber}},\ }\href
  {https://doi.org/10.1038/s41467-020-16069-z} {\bibfield  {journal} {\bibinfo
  {journal} {Nat. Commun.}\ }\textbf {\bibinfo {volume} {11}},\ \bibinfo
  {pages} {2167} (\bibinfo {year} {2020})}\BibitemShut {NoStop}%
\bibitem [{\citenamefont {Zhang}\ \emph {et~al.}(2020)\citenamefont {Zhang},
  \citenamefont {Zhang}, \citenamefont {Wu}, \citenamefont {Wang},
  \citenamefont {Gogna}, \citenamefont {Hou}, \citenamefont {Watanabe},
  \citenamefont {Taniguchi}, \citenamefont {Kulkarni}, \citenamefont {Kuo},
  \citenamefont {Forrest},\ and\ \citenamefont {Deng}}]{zhang2020}%
  \BibitemOpen
  \bibfield  {author} {\bibinfo {author} {\bibfnamefont {L.}~\bibnamefont
  {Zhang}}, \bibinfo {author} {\bibfnamefont {Z.}~\bibnamefont {Zhang}},
  \bibinfo {author} {\bibfnamefont {F.}~\bibnamefont {Wu}}, \bibinfo {author}
  {\bibfnamefont {D.}~\bibnamefont {Wang}}, \bibinfo {author} {\bibfnamefont
  {R.}~\bibnamefont {Gogna}}, \bibinfo {author} {\bibfnamefont
  {S.}~\bibnamefont {Hou}}, \bibinfo {author} {\bibfnamefont {K.}~\bibnamefont
  {Watanabe}}, \bibinfo {author} {\bibfnamefont {T.}~\bibnamefont {Taniguchi}},
  \bibinfo {author} {\bibfnamefont {K.}~\bibnamefont {Kulkarni}}, \bibinfo
  {author} {\bibfnamefont {T.}~\bibnamefont {Kuo}}, \bibinfo {author}
  {\bibfnamefont {S.~R.}\ \bibnamefont {Forrest}},\ and\ \bibinfo {author}
  {\bibfnamefont {H.}~\bibnamefont {Deng}},\ }\href
  {https://doi.org/10.1038/s41467-020-19466-6} {\bibfield  {journal} {\bibinfo
  {journal} {Nat. Commun.}\ }\textbf {\bibinfo {volume} {11}},\ \bibinfo
  {pages} {5888} (\bibinfo {year} {2020})}\BibitemShut {NoStop}%
\bibitem [{\citenamefont {Tang}\ \emph {et~al.}(2021)\citenamefont {Tang},
  \citenamefont {Gu}, \citenamefont {Liu}, \citenamefont {Watanabe},
  \citenamefont {Taniguchi}, \citenamefont {Hone}, \citenamefont {Mak},\ and\
  \citenamefont {Shan}}]{tang2020}%
  \BibitemOpen
  \bibfield  {author} {\bibinfo {author} {\bibfnamefont {Y.}~\bibnamefont
  {Tang}}, \bibinfo {author} {\bibfnamefont {J.}~\bibnamefont {Gu}}, \bibinfo
  {author} {\bibfnamefont {S.}~\bibnamefont {Liu}}, \bibinfo {author}
  {\bibfnamefont {K.}~\bibnamefont {Watanabe}}, \bibinfo {author}
  {\bibfnamefont {T.}~\bibnamefont {Taniguchi}}, \bibinfo {author}
  {\bibfnamefont {J.}~\bibnamefont {Hone}}, \bibinfo {author} {\bibfnamefont
  {K.~F.}\ \bibnamefont {Mak}},\ and\ \bibinfo {author} {\bibfnamefont
  {J.}~\bibnamefont {Shan}},\ }\href
  {https://doi.org/10.1038/s41565-020-00783-2} {\bibfield  {journal} {\bibinfo
  {journal} {Nat. Nanotechnol.}\ }\textbf {\bibinfo {volume} {16}},\ \bibinfo
  {pages} {52} (\bibinfo {year} {2021})}\BibitemShut {NoStop}%
\bibitem [{\citenamefont {McDonnell}\ \emph {et~al.}(2021)\citenamefont
  {McDonnell}, \citenamefont {Viner}, \citenamefont {Ruiz-Tijerina},
  \citenamefont {Rivera}, \citenamefont {Xu}, \citenamefont {Fal'ko},\ and\
  \citenamefont {Smith}}]{mcdonnell2021}%
  \BibitemOpen
  \bibfield  {author} {\bibinfo {author} {\bibfnamefont {L.~P.}\ \bibnamefont
  {McDonnell}}, \bibinfo {author} {\bibfnamefont {J.~J.~S.}\ \bibnamefont
  {Viner}}, \bibinfo {author} {\bibfnamefont {D.~A.}\ \bibnamefont
  {Ruiz-Tijerina}}, \bibinfo {author} {\bibfnamefont {P.}~\bibnamefont
  {Rivera}}, \bibinfo {author} {\bibfnamefont {X.}~\bibnamefont {Xu}}, \bibinfo
  {author} {\bibfnamefont {V.~I.}\ \bibnamefont {Fal'ko}},\ and\ \bibinfo
  {author} {\bibfnamefont {D.~C.}\ \bibnamefont {Smith}},\ }\href
  {https://doi.org/10.1088/2053-1583/abe778} {\bibfield  {journal} {\bibinfo
  {journal} {2D Mater.}\ }\textbf {\bibinfo {volume} {8}},\ \bibinfo {pages}
  {035009} (\bibinfo {year} {2021})}\BibitemShut {NoStop}%
\bibitem [{\citenamefont {Roch}\ \emph {et~al.}(2019)\citenamefont {Roch},
  \citenamefont {Froehlicher}, \citenamefont {Leisgang}, \citenamefont {Makk},
  \citenamefont {Watanabe}, \citenamefont {Taniguchi},\ and\ \citenamefont
  {Warburton}}]{roch2019}%
  \BibitemOpen
  \bibfield  {author} {\bibinfo {author} {\bibfnamefont {J.~G.}\ \bibnamefont
  {Roch}}, \bibinfo {author} {\bibfnamefont {G.}~\bibnamefont {Froehlicher}},
  \bibinfo {author} {\bibfnamefont {N.}~\bibnamefont {Leisgang}}, \bibinfo
  {author} {\bibfnamefont {P.}~\bibnamefont {Makk}}, \bibinfo {author}
  {\bibfnamefont {K.}~\bibnamefont {Watanabe}}, \bibinfo {author}
  {\bibfnamefont {T.}~\bibnamefont {Taniguchi}},\ and\ \bibinfo {author}
  {\bibfnamefont {R.~J.}\ \bibnamefont {Warburton}},\ }\href
  {https://doi.org/10.1038/s41565-019-0397-y} {\bibfield  {journal} {\bibinfo
  {journal} {Nat. Nanotechnol.}\ }\textbf {\bibinfo {volume} {14}},\ \bibinfo
  {pages} {432} (\bibinfo {year} {2019})}\BibitemShut {NoStop}%
\bibitem [{\citenamefont {Gerber}\ \emph {et~al.}(2019)\citenamefont {Gerber},
  \citenamefont {Courtade}, \citenamefont {Shree}, \citenamefont {Robert},
  \citenamefont {Taniguchi}, \citenamefont {Watanabe}, \citenamefont
  {Balocchi}, \citenamefont {Renucci}, \citenamefont {Lagarde}, \citenamefont
  {Marie},\ and\ \citenamefont {Urbaszek}}]{gerber2019}%
  \BibitemOpen
  \bibfield  {author} {\bibinfo {author} {\bibfnamefont {I.~C.}\ \bibnamefont
  {Gerber}}, \bibinfo {author} {\bibfnamefont {E.}~\bibnamefont {Courtade}},
  \bibinfo {author} {\bibfnamefont {S.}~\bibnamefont {Shree}}, \bibinfo
  {author} {\bibfnamefont {C.}~\bibnamefont {Robert}}, \bibinfo {author}
  {\bibfnamefont {T.}~\bibnamefont {Taniguchi}}, \bibinfo {author}
  {\bibfnamefont {K.}~\bibnamefont {Watanabe}}, \bibinfo {author}
  {\bibfnamefont {A.}~\bibnamefont {Balocchi}}, \bibinfo {author}
  {\bibfnamefont {P.}~\bibnamefont {Renucci}}, \bibinfo {author} {\bibfnamefont
  {D.}~\bibnamefont {Lagarde}}, \bibinfo {author} {\bibfnamefont
  {X.}~\bibnamefont {Marie}},\ and\ \bibinfo {author} {\bibfnamefont
  {B.}~\bibnamefont {Urbaszek}},\ }\href
  {https://doi.org/10.1103/PhysRevB.99.035443} {\bibfield  {journal} {\bibinfo
  {journal} {Phys. Rev. B}\ }\textbf {\bibinfo {volume} {99}},\ \bibinfo
  {pages} {035443} (\bibinfo {year} {2019})}\BibitemShut {NoStop}%
\bibitem [{\citenamefont {Pisoni}\ \emph {et~al.}(2019)\citenamefont {Pisoni},
  \citenamefont {Davatz}, \citenamefont {Watanabe}, \citenamefont {Taniguchi},
  \citenamefont {Ihn},\ and\ \citenamefont {Ensslin}}]{pisoni2019}%
  \BibitemOpen
  \bibfield  {author} {\bibinfo {author} {\bibfnamefont {R.}~\bibnamefont
  {Pisoni}}, \bibinfo {author} {\bibfnamefont {T.}~\bibnamefont {Davatz}},
  \bibinfo {author} {\bibfnamefont {K.}~\bibnamefont {Watanabe}}, \bibinfo
  {author} {\bibfnamefont {T.}~\bibnamefont {Taniguchi}}, \bibinfo {author}
  {\bibfnamefont {T.}~\bibnamefont {Ihn}},\ and\ \bibinfo {author}
  {\bibfnamefont {K.}~\bibnamefont {Ensslin}},\ }\href
  {https://doi.org/10.1103/PhysRevLett.123.117702} {\bibfield  {journal}
  {\bibinfo  {journal} {Phys. Rev. Lett.}\ }\textbf {\bibinfo {volume} {123}},\
  \bibinfo {pages} {117702} (\bibinfo {year} {2019})}\BibitemShut {NoStop}%
\bibitem [{\citenamefont {Leisgang}\ \emph {et~al.}(2020)\citenamefont
  {Leisgang}, \citenamefont {Shree}, \citenamefont {Paradisanos}, \citenamefont
  {Sponfeldner}, \citenamefont {Robert}, \citenamefont {Lagarde}, \citenamefont
  {Balocchi}, \citenamefont {Watanabe}, \citenamefont {Taniguchi},
  \citenamefont {Marie}, \citenamefont {Warburton}, \citenamefont {Gerber},\
  and\ \citenamefont {Urbaszek}}]{leisgang2020}%
  \BibitemOpen
  \bibfield  {author} {\bibinfo {author} {\bibfnamefont {N.}~\bibnamefont
  {Leisgang}}, \bibinfo {author} {\bibfnamefont {S.}~\bibnamefont {Shree}},
  \bibinfo {author} {\bibfnamefont {I.}~\bibnamefont {Paradisanos}}, \bibinfo
  {author} {\bibfnamefont {L.}~\bibnamefont {Sponfeldner}}, \bibinfo {author}
  {\bibfnamefont {C.}~\bibnamefont {Robert}}, \bibinfo {author} {\bibfnamefont
  {D.}~\bibnamefont {Lagarde}}, \bibinfo {author} {\bibfnamefont
  {A.}~\bibnamefont {Balocchi}}, \bibinfo {author} {\bibfnamefont
  {K.}~\bibnamefont {Watanabe}}, \bibinfo {author} {\bibfnamefont
  {T.}~\bibnamefont {Taniguchi}}, \bibinfo {author} {\bibfnamefont
  {X.}~\bibnamefont {Marie}}, \bibinfo {author} {\bibfnamefont {R.~J.}\
  \bibnamefont {Warburton}}, \bibinfo {author} {\bibfnamefont {I.~C.}\
  \bibnamefont {Gerber}},\ and\ \bibinfo {author} {\bibfnamefont
  {B.}~\bibnamefont {Urbaszek}},\ }\href
  {https://doi.org/10.1038/s41565-020-0750-1} {\bibfield  {journal} {\bibinfo
  {journal} {Nat. Nanotechnol.}\ }\textbf {\bibinfo {volume} {15}},\ \bibinfo
  {pages} {901} (\bibinfo {year} {2020})}\BibitemShut {NoStop}%
\bibitem [{\citenamefont {Lorchat}\ \emph {et~al.}(2021)\citenamefont
  {Lorchat}, \citenamefont {Selig}, \citenamefont {Katsch}, \citenamefont
  {Yumigeta}, \citenamefont {Tongay}, \citenamefont {Knorr}, \citenamefont
  {Schneider},\ and\ \citenamefont {Höfling}}]{lorchat2020}%
  \BibitemOpen
  \bibfield  {author} {\bibinfo {author} {\bibfnamefont {E.}~\bibnamefont
  {Lorchat}}, \bibinfo {author} {\bibfnamefont {M.}~\bibnamefont {Selig}},
  \bibinfo {author} {\bibfnamefont {F.}~\bibnamefont {Katsch}}, \bibinfo
  {author} {\bibfnamefont {K.}~\bibnamefont {Yumigeta}}, \bibinfo {author}
  {\bibfnamefont {S.}~\bibnamefont {Tongay}}, \bibinfo {author} {\bibfnamefont
  {A.}~\bibnamefont {Knorr}}, \bibinfo {author} {\bibfnamefont
  {C.}~\bibnamefont {Schneider}},\ and\ \bibinfo {author} {\bibfnamefont
  {S.}~\bibnamefont {Höfling}},\ }\href
  {https://doi.org/10.1103/PhysRevLett.126.037401} {\bibfield  {journal}
  {\bibinfo  {journal} {Phys. Rev. Lett.}\ }\textbf {\bibinfo {volume} {126}},\
  \bibinfo {pages} {037401} (\bibinfo {year} {2021})}\BibitemShut {NoStop}%
\bibitem [{\citenamefont {Peimyoo}\ \emph {et~al.}(2021)\citenamefont
  {Peimyoo}, \citenamefont {Deilmann}, \citenamefont {Withers}, \citenamefont
  {Escolar}, \citenamefont {Nutting}, \citenamefont {Taniguchi}, \citenamefont
  {Watanabe}, \citenamefont {Taghizadeh}, \citenamefont {Craciun},
  \citenamefont {Thygesen},\ and\ \citenamefont {Russo}}]{peimyoo2021}%
  \BibitemOpen
  \bibfield  {author} {\bibinfo {author} {\bibfnamefont {N.}~\bibnamefont
  {Peimyoo}}, \bibinfo {author} {\bibfnamefont {T.}~\bibnamefont {Deilmann}},
  \bibinfo {author} {\bibfnamefont {F.}~\bibnamefont {Withers}}, \bibinfo
  {author} {\bibfnamefont {J.}~\bibnamefont {Escolar}}, \bibinfo {author}
  {\bibfnamefont {D.}~\bibnamefont {Nutting}}, \bibinfo {author} {\bibfnamefont
  {T.}~\bibnamefont {Taniguchi}}, \bibinfo {author} {\bibfnamefont
  {K.}~\bibnamefont {Watanabe}}, \bibinfo {author} {\bibfnamefont
  {A.}~\bibnamefont {Taghizadeh}}, \bibinfo {author} {\bibfnamefont {M.~F.}\
  \bibnamefont {Craciun}}, \bibinfo {author} {\bibfnamefont {K.~S.}\
  \bibnamefont {Thygesen}},\ and\ \bibinfo {author} {\bibfnamefont
  {S.}~\bibnamefont {Russo}},\ }\bibfield  {journal} {\bibinfo  {journal} {Nat.
  Nanotechnol.}\ }\href {https://doi.org/10.1038/s41565-021-00916-1}
  {10.1038/s41565-021-00916-1} (\bibinfo {year} {2021})\BibitemShut {NoStop}%
\bibitem [{\citenamefont {Lindberg}\ and\ \citenamefont
  {Koch}(1988)}]{lindberg1988}%
  \BibitemOpen
  \bibfield  {author} {\bibinfo {author} {\bibfnamefont {M.}~\bibnamefont
  {Lindberg}}\ and\ \bibinfo {author} {\bibfnamefont {S.~W.}\ \bibnamefont
  {Koch}},\ }\href {https://doi.org/10.1103/PhysRevB.38.3342} {\bibfield
  {journal} {\bibinfo  {journal} {Phys. Rev. B}\ }\textbf {\bibinfo {volume}
  {38}},\ \bibinfo {pages} {3342} (\bibinfo {year} {1988})}\BibitemShut
  {NoStop}%
\bibitem [{\citenamefont {Koch}\ \emph {et~al.}(2006)\citenamefont {Koch},
  \citenamefont {Kira}, \citenamefont {Khitrova},\ and\ \citenamefont
  {Gibbs}}]{koch2006}%
  \BibitemOpen
  \bibfield  {author} {\bibinfo {author} {\bibfnamefont {S.~W.}\ \bibnamefont
  {Koch}}, \bibinfo {author} {\bibfnamefont {M.}~\bibnamefont {Kira}}, \bibinfo
  {author} {\bibfnamefont {G.}~\bibnamefont {Khitrova}},\ and\ \bibinfo
  {author} {\bibfnamefont {H.~M.}\ \bibnamefont {Gibbs}},\ }\href
  {https://doi.org/10.1038/nmat1658} {\bibfield  {journal} {\bibinfo  {journal}
  {Nat. Mater.}\ }\textbf {\bibinfo {volume} {5}},\ \bibinfo {pages} {523}
  (\bibinfo {year} {2006})}\BibitemShut {NoStop}%
\bibitem [{\citenamefont {Haug}\ and\ \citenamefont {Koch}(2009)}]{haug2009}%
  \BibitemOpen
  \bibfield  {author} {\bibinfo {author} {\bibfnamefont {H.}~\bibnamefont
  {Haug}}\ and\ \bibinfo {author} {\bibfnamefont {S.~W.}\ \bibnamefont
  {Koch}},\ }\href {https://doi.org/10.1142/7184} {\emph {\bibinfo {title}
  {Quantum {Theory} of the {Optical} and {Electronic} {Properties} of
  {Semiconductors}}}}\ (\bibinfo  {publisher} {World Scientific},\ \bibinfo
  {year} {2009})\BibitemShut {NoStop}%
\bibitem [{\citenamefont {Klingshirn}(2012)}]{klingshirn2012}%
  \BibitemOpen
  \bibfield  {author} {\bibinfo {author} {\bibfnamefont {C.~F.}\ \bibnamefont
  {Klingshirn}},\ }\href {https://doi.org/10.1007/978-3-642-28362-8} {\emph
  {\bibinfo {title} {Semiconductor {Optics}}}}\ (\bibinfo  {publisher}
  {Springer},\ \bibinfo {address} {Berlin, Heidelberg},\ \bibinfo {year}
  {2012})\BibitemShut {NoStop}%
\bibitem [{\citenamefont {Karrai}\ and\ \citenamefont
  {J.~Warburton}(2003)}]{karrai2003}%
  \BibitemOpen
  \bibfield  {author} {\bibinfo {author} {\bibfnamefont {K.}~\bibnamefont
  {Karrai}}\ and\ \bibinfo {author} {\bibfnamefont {R.}~\bibnamefont
  {J.~Warburton}},\ }\href {https://doi.org/10.1016/j.spmi.2004.02.007}
  {\bibfield  {journal} {\bibinfo  {journal} {Superlattices and Microstruct.}\
  }\textbf {\bibinfo {volume} {33}},\ \bibinfo {pages} {311} (\bibinfo {year}
  {2003})}\BibitemShut {NoStop}%
\bibitem [{\citenamefont {Deilmann}\ and\ \citenamefont
  {Thygesen}(2018)}]{deilmann2018}%
  \BibitemOpen
  \bibfield  {author} {\bibinfo {author} {\bibfnamefont {T.}~\bibnamefont
  {Deilmann}}\ and\ \bibinfo {author} {\bibfnamefont {K.~S.}\ \bibnamefont
  {Thygesen}},\ }\href {https://doi.org/10.1021/acs.nanolett.8b00438}
  {\bibfield  {journal} {\bibinfo  {journal} {Nano Lett.}\ }\textbf {\bibinfo
  {volume} {18}},\ \bibinfo {pages} {2984} (\bibinfo {year}
  {2018})}\BibitemShut {NoStop}%
\bibitem [{\citenamefont {Miller}\ \emph {et~al.}(1984)\citenamefont {Miller},
  \citenamefont {Chemla}, \citenamefont {Damen}, \citenamefont {Gossard},
  \citenamefont {Wiegmann}, \citenamefont {Wood},\ and\ \citenamefont
  {Burrus}}]{miller1984}%
  \BibitemOpen
  \bibfield  {author} {\bibinfo {author} {\bibfnamefont {D.~A.~B.}\
  \bibnamefont {Miller}}, \bibinfo {author} {\bibfnamefont {D.~S.}\
  \bibnamefont {Chemla}}, \bibinfo {author} {\bibfnamefont {T.~C.}\
  \bibnamefont {Damen}}, \bibinfo {author} {\bibfnamefont {A.~C.}\ \bibnamefont
  {Gossard}}, \bibinfo {author} {\bibfnamefont {W.}~\bibnamefont {Wiegmann}},
  \bibinfo {author} {\bibfnamefont {T.~H.}\ \bibnamefont {Wood}},\ and\
  \bibinfo {author} {\bibfnamefont {C.~A.}\ \bibnamefont {Burrus}},\ }\href
  {https://doi.org/10.1103/PhysRevLett.53.2173} {\bibfield  {journal} {\bibinfo
   {journal} {Phys. Rev. Lett.}\ }\textbf {\bibinfo {volume} {53}},\ \bibinfo
  {pages} {2173} (\bibinfo {year} {1984})}\BibitemShut {NoStop}%
\bibitem [{\citenamefont {Guo}\ \emph {et~al.}(2019)\citenamefont {Guo},
  \citenamefont {Wu}, \citenamefont {Cao}, \citenamefont {Monahan},
  \citenamefont {Lee}, \citenamefont {Louie},\ and\ \citenamefont
  {Fleming}}]{guo2019}%
  \BibitemOpen
  \bibfield  {author} {\bibinfo {author} {\bibfnamefont {L.}~\bibnamefont
  {Guo}}, \bibinfo {author} {\bibfnamefont {M.}~\bibnamefont {Wu}}, \bibinfo
  {author} {\bibfnamefont {T.}~\bibnamefont {Cao}}, \bibinfo {author}
  {\bibfnamefont {D.~M.}\ \bibnamefont {Monahan}}, \bibinfo {author}
  {\bibfnamefont {Y.-H.}\ \bibnamefont {Lee}}, \bibinfo {author} {\bibfnamefont
  {S.~G.}\ \bibnamefont {Louie}},\ and\ \bibinfo {author} {\bibfnamefont
  {G.~R.}\ \bibnamefont {Fleming}},\ }\href
  {https://doi.org/10.1038/s41567-018-0362-y} {\bibfield  {journal} {\bibinfo
  {journal} {Nat. Phys.}\ }\textbf {\bibinfo {volume} {15}},\ \bibinfo {pages}
  {228} (\bibinfo {year} {2019})}\BibitemShut {NoStop}%
\bibitem [{\citenamefont {Löwdin}(1951)}]{lowdin1951}%
  \BibitemOpen
  \bibfield  {author} {\bibinfo {author} {\bibfnamefont {P.}~\bibnamefont
  {Löwdin}},\ }\href {https://doi.org/10.1063/1.1748067} {\bibfield  {journal}
  {\bibinfo  {journal} {J. Chem. Phys.}\ }\textbf {\bibinfo {volume} {19}},\
  \bibinfo {pages} {1396} (\bibinfo {year} {1951})}\BibitemShut {NoStop}%
\bibitem [{\citenamefont {Wang}\ \emph {et~al.}(2017)\citenamefont {Wang},
  \citenamefont {Robert}, \citenamefont {Glazov}, \citenamefont {Cadiz},
  \citenamefont {Courtade}, \citenamefont {Amand}, \citenamefont {Lagarde},
  \citenamefont {Taniguchi}, \citenamefont {Watanabe}, \citenamefont
  {Urbaszek},\ and\ \citenamefont {Marie}}]{wang_2017}%
  \BibitemOpen
  \bibfield  {author} {\bibinfo {author} {\bibfnamefont {G.}~\bibnamefont
  {Wang}}, \bibinfo {author} {\bibfnamefont {C.}~\bibnamefont {Robert}},
  \bibinfo {author} {\bibfnamefont {M.}~\bibnamefont {Glazov}}, \bibinfo
  {author} {\bibfnamefont {F.}~\bibnamefont {Cadiz}}, \bibinfo {author}
  {\bibfnamefont {E.}~\bibnamefont {Courtade}}, \bibinfo {author}
  {\bibfnamefont {T.}~\bibnamefont {Amand}}, \bibinfo {author} {\bibfnamefont
  {D.}~\bibnamefont {Lagarde}}, \bibinfo {author} {\bibfnamefont
  {T.}~\bibnamefont {Taniguchi}}, \bibinfo {author} {\bibfnamefont
  {K.}~\bibnamefont {Watanabe}}, \bibinfo {author} {\bibfnamefont
  {B.}~\bibnamefont {Urbaszek}},\ and\ \bibinfo {author} {\bibfnamefont
  {X.}~\bibnamefont {Marie}},\ }\href
  {https://doi.org/10.1103/PhysRevLett.119.047401} {\bibfield  {journal}
  {\bibinfo  {journal} {Phys. Rev. Lett.}\ }\textbf {\bibinfo {volume} {119}},\
  \bibinfo {pages} {047401} (\bibinfo {year} {2017})}\BibitemShut {NoStop}%
\end{thebibliography}%


\begin{thebibliography}{31}%
\makeatletter
\providecommand \@ifxundefined [1]{%
 \@ifx{#1\undefined}
}%
\providecommand \@ifnum [1]{%
 \ifnum #1\expandafter \@firstoftwo
 \else \expandafter \@secondoftwo
 \fi
}%
\providecommand \@ifx [1]{%
 \ifx #1\expandafter \@firstoftwo
 \else \expandafter \@secondoftwo
 \fi
}%
\providecommand \natexlab [1]{#1}%
\providecommand \enquote  [1]{``#1''}%
\providecommand \bibnamefont  [1]{#1}%
\providecommand \bibfnamefont [1]{#1}%
\providecommand \citenamefont [1]{#1}%
\providecommand \href@noop [0]{\@secondoftwo}%
\providecommand \href [0]{\begingroup \@sanitize@url \@href}%
\providecommand \@href[1]{\@@startlink{#1}\@@href}%
\providecommand \@@href[1]{\endgroup#1\@@endlink}%
\providecommand \@sanitize@url [0]{\catcode `\\12\catcode `\$12\catcode
  `\&12\catcode `\#12\catcode `\^12\catcode `\_12\catcode `\%12\relax}%
\providecommand \@@startlink[1]{}%
\providecommand \@@endlink[0]{}%
\providecommand \url  [0]{\begingroup\@sanitize@url \@url }%
\providecommand \@url [1]{\endgroup\@href {#1}{\urlprefix }}%
\providecommand \urlprefix  [0]{URL }%
\providecommand \Eprint [0]{\href }%
\providecommand \doibase [0]{https://doi.org/}%
\providecommand \selectlanguage [0]{\@gobble}%
\providecommand \bibinfo  [0]{\@secondoftwo}%
\providecommand \bibfield  [0]{\@secondoftwo}%
\providecommand \translation [1]{[#1]}%
\providecommand \BibitemOpen [0]{}%
\providecommand \bibitemStop [0]{}%
\providecommand \bibitemNoStop [0]{.\EOS\space}%
\providecommand \EOS [0]{\spacefactor3000\relax}%
\providecommand \BibitemShut  [1]{\csname bibitem#1\endcsname}%
\let\auto@bib@innerbib\@empty
\bibitem [{\citenamefont {Zomer}\ \emph {et~al.}(2014)\citenamefont {Zomer},
  \citenamefont {Guimarães}, \citenamefont {Brant}, \citenamefont {Tombros},\
  and\ \citenamefont {van Wees}}]{zomer2014}%
  \BibitemOpen
  \bibfield  {author} {\bibinfo {author} {\bibfnamefont {P.~J.}\ \bibnamefont
  {Zomer}}, \bibinfo {author} {\bibfnamefont {M.~H.~D.}\ \bibnamefont
  {Guimarães}}, \bibinfo {author} {\bibfnamefont {J.~C.}\ \bibnamefont
  {Brant}}, \bibinfo {author} {\bibfnamefont {N.}~\bibnamefont {Tombros}},\
  and\ \bibinfo {author} {\bibfnamefont {B.~J.}\ \bibnamefont {van Wees}},\
  }\href {https://doi.org/10.1063/1.4886096} {\bibfield  {journal} {\bibinfo
  {journal} {Appl. Phys. Lett.}\ }\textbf {\bibinfo {volume} {105}},\ \bibinfo
  {pages} {013101} (\bibinfo {year} {2014})}\BibitemShut {NoStop}%
\bibitem [{\citenamefont {Taniguchi}\ and\ \citenamefont
  {Watanabe}(2007)}]{taniguchi2007}%
  \BibitemOpen
  \bibfield  {author} {\bibinfo {author} {\bibfnamefont {T.}~\bibnamefont
  {Taniguchi}}\ and\ \bibinfo {author} {\bibfnamefont {K.}~\bibnamefont
  {Watanabe}},\ }\href {https://doi.org/10.1016/j.jcrysgro.2006.12.061}
  {\bibfield  {journal} {\bibinfo  {journal} {J. Cryst. Growth}\ }\textbf
  {\bibinfo {volume} {303}},\ \bibinfo {pages} {525} (\bibinfo {year}
  {2007})}\BibitemShut {NoStop}%
\bibitem [{\citenamefont {Leisgang}\ \emph {et~al.}(2020)\citenamefont
  {Leisgang}, \citenamefont {Shree}, \citenamefont {Paradisanos}, \citenamefont
  {Sponfeldner}, \citenamefont {Robert}, \citenamefont {Lagarde}, \citenamefont
  {Balocchi}, \citenamefont {Watanabe}, \citenamefont {Taniguchi},
  \citenamefont {Marie}, \citenamefont {Warburton}, \citenamefont {Gerber},\
  and\ \citenamefont {Urbaszek}}]{leisgang2020}%
  \BibitemOpen
  \bibfield  {author} {\bibinfo {author} {\bibfnamefont {N.}~\bibnamefont
  {Leisgang}}, \bibinfo {author} {\bibfnamefont {S.}~\bibnamefont {Shree}},
  \bibinfo {author} {\bibfnamefont {I.}~\bibnamefont {Paradisanos}}, \bibinfo
  {author} {\bibfnamefont {L.}~\bibnamefont {Sponfeldner}}, \bibinfo {author}
  {\bibfnamefont {C.}~\bibnamefont {Robert}}, \bibinfo {author} {\bibfnamefont
  {D.}~\bibnamefont {Lagarde}}, \bibinfo {author} {\bibfnamefont
  {A.}~\bibnamefont {Balocchi}}, \bibinfo {author} {\bibfnamefont
  {K.}~\bibnamefont {Watanabe}}, \bibinfo {author} {\bibfnamefont
  {T.}~\bibnamefont {Taniguchi}}, \bibinfo {author} {\bibfnamefont
  {X.}~\bibnamefont {Marie}}, \bibinfo {author} {\bibfnamefont {R.~J.}\
  \bibnamefont {Warburton}}, \bibinfo {author} {\bibfnamefont {I.~C.}\
  \bibnamefont {Gerber}},\ and\ \bibinfo {author} {\bibfnamefont
  {B.}~\bibnamefont {Urbaszek}},\ }\href
  {https://doi.org/10.1038/s41565-020-0750-1} {\bibfield  {journal} {\bibinfo
  {journal} {Nat. Nanotechnol.}\ }\textbf {\bibinfo {volume} {15}},\ \bibinfo
  {pages} {901} (\bibinfo {year} {2020})}\BibitemShut {NoStop}%
\bibitem [{\citenamefont {Laturia}\ \emph {et~al.}(2018)\citenamefont
  {Laturia}, \citenamefont {Put},\ and\ \citenamefont
  {Vandenberghe}}]{laturia2018}%
  \BibitemOpen
  \bibfield  {author} {\bibinfo {author} {\bibfnamefont {A.}~\bibnamefont
  {Laturia}}, \bibinfo {author} {\bibfnamefont {M.~L. V.~d.}\ \bibnamefont
  {Put}},\ and\ \bibinfo {author} {\bibfnamefont {W.~G.}\ \bibnamefont
  {Vandenberghe}},\ }\href {https://doi.org/10.1038/s41699-018-0050-x}
  {\bibfield  {journal} {\bibinfo  {journal} {npj 2D Mater. Appl.}\ }\textbf
  {\bibinfo {volume} {2}},\ \bibinfo {pages} {6} (\bibinfo {year}
  {2018})}\BibitemShut {NoStop}%
\bibitem [{\citenamefont {Roch}\ \emph {et~al.}(2019)\citenamefont {Roch},
  \citenamefont {Froehlicher}, \citenamefont {Leisgang}, \citenamefont {Makk},
  \citenamefont {Watanabe}, \citenamefont {Taniguchi},\ and\ \citenamefont
  {Warburton}}]{roch2019}%
  \BibitemOpen
  \bibfield  {author} {\bibinfo {author} {\bibfnamefont {J.~G.}\ \bibnamefont
  {Roch}}, \bibinfo {author} {\bibfnamefont {G.}~\bibnamefont {Froehlicher}},
  \bibinfo {author} {\bibfnamefont {N.}~\bibnamefont {Leisgang}}, \bibinfo
  {author} {\bibfnamefont {P.}~\bibnamefont {Makk}}, \bibinfo {author}
  {\bibfnamefont {K.}~\bibnamefont {Watanabe}}, \bibinfo {author}
  {\bibfnamefont {T.}~\bibnamefont {Taniguchi}},\ and\ \bibinfo {author}
  {\bibfnamefont {R.~J.}\ \bibnamefont {Warburton}},\ }\href
  {https://doi.org/10.1038/s41565-019-0397-y} {\bibfield  {journal} {\bibinfo
  {journal} {Nat. Nanotechnol.}\ }\textbf {\bibinfo {volume} {14}},\ \bibinfo
  {pages} {432} (\bibinfo {year} {2019})}\BibitemShut {NoStop}%
\bibitem [{\citenamefont {Back}\ \emph {et~al.}(2017)\citenamefont {Back},
  \citenamefont {Sidler}, \citenamefont {Cotlet}, \citenamefont {Srivastava},
  \citenamefont {Takemura}, \citenamefont {Kroner},\ and\ \citenamefont
  {Imamoğlu}}]{back2017}%
  \BibitemOpen
  \bibfield  {author} {\bibinfo {author} {\bibfnamefont {P.}~\bibnamefont
  {Back}}, \bibinfo {author} {\bibfnamefont {M.}~\bibnamefont {Sidler}},
  \bibinfo {author} {\bibfnamefont {O.}~\bibnamefont {Cotlet}}, \bibinfo
  {author} {\bibfnamefont {A.}~\bibnamefont {Srivastava}}, \bibinfo {author}
  {\bibfnamefont {N.}~\bibnamefont {Takemura}}, \bibinfo {author}
  {\bibfnamefont {M.}~\bibnamefont {Kroner}},\ and\ \bibinfo {author}
  {\bibfnamefont {A.}~\bibnamefont {Imamoğlu}},\ }\href
  {https://doi.org/10.1103/PhysRevLett.118.237404} {\bibfield  {journal}
  {\bibinfo  {journal} {Phys. Rev. Lett.}\ }\textbf {\bibinfo {volume} {118}},\
  \bibinfo {pages} {237404} (\bibinfo {year} {2017})}\BibitemShut {NoStop}%
\bibitem [{\citenamefont {Frimmer}\ and\ \citenamefont
  {Novotny}(2014)}]{frimmer2014}%
  \BibitemOpen
  \bibfield  {author} {\bibinfo {author} {\bibfnamefont {M.}~\bibnamefont
  {Frimmer}}\ and\ \bibinfo {author} {\bibfnamefont {L.}~\bibnamefont
  {Novotny}},\ }\href {https://doi.org/10.1119/1.4878621} {\bibfield  {journal}
  {\bibinfo  {journal} {Am. J. Phys.}\ }\textbf {\bibinfo {volume} {82}},\
  \bibinfo {pages} {947} (\bibinfo {year} {2014})}\BibitemShut {NoStop}%
\bibitem [{\citenamefont {Karrai}\ and\ \citenamefont
  {J.~Warburton}(2003)}]{karrai2003}%
  \BibitemOpen
  \bibfield  {author} {\bibinfo {author} {\bibfnamefont {K.}~\bibnamefont
  {Karrai}}\ and\ \bibinfo {author} {\bibfnamefont {R.}~\bibnamefont
  {J.~Warburton}},\ }\href {https://doi.org/10.1016/j.spmi.2004.02.007}
  {\bibfield  {journal} {\bibinfo  {journal} {Superlattices and Microstruct.}\
  }\textbf {\bibinfo {volume} {33}},\ \bibinfo {pages} {311} (\bibinfo {year}
  {2003})}\BibitemShut {NoStop}%
\bibitem [{\citenamefont {Kira}\ and\ \citenamefont {Koch}(2006)}]{kira2006}%
  \BibitemOpen
  \bibfield  {author} {\bibinfo {author} {\bibfnamefont {M.}~\bibnamefont
  {Kira}}\ and\ \bibinfo {author} {\bibfnamefont {S.~W.}\ \bibnamefont
  {Koch}},\ }\href {https://doi.org/10.1016/j.pquantelec.2006.12.002}
  {\bibfield  {journal} {\bibinfo  {journal} {Prog. Quantum Electron.}\
  }\textbf {\bibinfo {volume} {30}},\ \bibinfo {pages} {155} (\bibinfo {year}
  {2006})}\BibitemShut {NoStop}%
\bibitem [{\citenamefont {Cohen-Tannoudji}\ \emph {et~al.}(2005)\citenamefont
  {Cohen-Tannoudji}, \citenamefont {Diu},\ and\ \citenamefont
  {Laloë}}]{cohen-tannoudji2005}%
  \BibitemOpen
  \bibfield  {author} {\bibinfo {author} {\bibfnamefont {C.}~\bibnamefont
  {Cohen-Tannoudji}}, \bibinfo {author} {\bibfnamefont {B.}~\bibnamefont
  {Diu}},\ and\ \bibinfo {author} {\bibfnamefont {F.}~\bibnamefont {Laloë}},\
  }\href@noop {} {\emph {\bibinfo {title} {Quantum mechanics. {Vol}. 2}}}\
  (\bibinfo  {publisher} {Wiley},\ \bibinfo {address} {New York},\ \bibinfo
  {year} {2005})\BibitemShut {NoStop}%
\bibitem [{\citenamefont {Klingshirn}(2012)}]{klingshirn2012}%
  \BibitemOpen
  \bibfield  {author} {\bibinfo {author} {\bibfnamefont {C.~F.}\ \bibnamefont
  {Klingshirn}},\ }\href {https://doi.org/10.1007/978-3-642-28362-8} {\emph
  {\bibinfo {title} {Semiconductor {Optics}}}}\ (\bibinfo  {publisher}
  {Springer},\ \bibinfo {address} {Berlin, Heidelberg},\ \bibinfo {year}
  {2012})\BibitemShut {NoStop}%
\bibitem [{\citenamefont {Lindberg}\ and\ \citenamefont
  {Koch}(1988)}]{lindberg1988}%
  \BibitemOpen
  \bibfield  {author} {\bibinfo {author} {\bibfnamefont {M.}~\bibnamefont
  {Lindberg}}\ and\ \bibinfo {author} {\bibfnamefont {S.~W.}\ \bibnamefont
  {Koch}},\ }\href {https://doi.org/10.1103/PhysRevB.38.3342} {\bibfield
  {journal} {\bibinfo  {journal} {Phys. Rev. B}\ }\textbf {\bibinfo {volume}
  {38}},\ \bibinfo {pages} {3342} (\bibinfo {year} {1988})}\BibitemShut
  {NoStop}%
\bibitem [{\citenamefont {Haug}\ and\ \citenamefont {Koch}(2009)}]{haug2009}%
  \BibitemOpen
  \bibfield  {author} {\bibinfo {author} {\bibfnamefont {H.}~\bibnamefont
  {Haug}}\ and\ \bibinfo {author} {\bibfnamefont {S.~W.}\ \bibnamefont
  {Koch}},\ }\href {https://doi.org/10.1142/7184} {\emph {\bibinfo {title}
  {Quantum {Theory} of the {Optical} and {Electronic} {Properties} of
  {Semiconductors}}}}\ (\bibinfo  {publisher} {World Scientific},\ \bibinfo
  {year} {2009})\BibitemShut {NoStop}%
\bibitem [{\citenamefont {Koch}\ \emph {et~al.}(2006)\citenamefont {Koch},
  \citenamefont {Kira}, \citenamefont {Khitrova},\ and\ \citenamefont
  {Gibbs}}]{koch2006}%
  \BibitemOpen
  \bibfield  {author} {\bibinfo {author} {\bibfnamefont {S.~W.}\ \bibnamefont
  {Koch}}, \bibinfo {author} {\bibfnamefont {M.}~\bibnamefont {Kira}}, \bibinfo
  {author} {\bibfnamefont {G.}~\bibnamefont {Khitrova}},\ and\ \bibinfo
  {author} {\bibfnamefont {H.~M.}\ \bibnamefont {Gibbs}},\ }\href
  {https://doi.org/10.1038/nmat1658} {\bibfield  {journal} {\bibinfo  {journal}
  {Nat. Mater.}\ }\textbf {\bibinfo {volume} {5}},\ \bibinfo {pages} {523}
  (\bibinfo {year} {2006})}\BibitemShut {NoStop}%
\bibitem [{\citenamefont {Miller}\ \emph {et~al.}(1984)\citenamefont {Miller},
  \citenamefont {Chemla}, \citenamefont {Damen}, \citenamefont {Gossard},
  \citenamefont {Wiegmann}, \citenamefont {Wood},\ and\ \citenamefont
  {Burrus}}]{miller1984}%
  \BibitemOpen
  \bibfield  {author} {\bibinfo {author} {\bibfnamefont {D.~A.~B.}\
  \bibnamefont {Miller}}, \bibinfo {author} {\bibfnamefont {D.~S.}\
  \bibnamefont {Chemla}}, \bibinfo {author} {\bibfnamefont {T.~C.}\
  \bibnamefont {Damen}}, \bibinfo {author} {\bibfnamefont {A.~C.}\ \bibnamefont
  {Gossard}}, \bibinfo {author} {\bibfnamefont {W.}~\bibnamefont {Wiegmann}},
  \bibinfo {author} {\bibfnamefont {T.~H.}\ \bibnamefont {Wood}},\ and\
  \bibinfo {author} {\bibfnamefont {C.~A.}\ \bibnamefont {Burrus}},\ }\href
  {https://doi.org/10.1103/PhysRevLett.53.2173} {\bibfield  {journal} {\bibinfo
   {journal} {Phys. Rev. Lett.}\ }\textbf {\bibinfo {volume} {53}},\ \bibinfo
  {pages} {2173} (\bibinfo {year} {1984})}\BibitemShut {NoStop}%
\bibitem [{\citenamefont {Roch}\ \emph {et~al.}(2018)\citenamefont {Roch},
  \citenamefont {Leisgang}, \citenamefont {Froehlicher}, \citenamefont {Makk},
  \citenamefont {Watanabe}, \citenamefont {Taniguchi}, \citenamefont
  {Schönenberger},\ and\ \citenamefont {Warburton}}]{roch2018}%
  \BibitemOpen
  \bibfield  {author} {\bibinfo {author} {\bibfnamefont {J.~G.}\ \bibnamefont
  {Roch}}, \bibinfo {author} {\bibfnamefont {N.}~\bibnamefont {Leisgang}},
  \bibinfo {author} {\bibfnamefont {G.}~\bibnamefont {Froehlicher}}, \bibinfo
  {author} {\bibfnamefont {P.}~\bibnamefont {Makk}}, \bibinfo {author}
  {\bibfnamefont {K.}~\bibnamefont {Watanabe}}, \bibinfo {author}
  {\bibfnamefont {T.}~\bibnamefont {Taniguchi}}, \bibinfo {author}
  {\bibfnamefont {C.}~\bibnamefont {Schönenberger}},\ and\ \bibinfo {author}
  {\bibfnamefont {R.~J.}\ \bibnamefont {Warburton}},\ }\href
  {https://doi.org/10.1021/acs.nanolett.7b04553} {\bibfield  {journal}
  {\bibinfo  {journal} {Nano Lett.}\ }\textbf {\bibinfo {volume} {18}},\
  \bibinfo {pages} {1070} (\bibinfo {year} {2018})}\BibitemShut {NoStop}%
\bibitem [{\citenamefont {Gerber}\ \emph {et~al.}(2019)\citenamefont {Gerber},
  \citenamefont {Courtade}, \citenamefont {Shree}, \citenamefont {Robert},
  \citenamefont {Taniguchi}, \citenamefont {Watanabe}, \citenamefont
  {Balocchi}, \citenamefont {Renucci}, \citenamefont {Lagarde}, \citenamefont
  {Marie},\ and\ \citenamefont {Urbaszek}}]{gerber2019}%
  \BibitemOpen
  \bibfield  {author} {\bibinfo {author} {\bibfnamefont {I.~C.}\ \bibnamefont
  {Gerber}}, \bibinfo {author} {\bibfnamefont {E.}~\bibnamefont {Courtade}},
  \bibinfo {author} {\bibfnamefont {S.}~\bibnamefont {Shree}}, \bibinfo
  {author} {\bibfnamefont {C.}~\bibnamefont {Robert}}, \bibinfo {author}
  {\bibfnamefont {T.}~\bibnamefont {Taniguchi}}, \bibinfo {author}
  {\bibfnamefont {K.}~\bibnamefont {Watanabe}}, \bibinfo {author}
  {\bibfnamefont {A.}~\bibnamefont {Balocchi}}, \bibinfo {author}
  {\bibfnamefont {P.}~\bibnamefont {Renucci}}, \bibinfo {author} {\bibfnamefont
  {D.}~\bibnamefont {Lagarde}}, \bibinfo {author} {\bibfnamefont
  {X.}~\bibnamefont {Marie}},\ and\ \bibinfo {author} {\bibfnamefont
  {B.}~\bibnamefont {Urbaszek}},\ }\href
  {https://doi.org/10.1103/PhysRevB.99.035443} {\bibfield  {journal} {\bibinfo
  {journal} {Phys. Rev. B}\ }\textbf {\bibinfo {volume} {99}},\ \bibinfo
  {pages} {035443} (\bibinfo {year} {2019})}\BibitemShut {NoStop}%
\bibitem [{\citenamefont {Pedersen}(2016)}]{pedersen2016}%
  \BibitemOpen
  \bibfield  {author} {\bibinfo {author} {\bibfnamefont {T.~G.}\ \bibnamefont
  {Pedersen}},\ }\href {https://doi.org/10.1103/PhysRevB.94.125424} {\bibfield
  {journal} {\bibinfo  {journal} {Phys. Rev. B}\ }\textbf {\bibinfo {volume}
  {94}},\ \bibinfo {pages} {125424} (\bibinfo {year} {2016})}\BibitemShut
  {NoStop}%
\bibitem [{\citenamefont {Kresse}\ and\ \citenamefont
  {Hafner}(1993)}]{Kresse:1993a}%
  \BibitemOpen
  \bibfield  {author} {\bibinfo {author} {\bibfnamefont {G.}~\bibnamefont
  {Kresse}}\ and\ \bibinfo {author} {\bibfnamefont {J.}~\bibnamefont
  {Hafner}},\ }\href {https://doi.org/10.1103/PhysRevB.47.558} {\bibfield
  {journal} {\bibinfo  {journal} {Phys. Rev. B}\ }\textbf {\bibinfo {volume}
  {47}},\ \bibinfo {pages} {558} (\bibinfo {year} {1993})}\BibitemShut
  {NoStop}%
\bibitem [{\citenamefont {Kresse}\ and\ \citenamefont
  {Furthm\"uller}(1996)}]{Kresse:1996a}%
  \BibitemOpen
  \bibfield  {author} {\bibinfo {author} {\bibfnamefont {G.}~\bibnamefont
  {Kresse}}\ and\ \bibinfo {author} {\bibfnamefont {J.}~\bibnamefont
  {Furthm\"uller}},\ }\href {https://doi.org/10.1103/PhysRevB.54.11169}
  {\bibfield  {journal} {\bibinfo  {journal} {Phys. Rev. B}\ }\textbf {\bibinfo
  {volume} {54}},\ \bibinfo {pages} {11169} (\bibinfo {year}
  {1996})}\BibitemShut {NoStop}%
\bibitem [{\citenamefont {Bl\"ochl}(1994)}]{blochl:prb:94}%
  \BibitemOpen
  \bibfield  {author} {\bibinfo {author} {\bibfnamefont {P.~E.}\ \bibnamefont
  {Bl\"ochl}},\ }\href {https://doi.org/10.1103/PhysRevB.50.17953} {\bibfield
  {journal} {\bibinfo  {journal} {Phys. Rev. B}\ }\textbf {\bibinfo {volume}
  {50}},\ \bibinfo {pages} {17953} (\bibinfo {year} {1994})}\BibitemShut
  {NoStop}%
\bibitem [{\citenamefont {Kresse}\ and\ \citenamefont
  {Joubert}(1999)}]{kresse:prb:99a}%
  \BibitemOpen
  \bibfield  {author} {\bibinfo {author} {\bibfnamefont {G.}~\bibnamefont
  {Kresse}}\ and\ \bibinfo {author} {\bibfnamefont {D.}~\bibnamefont
  {Joubert}},\ }\href {https://doi.org/10.1103/PhysRevB.59.1758} {\bibfield
  {journal} {\bibinfo  {journal} {Phys. Rev. B}\ }\textbf {\bibinfo {volume}
  {59}},\ \bibinfo {pages} {1758} (\bibinfo {year} {1999})}\BibitemShut
  {NoStop}%
\bibitem [{\citenamefont {Grimme}\ \emph {et~al.}(2010)\citenamefont {Grimme},
  \citenamefont {Antony}, \citenamefont {Ehrlich},\ and\ \citenamefont
  {Krieg}}]{Grimme:2010ij}%
  \BibitemOpen
  \bibfield  {author} {\bibinfo {author} {\bibfnamefont {S.}~\bibnamefont
  {Grimme}}, \bibinfo {author} {\bibfnamefont {J.}~\bibnamefont {Antony}},
  \bibinfo {author} {\bibfnamefont {S.}~\bibnamefont {Ehrlich}},\ and\ \bibinfo
  {author} {\bibfnamefont {H.}~\bibnamefont {Krieg}},\ }\href
  {https://doi.org/10.1063/1.3382344} {\bibfield  {journal} {\bibinfo
  {journal} {J. Chem. Phys.}\ }\textbf {\bibinfo {volume} {132}},\ \bibinfo
  {pages} {154104} (\bibinfo {year} {2010})}\BibitemShut {NoStop}%
\bibitem [{\citenamefont {Heyd}\ and\ \citenamefont
  {Scuseria}(2004)}]{heyd:jcp:04_a}%
  \BibitemOpen
  \bibfield  {author} {\bibinfo {author} {\bibfnamefont {J.}~\bibnamefont
  {Heyd}}\ and\ \bibinfo {author} {\bibfnamefont {G.~E.}\ \bibnamefont
  {Scuseria}},\ }\href {https://doi.org/10.1063/1.1668634} {\bibfield
  {journal} {\bibinfo  {journal} {J. Chem. Phys.}\ }\textbf {\bibinfo {volume}
  {120}},\ \bibinfo {pages} {7274} (\bibinfo {year} {2004})}\BibitemShut
  {NoStop}%
\bibitem [{\citenamefont {Heyd}\ \emph {et~al.}(2005)\citenamefont {Heyd},
  \citenamefont {Peralta}, \citenamefont {Scuseria},\ and\ \citenamefont
  {Martin}}]{heyd:jcp:05}%
  \BibitemOpen
  \bibfield  {author} {\bibinfo {author} {\bibfnamefont {J.}~\bibnamefont
  {Heyd}}, \bibinfo {author} {\bibfnamefont {J.~E.}\ \bibnamefont {Peralta}},
  \bibinfo {author} {\bibfnamefont {G.~E.}\ \bibnamefont {Scuseria}},\ and\
  \bibinfo {author} {\bibfnamefont {R.~L.}\ \bibnamefont {Martin}},\ }\href
  {https://doi.org/10.1063/1.2085170} {\bibfield  {journal} {\bibinfo
  {journal} {J. Chem. Phys.}\ }\textbf {\bibinfo {volume} {123}},\ \bibinfo
  {pages} {174101} (\bibinfo {year} {2005})}\BibitemShut {NoStop}%
\bibitem [{\citenamefont {Paier}\ \emph {et~al.}(2006)\citenamefont {Paier},
  \citenamefont {Marsman}, \citenamefont {Hummer}, \citenamefont {Kresse},
  \citenamefont {Gerber},\ and\ \citenamefont {\'Angy\'an}}]{paier:jcp:06}%
  \BibitemOpen
  \bibfield  {author} {\bibinfo {author} {\bibfnamefont {J.}~\bibnamefont
  {Paier}}, \bibinfo {author} {\bibfnamefont {M.}~\bibnamefont {Marsman}},
  \bibinfo {author} {\bibfnamefont {K.}~\bibnamefont {Hummer}}, \bibinfo
  {author} {\bibfnamefont {G.}~\bibnamefont {Kresse}}, \bibinfo {author}
  {\bibfnamefont {I.~C.}\ \bibnamefont {Gerber}},\ and\ \bibinfo {author}
  {\bibfnamefont {J.~G.}\ \bibnamefont {\'Angy\'an}},\ }\href
  {https://doi.org/10.1063/1.2187006} {\bibfield  {journal} {\bibinfo
  {journal} {J. Chem. Phys.}\ }\textbf {\bibinfo {volume} {124}},\ \bibinfo
  {pages} {154709} (\bibinfo {year} {2006})}\BibitemShut {NoStop}%
\bibitem [{\citenamefont {Shishkin}\ and\ \citenamefont
  {Kresse}(2006)}]{Shishkin:2006a}%
  \BibitemOpen
  \bibfield  {author} {\bibinfo {author} {\bibfnamefont {M.}~\bibnamefont
  {Shishkin}}\ and\ \bibinfo {author} {\bibfnamefont {G.}~\bibnamefont
  {Kresse}},\ }\href {https://doi.org/10.1103/PhysRevB.74.035101} {\bibfield
  {journal} {\bibinfo  {journal} {Phys. Rev. B}\ }\textbf {\bibinfo {volume}
  {74}},\ \bibinfo {pages} {035101} (\bibinfo {year} {2006})}\BibitemShut
  {NoStop}%
\bibitem [{\citenamefont {Mostofi}\ \emph {et~al.}(2008)\citenamefont
  {Mostofi}, \citenamefont {Yates}, \citenamefont {Lee}, \citenamefont {Souza},
  \citenamefont {Vanderbilt},\ and\ \citenamefont {Marzari}}]{Mostofi:2008ff}%
  \BibitemOpen
  \bibfield  {author} {\bibinfo {author} {\bibfnamefont {A.~A.}\ \bibnamefont
  {Mostofi}}, \bibinfo {author} {\bibfnamefont {J.~R.}\ \bibnamefont {Yates}},
  \bibinfo {author} {\bibfnamefont {Y.-S.}\ \bibnamefont {Lee}}, \bibinfo
  {author} {\bibfnamefont {I.}~\bibnamefont {Souza}}, \bibinfo {author}
  {\bibfnamefont {D.}~\bibnamefont {Vanderbilt}},\ and\ \bibinfo {author}
  {\bibfnamefont {N.}~\bibnamefont {Marzari}},\ }\href
  {https://doi.org/10.1016/j.cpc.2007.11.016} {\bibfield  {journal} {\bibinfo
  {journal} {Comput. Phys. Commun.}\ }\textbf {\bibinfo {volume} {178}},\
  \bibinfo {pages} {685} (\bibinfo {year} {2008})}\BibitemShut {NoStop}%
\bibitem [{\citenamefont {Hanke}\ and\ \citenamefont
  {Sham}(1979)}]{Hanke:1979to}%
  \BibitemOpen
  \bibfield  {author} {\bibinfo {author} {\bibfnamefont {W.}~\bibnamefont
  {Hanke}}\ and\ \bibinfo {author} {\bibfnamefont {L.~J.}\ \bibnamefont
  {Sham}},\ }\href {https://doi.org/10.1103/PhysRevLett.43.387} {\bibfield
  {journal} {\bibinfo  {journal} {Phys. Rev. Lett.}\ }\textbf {\bibinfo
  {volume} {43}},\ \bibinfo {pages} {387} (\bibinfo {year} {1979})}\BibitemShut
  {NoStop}%
\bibitem [{\citenamefont {Rohlfing}\ and\ \citenamefont
  {Louie}(1998)}]{Rohlfing:1998vb}%
  \BibitemOpen
  \bibfield  {author} {\bibinfo {author} {\bibfnamefont {M.}~\bibnamefont
  {Rohlfing}}\ and\ \bibinfo {author} {\bibfnamefont {S.~G.}\ \bibnamefont
  {Louie}},\ }\href {https://doi.org/10.1103/PhysRevLett.81.2312} {\bibfield
  {journal} {\bibinfo  {journal} {Phys. Rev. Lett.}\ }\textbf {\bibinfo
  {volume} {81}},\ \bibinfo {pages} {2312} (\bibinfo {year}
  {1998})}\BibitemShut {NoStop}%
\bibitem [{\citenamefont {Guo}\ \emph {et~al.}(2019)\citenamefont {Guo},
  \citenamefont {Wu}, \citenamefont {Cao}, \citenamefont {Monahan},
  \citenamefont {Lee}, \citenamefont {Louie},\ and\ \citenamefont
  {Fleming}}]{guo2019}%
  \BibitemOpen
  \bibfield  {author} {\bibinfo {author} {\bibfnamefont {L.}~\bibnamefont
  {Guo}}, \bibinfo {author} {\bibfnamefont {M.}~\bibnamefont {Wu}}, \bibinfo
  {author} {\bibfnamefont {T.}~\bibnamefont {Cao}}, \bibinfo {author}
  {\bibfnamefont {D.~M.}\ \bibnamefont {Monahan}}, \bibinfo {author}
  {\bibfnamefont {Y.-H.}\ \bibnamefont {Lee}}, \bibinfo {author} {\bibfnamefont
  {S.~G.}\ \bibnamefont {Louie}},\ and\ \bibinfo {author} {\bibfnamefont
  {G.~R.}\ \bibnamefont {Fleming}},\ }\href
  {https://doi.org/10.1038/s41567-018-0362-y} {\bibfield  {journal} {\bibinfo
  {journal} {Nat. Phys.}\ }\textbf {\bibinfo {volume} {15}},\ \bibinfo {pages}
  {228} (\bibinfo {year} {2019})}\BibitemShut {NoStop}%
\end{thebibliography}%

\clearpage 
\end{document}


\title{Capacitively-coupled and inductively-coupled excitons in bilayer MoS$_2$\\ \vspace{10PT}
\small{Supplementary Information (SI)}}

\author{Lukas Sponfeldner} 
\affiliation{Department of Physics, University of Basel, Klingelbergstrasse 82, CH-4056 Basel, Switzerland}

\author{Nadine Leisgang}
\affiliation{Department of Physics, University of Basel, Klingelbergstrasse 82, CH-4056 Basel, Switzerland}

\author{Shivangi Shree}
\affiliation{Universit\'e de Toulouse, INSA-CNRS-UPS, LPCNO, 135 Avenue Rangueil, 31077 Toulouse, France}

\author{Ioannis Paradisanos}
\affiliation{Universit\'e de Toulouse, INSA-CNRS-UPS, LPCNO, 135 Avenue Rangueil, 31077 Toulouse, France}

\author{Kenji Watanabe}
\affiliation{Research Center for Functional Materials, National Institute for Materials Science, 1-1 Namiki, Tsukuba 305-0044, Japan}

\author{Takashi Taniguchi}
\affiliation{International Center for Materials Nanoarchitectonics, National Institute for Materials Science, 1-1 Namiki, Tsukuba 305-0044, Japan}

\author{Cedric Robert}
\affiliation{Universit\'e de Toulouse, INSA-CNRS-UPS, LPCNO, 135 Avenue Rangueil, 31077 Toulouse, France}

\author{Delphine Lagarde}
\affiliation{Universit\'e de Toulouse, INSA-CNRS-UPS, LPCNO, 135 Avenue Rangueil, 31077 Toulouse, France}

\author{Andrea Balocchi}
\affiliation{Universit\'e de Toulouse, INSA-CNRS-UPS, LPCNO, 135 Avenue Rangueil, 31077 Toulouse, France}

\author{Xavier Marie}
\affiliation{Universit\'e de Toulouse, INSA-CNRS-UPS, LPCNO, 135 Avenue Rangueil, 31077 Toulouse, France}

\author{Iann C. Gerber}
\affiliation{Universit\'e de Toulouse, INSA-CNRS-UPS, LPCNO, 135 Avenue Rangueil, 31077 Toulouse, France}

\author{Bernhard Urbaszek}
\affiliation{Universit\'e de Toulouse, INSA-CNRS-UPS, LPCNO, 135 Avenue Rangueil, 31077 Toulouse, France}

\author{Richard J. Warburton}
\affiliation{Department of Physics, University of Basel, Klingelbergstrasse 82, CH-4056 Basel, Switzerland}

\maketitle


\section{Sample fabrication}

The studied van der Waals heterostructure was assembled with a dry-transfer technique \cite{zomer2014}. The individual layers were mechanically exfoliated from bulk crystals (natural MoS$_2$ crystal from SPI Supplies, synthetic h-BN \cite{taniguchi2007}, and natural graphite from Graphene Supermarket) onto SiO$_2$ (300~nm)/Si substrates. The flakes were picked up and stacked with a stamp made of polydimethylsiloxane (PDMS) covered with a thin layer of polycarbonate (PC). The naturally 2H-stacked bilayer MoS$_2$ and few-layer graphene (FLG) layers were contacted by Ti/Au (5~nm/45~nm) electrodes defined by electron-beam lithography. Using atomic force microscopy (AFM), the top and bottom h-BN thicknesses were determined to be $d_\text{T}=16.2$~nm and $d_\text{B}=21.5$~nm, respectively. The studied sample is the same as device 1 in Ref.~\cite{leisgang2020} albeit in a second cooldown cycle. An exception is the measurement shown in Main~Fig.~2b and SI~Fig.~\ref{bl1}b which was recorded in the first cooldown cycle.\\

The dual-gating scheme allows for independent control of the electric field perpendicular to the MoS$_2$ plane $F_z$ and the total charge carrier density $n$ in the MoS$_2$ bilayer. $F_z$ and $n$ can be changed by applying voltages $V_\text{TG}$ and $V_\text{BG}$ to the top and bottom FLG gates while the MoS$_2$ bilayer is electrically grounded. Assuming no free charge carriers in the bilayer, the electric field across the MoS$_2$ bilayer can be written as 
\begin{equation}
F_z=\frac{1}{2}\frac{\epsilon_\text{hBN}}{\epsilon_\text{BL}d_\text{B}}\left(V_\text{TG}-\frac{d_\text{T}}{d_\text{B}}V_\text{BG}\right) \ ,
\label{Fz}
\end{equation}
with the dielectric constant of h-BN $\epsilon_\text{hBN}\approx3.76$ and of bilayer MoS$_2$ $\epsilon_\text{BL}\approx6.8$ \cite{laturia2018}. For a more detailed description of the electrostatic model see Supplementary Information of Ref. \cite{leisgang2020}. In the experimental electric field sweeps shown in the main paper, the applied voltages only change $F_z$. The carrier density in the MoS$_2$ bilayer stays constant and at a low value. This can be deduced from the stable absorption strength at the A:1s resonance.

\begin{figure}[b!]
\centering
\includegraphics[width=86mm]{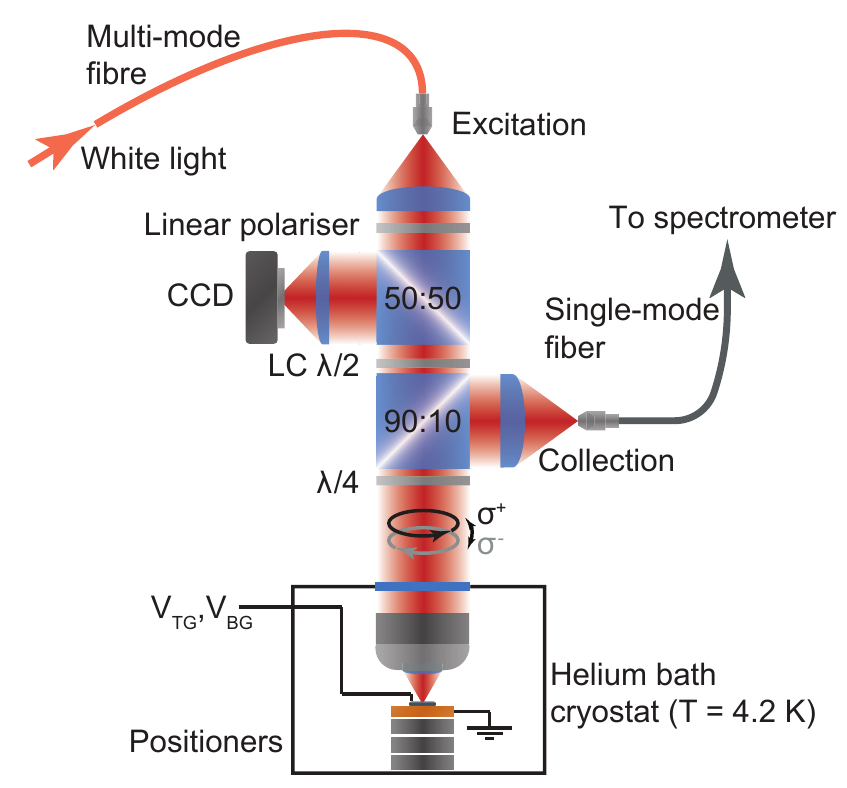}
\caption{Schematic of the experimental setup.}
\label{setup}
\end{figure}

\section{Experimental setup}
Absorption measurements were conducted with the setup sketched in SI~Fig.~\ref{setup}. A home-built confocal microscope facilitates a local measurement of the sample's optical susceptibility at a temperature of 4.2~K. White (Osram warm white) light is coupled into a multi-mode fiber and routed to the excitation arm of the microscope. A combination of a linear polariser and a quarter wave plate ($\lambda$/4) is used to create a circularly polarised excitation beam. With the help of a computer controlled liquid-crystal (LC) variable retarder, the excitation light can be either set to the $\sigma^+$ or the $\sigma^-$ circular polarisation state. The incident beam is then focused on the sample surface using a microscope objective (NA = 0.45). Piezoelectric nano-positioners are used to control the position of the spot on the sample. The reflected signal from the sample is collected through the same microscope objective and coupled into a single-mode fibre at the end of the collection arm of the confocal microscope. The collection fiber is connected to a spectrometer where the reflected signal is dispersed by a 600~g/mm (IE-B interaction measurement), a 1500~g/mm grating (IE-A interaction measurement), or a 300~g/mm grating (Main~Fig.~2b and SI~Fig.~\ref{bl1}b)  and detected by a liquid nitrogen-cooled charge coupled device (CCD) camera. \\

The absorption of the bilayer MoS$_2$ can be described by the imaginary part of the optical susceptibility Im$(\chi)$. Im$(\chi)$ is determined from the differential reflectivity $(R-R_0)/R_0=\Delta R/R_0$ using a Kramers-Kronig relation (see Supplementary Information of Ref. \cite{roch2019} for a detailed calculation). $R$ is the reflected signal from the bilayer MoS$_2$ and $R_0$ is a reference reflectivity spectrum. For $R_0$ we interpolate the raw reflectivity spectrum at high electron densities where the total absorption strength is spread over a large spectral range \cite{back2017}.

\section{Exciton interaction model}

\begin{figure}[t!]
\centering
\includegraphics[width=86mm]{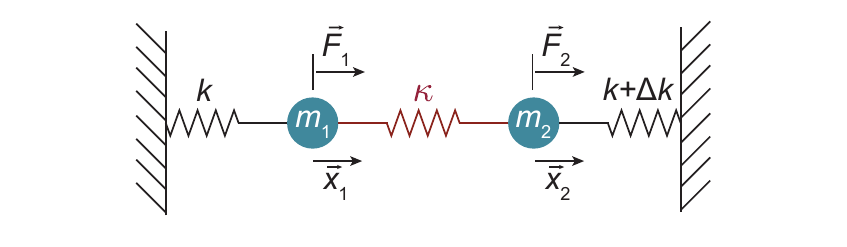}
\caption{Sketch of two masses $m_{1}$ and $m_{2}$ coupled by a spring with spring constant $\kappa$. $m_1$ has a spring constant of $k$, $m_2$ has a tunable spring constant of $k+\Delta k$. Each mass is driven by an external force $\vec{F}_{1,2}$ with different amplitudes. $\vec{x}_{1,2}$ describe the displacements of each mass.}
\label{masses}
\end{figure}

The excitonic interactions are modelled as the coupling of two driven oscillating optical dipoles. The dipole oscillation can be described by the oscillation of the dipole extension $\vec{r}$ or by the oscillation of the charge $\vec{q}$. One can think of two frameworks which provide a classical analogue. In a mechanical system, two masses represent the dipoles, a spring represents the coupling. In an electrical system, two RLC circuits represent the dipoles, an impedance (either capacitor or inductor) represents the coupling. Both, the mechanical and the electrical system, can be described by equivalent equations of motion as shown below.

\subsection{Equations of motion}

First, we look at a mechanical system. We consider two driven and damped harmonic oscillators with a mass $m_1$ and $m_2$ and a spring constant of $k_1=k$ and $k_2=k+\Delta k$, respectively. The term $\Delta k$ allows for the detuning of one oscillator with respect to the other. The two oscillators are coupled by a spring with spring constant $\kappa$. The damping of each oscillator is described by $\gamma_1$ and $\gamma_2$. The oscillators are externally driven by the forces $\vec{F}_1$ and $\vec{F}_2$, respectively. A sketch of the mechanical system is shown in SI~Fig.~\ref{masses}.\\

In the mechanical system, an oscillating force $\vec{F}_{1,2}$ induces a displacement $\vec{x}_{1,2}$ of the two masses $m_{1,2}$. For the following derivation, we assume $m_1=m_2=m$ and take $\vec{F}_{1,2}=F_{1,2}\vec{E}_0 \operatorname{Re}(\text{e}^{i\omega t})$. $F_{1,2}$ is the mechanical oscillator strength of each mass. The matrix form of the equations of motion for this system is
\renewcommand\arraystretch{2}
\begin{widetext}
\begin{equation}
\begin{bmatrix} 
\dfrac{d^2}{dt^2}+\gamma_1 \dfrac{d}{dt}+\dfrac{k+\kappa}{m} & -\dfrac{\kappa}{m} \\
 -\dfrac{\kappa}{m} & \dfrac{d^2}{dt^2}+\gamma_2 \dfrac{d}{dt}+\dfrac{k+\kappa}{m}+\dfrac{\Delta k}{m}
\end{bmatrix}
\begin{bmatrix} 
\vec{x}_1 \\
\vec{x}_2
\end{bmatrix}=\begin{bmatrix} 
\dfrac{F_1}{m}\vec{E}_0 \operatorname{Re}(\text{e}^{i\omega t})\\
\dfrac{F_2}{m}\vec{E}_0 \operatorname{Re}(\text{e}^{i\omega t})
\end{bmatrix}\coloneqq\begin{bmatrix} 
\vec{L}_1 \\
\vec{L}_2
\end{bmatrix}
\ .
\label{EOM}
\end{equation}
\end{widetext}

\subsection{Decoupling the equations of motion}

For an analytical solution to equation~\ref{EOM}, we need to transform the equations of motion into the eigenvector basis. The two eigenmodes $\vec{r}_{1,2}$ are given by \cite{frimmer2014}
\begin{equation}
\begin{bmatrix} 
\vec{x}_1 \\
\vec{x}_2
\end{bmatrix}= \textbf{U}^{-1}\begin{bmatrix} 
\vec{r}_1 \\
\vec{r}_2
\end{bmatrix}
\ ,
\label{transform}
\end{equation}
where the rows of the transformation matrix $\textbf{U}$ are the normalised eigenvectors of the matrix in equation~\ref{EOM}. For simplicity, only the non-normalised eigenvectors will be shown in the following derivation. All calculations and plots shown in the main text and the Supplement use the normalised eigenvectors. The eigenvectors take the form

\begin{align}
\vec{v}_1 &=
\begin{bmatrix} 
1 \\
-\dfrac{1}{2} \dfrac{\Delta k}{\kappa}+\sgn \left(\Delta k\right) \dfrac{\sqrt{\Delta k^2+4 \kappa^2}}{2\kappa} 
\end{bmatrix} \ ,   \nonumber  \\ 
\vec{v}_2 &=
\begin{bmatrix} 
+\dfrac{1}{2} \dfrac{\Delta k}{\kappa}-\sgn \left(\Delta k\right)\dfrac{\sqrt{\Delta k^2+4 \kappa^2}}{2\kappa}  \\
1
\end{bmatrix}
\ .
\label{eigenvectors2}
\end{align}
Introducing the ratio $\rho$ between the detuning and coupling constant $\rho=\frac{\Delta k}{\kappa}$, we can write
\begin{align}
\vec{v}_1=
\begin{bmatrix} 
1 \\
-\dfrac{1}{2}\rho+\sgn \left(\rho\right)\sqrt{\dfrac{1}{4}\rho^2+1}
\end{bmatrix} \ , \\
\vec{v}_2=
\begin{bmatrix} 
+\dfrac{1}{2}\rho-\sgn \left(\rho\right)\sqrt{\dfrac{1}{4}\rho^2+1} \\
1
\end{bmatrix}
\ .
\label{eigenvectors}
\end{align}
We can write the transformation matrix \textbf{U} as

\begin{widetext}
\begin{equation}
\textbf{U}=
\begin{bmatrix} 
U_{11} & U_{12} \\
U_{21} & U_{22}
\end{bmatrix}=
\begin{bmatrix} 
1 &-\dfrac{1}{2}\rho+\sgn \left(\rho\right)\sqrt{\dfrac{1}{4}\rho^2+1} \\
+\dfrac{1}{2}\rho-\sgn \left(\rho\right)\sqrt{\dfrac{1}{4}\rho^2+1} & 1
\end{bmatrix}
\ .
\label{Umatrix}
\end{equation}
\end{widetext}
The angular eigenfrequencies of the system are given by

\begin{equation}
\Omega_{\pm}=\left(\dfrac{k+\kappa}{m}+\frac{1}{2} \frac{\Delta k}{m} \mp \frac{1}{2} \sqrt{\left(\dfrac{\Delta k}{m}\right)^2+4\left(\dfrac{\kappa}{m}\right)^2} \right)^{\frac{1}{2}}
\ .
\label{eigenfrequencies}
\end{equation}
It is important to note that $\Omega_{\pm}$ do not depend on $\vec{F}_{1,2}$.\\

After transforming equation~\ref{EOM} with the matrix \textbf{U}, we find two uncoupled differential equations for each of the eigenmode displacements $\vec{r}_1$ and $\vec{r}_2$:

\begin{align} 
\left(\frac{d^2}{dt^2}+\gamma_1 \frac{d}{dt}+\Omega_+^2  \right) \vec{r}_1 &= \dfrac{\vec{E_0}}{m}\left(U_{11}F_1+U_{12}F_2\right) \operatorname{Re}(\text{e}^{i\omega t})\ , \nonumber \\
\left(\frac{d^2}{dt^2}+\gamma_2 \frac{d}{dt}+\Omega_-^2  \right) \vec{r}_2 &= \dfrac{\vec{E_0}}{m}\left(U_{21}F_1+U_{22}F_2\right) \operatorname{Re}(\text{e}^{i\omega t})
\ .
\label{uncoupled}
\end{align}

\subsection{Frequency response of the coupled masses}

Solving equation~\ref{uncoupled} for the eigenmodes $\vec{r}_{1}$ and $\vec{r}_{2}$ gives

\begin{align}
\vec{r}_1  &= \dfrac{\vec{E}_0}{m} \dfrac{\left(U_{11}F_1+U_{12}F_2\right) }{-\omega^2+i \gamma_1\omega+\Omega_+^2} \: \operatorname{Re}(\text{e}^{i\omega t})\ , \nonumber \\
\vec{r}_2 &= \dfrac{\vec{E}_0}{m} \dfrac{\left(U_{21}F_1+U_{22}F_2\right)}{-\omega^2+i \gamma_2\omega+\Omega_-^2} \: \operatorname{Re}(\text{e}^{i\omega t})
\ .
\label{z}
\end{align}
The mechanical response $f_{1,2}$ of each eigenmode is modified as a function of the ratio $\rho$ through $U_{ij}$:
\begin{equation}
\begin{bmatrix} 
f_1   \\
f_2
\end{bmatrix}=\begin{bmatrix} 
U_{11}F_1+U_{12}F_2  \\
U_{21}F_1+U_{22}F_2
\end{bmatrix}
\ .
\label{fComparison}
\end{equation}

For large detuning compared to the coupling strength ($\rho \gg 1$), the mechanical oscillator strength $F_{1,2}$ is recovered for each mass:
\begin{equation}
f_1 (\left|\Delta k\right| \gg1)=F_1 \: ,  \: f_2 (\left|\Delta k\right| \gg 1)=F_2 \ .
\label{largeRwrong}
\end{equation}

For a large positive detuning, the oscillator strengths shown in equation~\ref{largeRwrong} are correct. For a large negative detuning, the oscillator strengths of the eigenmodes are swapped. To accomodate this problem, we need to change the transformation matrix \textbf{U} upon changing the sign of the detuning $\Delta k$. Switching the rows in the transformation matrix \textbf{U} yields the correct oscillator strengths at large negative detuning. The responses at large detuning are then:

\begin{align}
f_1 (\left|\Delta k\right|\gg1)&=\left \{ \begin{matrix} F_1 \: ,  \: \Delta k >0 \\
F_2  \: ,  \: \Delta k < 0 \end{matrix} \right. \: , \: \nonumber \\
f_2 (\left|\Delta k\right| \gg1)&=\left \{ \begin{matrix} F_2 \: ,  \: \Delta k>0  \\
F_1  \: ,  \: \Delta k < 0 \end{matrix} \right. \ .
\label{largeR}
\end{align}

The full description of the transformation matrix \textbf{U} is then:
\renewcommand\arraystretch{2}
\begin{equation}
\textbf{U}(\Delta k >0)=
\begin{bmatrix} 
U_{11} & U_{12} \\
U_{21} & U_{22}
\end{bmatrix}
\ ,\;
\textbf{U}(\Delta k <0)=
\begin{bmatrix} 
U_{21} & U_{22} \\
U_{11} & U_{12}
\end{bmatrix}
\ .
\label{Ufull}
\end{equation}

\subsection{Calculation of the dissipated power}

The dissipated power $P$ in the system is given by the product of the velocity of the eigenmodes and the right hand side of the equations of motion, $\vec{L}$, in the eigenvector basis:
\begin{equation}
P(t)= \dot{\vec{r}}^\text{\;T} \textbf{U}\vec{L}
\ .
\label{power}
\end{equation}

The driving force oscillates as $\operatorname{Re}(\text{e}^{i\omega t})$. Therefore, only the real part of the solution for $\vec{r}_{1,2}$ is relevant. After taking the real part and calculating the derivative of $\vec{r}$ in time, we derive an analytical solution for the time-dependent dissipated power, $P$. We are interested in the steady-state solution of $P$. After averaging the total power $P$ over time, the total average power $\left<P\right>$ is given by
\begin{align}
\left<P\right> =& \frac{1}{2} \dfrac{\left|\vec{E}_0\right|^2}{m}\left(U_{11}F_1+U_{12}F_2\right)^2 \frac{\gamma_1\omega^2}{\gamma_1^2\omega^2+\left(\Omega_+^2-\omega^2\right)^2}+ \nonumber \\
&\frac{1}{2} \dfrac{\left|\vec{E}_0\right|^2}{m} \left(U_{21}F_1+U_{22}F_2\right)^2 \frac{\gamma_2\omega^2}{\gamma_2^2\omega^2+\left(\Omega_-^2-\omega^2\right)^2}
\ .
\label{powerAvg}
\end{align}
The total average power can be seen as the sum of the average power of each of the two eigenmodes:
\begin{align}
\left<P\right> &=\left<P_+\right> +\left<P_-\right>  
\ .
\label{powerAvg2}
\end{align}
Assuming that $\kappa\ll k$ then $\frac{k+\kappa}{m} \simeq \frac{k}{m}$. Using this we can rewrite the term $\left(\Omega_\pm^2-\omega^2\right)= \left(\Omega_\pm-\omega\right)\left(\Omega_\pm+\omega\right)\simeq \left(\Omega_\pm-\omega\right)2\omega$. Inserting this approximation into equation~\ref{powerAvg} yields:

\begin{align}
\left<P\right> \simeq  & \dfrac{\left|\vec{E}_0\right|^2}{4m}\left(U_{11}F_1+U_{12}F_2\right)^2 \frac{\frac{1}{2}\gamma_1}{\left(\frac{1}{2}\gamma_1\right)^2+\left(\Omega_+-\omega\right)^2}+ \nonumber \\
& \dfrac{\left|\vec{E}_0\right|^2}{4m}\left(U_{21}F_1+U_{22}F_2\right)^2 \frac{\frac{1}{2}\gamma_2}{\left(\frac{1}{2}\gamma_2\right)^2+\left(\Omega_--\omega\right)^2}
\ .
\label{powerApprox}
\end{align}
The average power is the sum of two Lorentzians centred around each of the eigenvalues $\Omega_\pm$. This is a significant result: the model predicts straightforward lineshapes.

\subsection{Linking the mechanical system to the optical response of coupled excitons}

The experimental results shown in the main paper study excitonic interactions. Therefore, we forge an analogy between the mechanical and optical systems. We show that excitonic coupling can be readily described by the mechanical model assuming a classical description of the driven excitons. \\

We model the interaction of inter- and intralayer excitons as two coupled oscillating optical dipoles driven by an external light field. The optical dipole representing the intralayer exciton (index 1) has a constant energy of $\hbar \omega_0$. The IE (index 2) is modelled by a dipole that has a tunable energy through the term $\hbar \Delta \omega$ with a total energy of $\hbar \omega_0+\hbar \Delta\omega$. The two dipoles are coupled to each other through a coupling constant $\kappa$. The detuning $\hbar \Delta\omega$ is a linear function of the applied electric field $F_z$ on account of the Stark effect. From experiments we know that the energy crossing point occurs at $F_z \neq 0$. Therefore,
\begin{equation}
\hbar \Delta\omega=\mu \left(F_z-F_{z,0}\right)  \ ,
\label{deltaK}
\end{equation}
where $\mu$ is the static dipole moment and $F_{z,0}$ is the field at which the two bare energies match. \\

From the experiment we know that the sign of $\mu$ is positive for IE$_1$ and negative for IE$_2$. The two dipoles are driven by the external light field $\vec{E}=\vec{E}_0 \operatorname{Re}(\text{e}^{i\omega t})$. The force of the field on the charge $e$ is $e\vec{E}$.  Each dipole has a different coupling strength $F_{1,2}$ to the light field and a dephasing constant $\gamma_{1,2}$. \\

The resulting equations of motion have the same form as the equations of motion of the mechanical system shown in equation~\ref{EOM}. Therefore, the calculations and solutions of the mechanical system also hold for the excitonic system. The eigenenergies of the system are given by
\begin{equation}
\hbar\Omega_{\pm} \simeq \hbar\left(\omega_0+\frac{1}{2}\Delta \omega \right)\mp \frac{1}{2} \sqrt{\left(\hbar\Delta\omega\right)^2+4\kappa^2}
\ ,
\label{eigenenergies}
\end{equation}
assuming $\hbar\omega_0+\kappa\simeq \hbar\omega_0$, as $\hbar\omega_0\gg\kappa$.

For the eigenvectors, the ratio $\rho$ is in this case defined as $\rho=\frac{\hbar\Delta\omega}{\kappa}$. The excitonic eigenmodes can be written as
\begin{align}
\vec{r}_{1,2}  = \dfrac{e\vec{E}_0}{m} \dfrac{\left(U_{11,21}F_1+U_{12,22}F_2\right) }{-\omega^2+i \gamma_{1,2}\omega+\Omega_{\pm}^2}\ .
\label{zx}
\end{align}
The result for the eigenmodes $\vec{r}_{1,2}$ from equation~\ref{zx} is related to the dipole moment via
\begin{equation}
\vec{p}_{1,2}=e\vec{r}_{1,2} \ .
\label{mu}
\end{equation}

So far we have described the response of two individual dipoles interacting with each other. In practice, an array of dipoles exists in a semiconductor. In a completely classical approach, the quantum well is treated as a two-dimensional array of optical dipoles \cite{karrai2003}. Following this approach, the linear absorption $\alpha(\omega)$ of a quantum well is given by \cite{karrai2003}
\begin{equation}
\alpha(\omega)= \frac{2\gamma \Gamma_0^\text{rad}}{(\omega_0-\omega)^2+(\gamma+ \Gamma_0^\text{rad})^2}\ .
\label{absorption}
\end{equation}
The definition of the radiative coupling constant $\Gamma_0^\text{rad}$ is
\begin{equation}
\Gamma_0^\text{rad}=  \frac{1}{4A}\frac{e^2 f}{\epsilon_0 c m n}\ ,
\label{gamma}
\end{equation}
where $f$ is the oscillator strength, $n$ is the refractive index of the quantum well, $c$ is the speed of light, and $1/A$ is the dipole area density. From equation~\ref{absorption} we see that the solution of the quantum well absorption is additionally broadened by $\Gamma_0^\text{rad}$ due to dipole-dipole interactions in the quantum well \cite{kira2006,karrai2003}.\\

The absorption of an array of dipoles shown in equation~\ref{absorption} has a similar form as the equation for the average dissipated power $\left<P\right>$ of two coupled masses in the mechanical system given in equation~\ref{powerApprox}. Both equations are equivalent if we take the experimental dephasing $\gamma_{1,2}$ to be equal to the dephasing of the dipole array $\gamma+\Gamma_0^\text{rad}$. By comparing the two equations we find a linear relation between the absorption Im($\chi$) of two coupled dipoles and the dissipated power $\left<P\right>$:
\begin{equation}
\text{Im}(\chi) = \frac{4 e^2}{A \epsilon_0 c  n \left|\vec{E}_0\right|^2}\left<P\right>\ .
\label{abs}
\end{equation}
The absorption strength $I_{1,2}$ of each coupled optical dipole is then dependent on the energy detuning $\hbar\Delta\omega$, the coupling $\kappa$, and the oscillator strengths $F_{1,2}$. For all calculations in the main paper and in this supplement, the prefactor $ \dfrac{4 \hbar e^2}{A \epsilon_0 c m n}$ is set to $1$~eV. $I_{1,2}$ is given by
\begin{equation}
\begin{bmatrix} 
I_1   \\
I_2
\end{bmatrix}=\begin{bmatrix} 
\left(U_{11}F_1+U_{12}F_2\right)^2  \\
\left(U_{21}F_1+U_{22}F_2\right)^2
\end{bmatrix}
\text{eV} \ .
\label{int}
\end{equation}
The absorption is then modelled as
\begin{align}
\text{Im}(\chi) =&I_1 \frac{\frac{1}{2}\hbar\gamma_1}{(\frac{1}{2}\hbar\gamma_1)^2+\left(\hbar\Omega_+ -\hbar\omega\right)^2} + \nonumber \\
&I_2 \frac{\frac{1}{2}\hbar\gamma_2}{(\frac{1}{2}\hbar\gamma_2)^2+\left(\hbar\Omega_--\hbar\omega\right)^2} 
\ .
\label{abs2}
\end{align}

\begin{figure*}[ht]
\centering
\includegraphics[width=172mm]{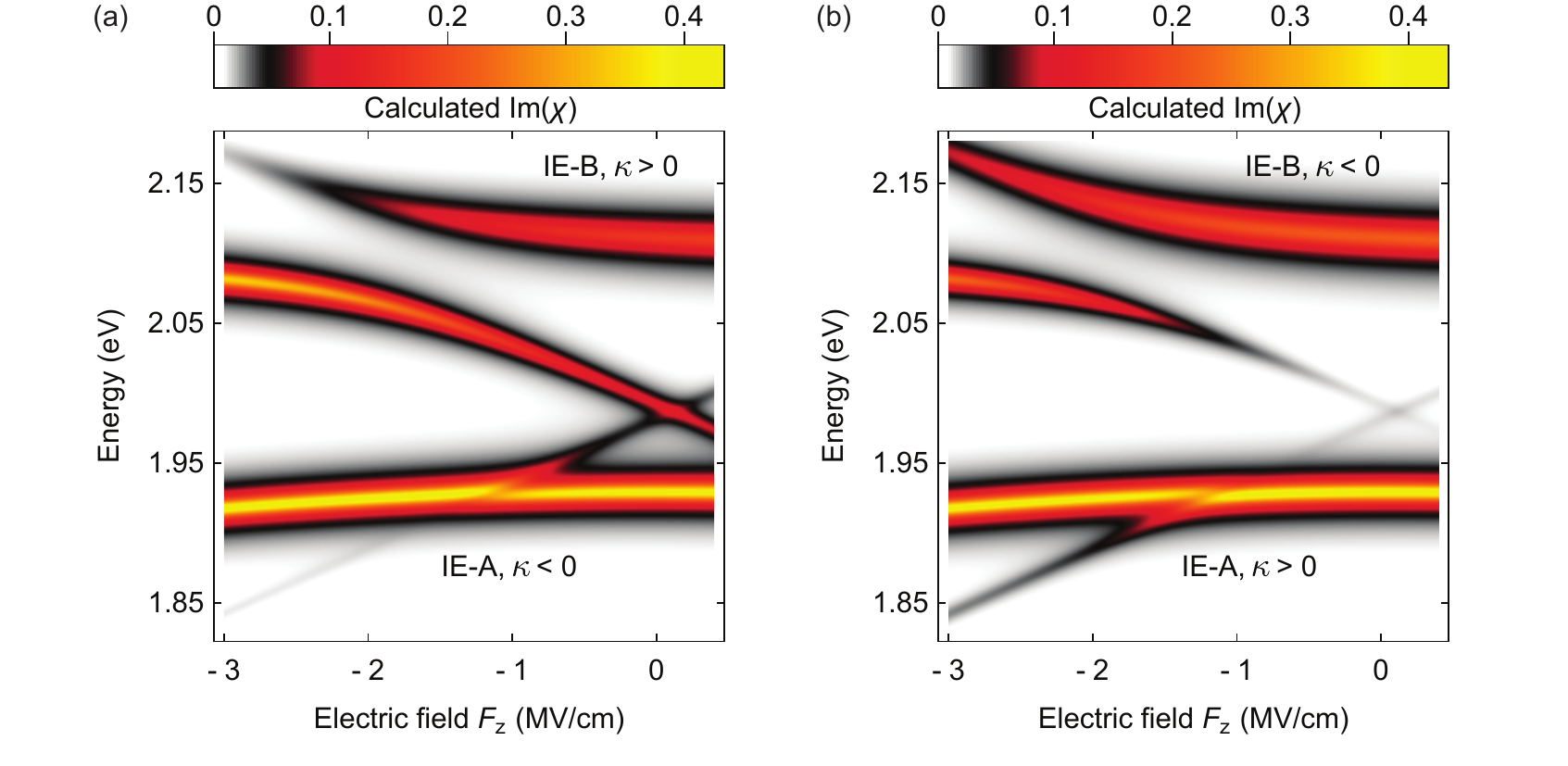}
\caption{Colour-map of the calculated absorption Im($\chi$) of the IE-A and IE-B interactions as a function of electric field according to equation~\ref{abs2}. Separate models are used to describe the IE-A and IE-B couplings. The Im($\chi$) is a sum of the two in order to visualise the (a) IE “cage" and (b) IE “anti-cage". The used parameters are the same as extracted from fits to the data shown in the main text in Main Fig.~2 and in SI~Fig.~\ref{energyFits}. The parameters are summarised in SI~Table~\ref{parameters}. The A-exciton energy is described with a quadratic Stark shift (see supplement section V). The coupling constants in (a) have the same sign as determined in the main text while each of the coupling constants in (b) have the opposite sign. The two colour-maps are interpolated to first order. }
\label{oppositeKappa}
\end{figure*}

\subsection{Semi-classical frequency response and absorption of excitons in a semiconductor}

In a semi-classical description, the coupling of a dipole to a light field is described by a quantity called the oscillator strength $f$. The oscillator strength is proportional to the expectation value of the dipole operator squared $\left|\left<\phi_n\left|\vec{p}\right|\phi_0\right>\right|^2$ \cite{cohen-tannoudji2005,klingshirn2012}. Usually $f$ is a sum over all $n$ resonances of the oscillator. In the following we will only consider the ground state transition from state $\left|\phi_0\right>$ to state $\left|\phi_1\right>$ with a resonance angular frequency $\omega_0$. The oscillator strength $f$ is defined as \cite{cohen-tannoudji2005,klingshirn2012}

\begin{equation}
f=\frac{2m\omega_0}{\hbar}\left|\left<\phi_1\left|\vec{r}\right|\phi_0\right>\right|^2=\frac{2m\omega_0}{\hbar e^2}\left|\left<\phi_1\left|\vec{p}\right|\phi_0\right>\right|^2  \ .
\label{oscillatorstrength}
\end{equation}

Then the semi-classical solution for the response of an oscillating dipole with a dephasing rate $\Gamma$ can be written as \cite{karrai2003,klingshirn2012}
\begin{equation}
\vec{r}=\frac{e\vec{E}_0}{m}\frac{f}{\omega_0^2-\omega^2-i \Gamma\omega}  \ .
\label{ExcitonResponse}
\end{equation}

In the linear optical regime, the optical susceptibility $\chi(\omega)$ relates the probe field $\vec{E}$ to the probe-induced macroscopic polarisation $P(\omega)$ \cite{kira2006}:\begin{equation}
P(\omega)=\epsilon_0 \chi(\omega)\vec{E} \ .
\label{linearChi}
\end{equation}

The Coulomb interaction between charges (exciton effects and screening) needs to be considered for a quantum mechanical description of the optical response of semicondcutors. These effects are described by the so-called semiconducting Bloch equations (SBEs) \cite{lindberg1988,klingshirn2012}. The SBEs are a set of coupled equations that include the microscopic polarisation $P_k$ as well as electron and hole distributions \cite{haug2009}. The index $k$ describes the momentum dependence of the charge carriers. The microscopic polarisation is related to the macroscopic polarisation as follows \cite{koch2006}:

\begin{equation}
P(\omega)=\left|\left<\phi_c\left|\vec{p}\right|\phi_v\right>\right| \sum_{k} P_k \ ,
\label{microscopic}
\end{equation}
where the indices $c$ and $v$ stand for valence and conduction band, respectively. \\

For a classical light field, the optical response of the semiconductor can be calculated by linking the SBEs to Maxwell's wave equation (MSBE) \cite{kira2006}. One can calculate the macroscopic polarisation as \cite{koch2006}
\begin{equation}
P(\omega)=2 \left|\left<\phi_c\left|\vec{p}\right|\phi_v\right>\right|^2 \sum_{\alpha} \frac{\left|\Psi_\alpha (r=0)\right|^2}{\hbar \omega_\alpha-\hbar\omega- i \hbar \gamma_\alpha} \, \vec{E} \ ,
\label{macroscopic}
\end{equation}
where $\Psi_\alpha$ is the excitonic wavefunction, $\omega_\alpha$ is the excitonic resonance angular frequency, and $\gamma_\alpha$ is the polarisation dephasing rate. We now consider the excitonic ground state $\Psi_0$ and rewrite equation~\ref{macroscopic} with the definition of the oscillator strength $f$ in equation~\ref{oscillatorstrength} to:
\begin{align} 
P(\omega)=\frac{e^2 \vec{E}}{m}  \frac{f}{\omega_0^2-\omega^2- i \gamma \omega} \ .
\label{macro}
\end{align}
Remarkably, the result for the macroscopic polarisation following the SBEs is the same as the result for the polarisation of an optical dipole driven by an oscillating field, compare equation~\ref{macro} with equation~\ref{zx} and equation~\ref{ExcitonResponse}.\\

M. Kira and S. W. Koch calculate the linear absorption $\alpha(\omega)$ of a quantum well following the MSBEs \cite{kira2006}. Their result for the absorption is the same as the absorption of an array of oscillating dipoles in a two-dimensional plane derived in a classical way shown in equation~\ref{absorption} \cite{karrai2003}.

\subsection{Influence of the sign of the coupling strength on the excitonic absorption}

As mentioned in the main text, comparing the calculated excitonic absorption according to equation~\ref{abs2} with the measured experimental absorption reveals the sign of the coupling strength. SI~Fig.~\ref{oppositeKappa}a shows the calculated average dissipated power of the IE-A and IE-B interactions with the same parameters as the calculations shown in Main~Fig.~2 and SI~Fig.~\ref{energyFits}. SI~Fig.~\ref{oppositeKappa}b shows the calculated dissipated power if the two coupling strengths have opposite sign. The two individual interactions are added together to highlight the IE “cage" (SI~Fig.~\ref{oppositeKappa}a) and IE “anti-cage" (SI~Fig.~\ref{oppositeKappa}b) behaviour. For the IE cage, the IE is weak “inside", i.e.\ at energies between the A-like and B-like excitons, and strong on the “outside"; the IE anti-cage describes the opposite behaviour. By comparing the two calculations with the measured absorption shown in Main~Fig.~2 it is clear that a negative coupling models the IE-A interaction, and a positive coupling models the IE-B interaction.\\

\begin{figure*}[t]
\centering
\includegraphics[width=172mm]{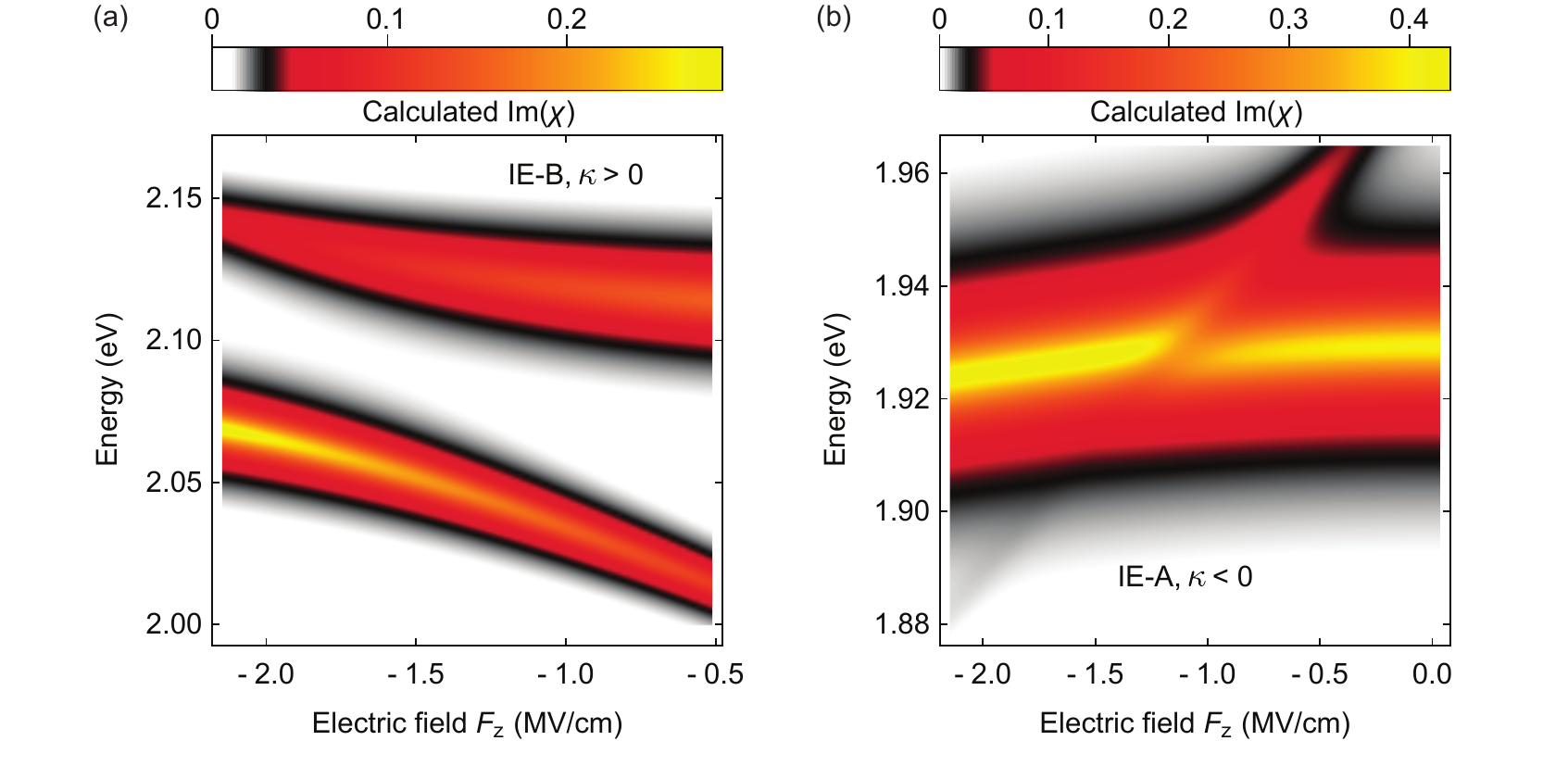}
\caption{Colour-maps of the calculated absorption Im($\chi$) of (a) the IE-B and (b) the IE-B interaction as a function of electric field according to equation~\ref{abs2}. The calculated absorption maps correspond directly to the measurements shown in Main~Fig.~2a and Main~Fig.~2c, respectively. The used parameters are the same as extracted from fits to the data shown in the main text in Main Fig.~2 and in SI~Fig.~\ref{energyFits}. The parameters are summarised in SI~Table~\ref{parameters}. The two colour-maps are interpolated to first order.}
\label{interactionMaps}
\end{figure*}

SI~Fig.~\ref{interactionMaps} shows the calculated absorption spectra for the IE-B and IE-A interaction corresponding to the measurement regions shown in Main~Fig.~2a and Main~Fig.~2c, respectively. The calculated maps reproduce the experimental data quite well.

\subsection{Coupled RLC circuits}

\begin{figure}[t!]
\centering
\includegraphics[width=86mm]{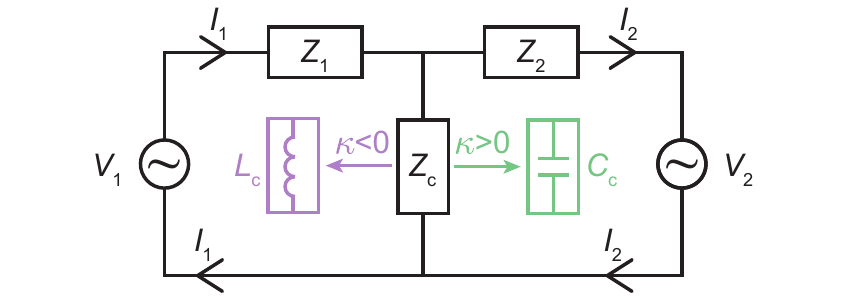}
\caption{Sketch of two RLC circuits coupled by a shared impedance $Z_c$. For a purely capacitive (inductive) coupling the coupling strength $\kappa$ is positive (negative). }
\label{rlc}
\end{figure}

Applying the interaction model to the two experimentally observed excitonic interactions yields a positive coupling constant for the IE-B interaction and a negative coupling constant for the IE-A interaction. A negative coupling constant in a mechanical framework is of course counter-intuitive. Therefore, we will look at an alternative, an electrical system consisting of two coupled RLC circuits (see SI~Fig.~\ref{rlc}). \\

First, we derive the equation of motion of a single RLC circuit. The complex impedance of a RLC circuit is given as
\begin{equation}
Z = \frac{1}{i\omega C}+R+i\omega L  \ ,
\label{impedance}
\end{equation}
where $C$ is the capacitance, $R$ is the resistance, and $L$ is the inductance. \\

The voltage can be related to the current $I$ as
\begin{equation}
V = Z I  \ .
\label{voltage}
\end{equation}
The current is equal to the change of the charge $q$ on the capacitor in time:
\begin{equation}
I = \frac{dq}{dt}  \ .
\label{current}
\end{equation}
We take the voltage driving the circuit as
\begin{equation}
V = V_0  \operatorname{Re}(\text{e}^{i\omega t})
\label{driving}
\end{equation}
and the corresponding charge as
\begin{equation}
q = q_0  \operatorname{Re}(\text{e}^{i\omega t}) .
\label{charge}
\end{equation}
We then calculate the equation of motion of the RLC circuit:
\begin{align}
V &=Z \frac{dq}{dt}  \nonumber \\
V_0\operatorname{Re}(\text{e}^{i\omega t})  &=  \left( \frac{1}{i \omega C}+R+i \omega L\right)i \omega q_0 \operatorname{Re}(\text{e}^{i\omega t}) \nonumber \\
V_0 &=  \left( \frac{1}{C}+i\omega R-\omega^2L\right) q_0
\ .
\label{eomelectric}
\end{align}

Equation~\ref{eomelectric} can be directly compared to the equation of motion of a mechanical oscillator system:
\begin{align}
F &=  \left( k+i \omega \gamma-\omega^2m\right) x
\ .
\label{eommechanic}
\end{align}
The inductance $L$ corresponds to the mass $m$, the resistance $R$ corresponds to the damping $\gamma$, and the inverse of the capacitance $\frac{1}{C}$ corresponds to the spring constant $k$. Alternatively we can write equation~\ref{eomelectric} as
\begin{equation}
\frac{V}{L} = \dfrac{d^2q}{dt^2}+\frac{R}{L}\frac{dq}{dt}+\frac{1}{LC}q\ .
\label{eomelectric2}
\end{equation}\\

Next, we look at two RLC circuits coupled together by a shared impedance $Z_c$ (see SI~Fig.~\ref{rlc}). Both circuits are driven by an AC voltage with different magnitude but with the same angular frequency and phase. We consider two distinct cases: $Z_c$ is either a capacitor or an inductor. Let us take $Z_c=\dfrac{1}{i \omega C_c}$ first. One can then write the equations of motion as:
\begin{align}
V_{1} &=L_{1} \dfrac{d^2q_{1}}{dt^2}+R_{1}\dfrac{dq_1}{dt}+\dfrac{1}{C_{1,\text{tot}}}q_1-\dfrac{1}{C_c}q_2\ , \\
V_{2} &=L_{2} \dfrac{d^2q_{2}}{dt^2}+R_{2}\dfrac{dq_2}{dt}+\dfrac{1}{C_{2,\text{tot}}}q_1-\dfrac{1}{C_c}q_1\ , 
\label{coupledeomC}
\end{align}
with $C_{1,\text{tot}}=\dfrac{C_{1}C_c}{C_{1}+C_c}$ and $C_{2,\text{tot}}=\dfrac{C_{2}C_c}{C_{2}+C_c}$ being the series capacitance of $C_{1,2}$ and $C_c$. The matrix form of this coupling is:

\begin{widetext}
\begin{equation}
\begin{bmatrix} 
V_1\\
V_2
\end{bmatrix}=
\begin{bmatrix} 
-\omega^2L_1+i \omega R_1+\dfrac{1}{C_{1,tot}} & -\dfrac{1}{C_c} \\
 -\dfrac{1}{C_c}  &-\omega^2L_2+i \omega R_2+\dfrac{1}{C_{2,tot}}
\end{bmatrix}
\begin{bmatrix} 
q_1 \\
q_2
\end{bmatrix}
\ .
\label{EOMC}
\end{equation}
\end{widetext}

We define the sign of the electrical coupling in the same way as for the mechanical system shown in equation~\ref{EOM}. For a capacitive coupling of the two RLC circuits, the coupling constant is positive. This means the IE-B interaction can be seen as a capacitive coupling.\\

Next, $Z_c$ is set to $i\omega L_c$. The matrix form of the equations of motion can be written as:
\begin{widetext}
\begin{equation}
\begin{bmatrix} 
V_1\\
V_2
\end{bmatrix}=
\begin{bmatrix} 
-\omega^2\left(L_1+L_c\right)+i \omega R_1+\dfrac{1}{C_{1}} & \omega^2L_c \\
 \omega^2L_c   &-\omega^2\left(L_2+L_c\right)+i \omega R_2+\dfrac{1}{C_{2}}
\end{bmatrix}
\begin{bmatrix} 
q_1 \\
q_2
\end{bmatrix}
\ .
\label{EOML}
\end{equation}
\end{widetext}
For an inductive coupling of the two circuits, the coupling constant is negative. The IE-A interaction can be described as an inductive coupling.

\subsection{Constructive and destructive interference}

\begin{figure}[t!]
\centering
\includegraphics[width=86mm]{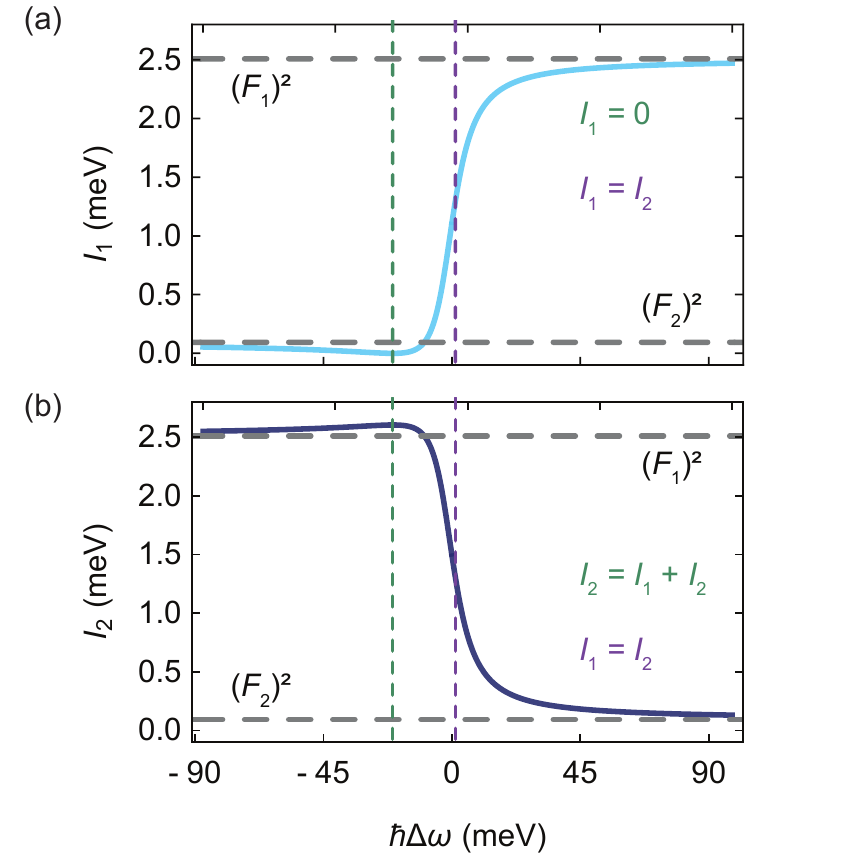}
\caption{Plot of the absorption strength (a) $I_1$ and (b) $I_2$ calculated according to equation~\ref{int} adjusted with the quadratic Stark shift of the A-exciton. The plots show the same curves as the fits to the measured data shown in Main~Fig.~2e as a function of detuning $\hbar\Delta\omega$. The parameters are summarised in SI~Table~\ref{parameters}. The horizontal dashed grey lines indicate the absorption strengths of the bare (i.e.\ uncoupled) excitons. The vertical dashed green lines indicate the detuning of the maximum/minimum absorption strength due to the constructive/destructive interference in each eigenmode. The vertical dashed purple lines indicate the detuning where the absorption strengths $I_1$ and $I_2$ are identical.}
\label{intensity}
\end{figure}

SI~Fig.~\ref{intensity} shows the calculated absorption strengths $I_{1,2}$ according to equation~\ref{int} using the same parameters as extracted for the IE-A interaction in the main text. Here, the absorption strengths are plotted as a function of energetic detuning $\hbar\Delta\omega$ as defined in equation~\ref{deltaK}. For large detuning, both absorption strengths approach the value of the respective oscillator strengths. This model also predicts destructive interference in eigenmode 1 (SI~Fig.~\ref{intensity}a), constructive interference in eigenmode 2 (SI~Fig.~\ref{intensity}b), for negative detuning and a negative coupling constant. The sign of $\kappa$ determines which eigenmode experiences destructive interference. For $\kappa>0$ ($\kappa<0$) the higher (lower) energy mode shows destructive interference. The maximum absorption strength in the eigenmode with constructive interference corresponds to the sum of the squared oscillator strengths $F_{1,2}$. The minimum absorption strength in the eigenmode with destructive interference is zero (see SI~Fig.~\ref{intensity}). \\

\begin{figure}[t!]
\centering
\includegraphics[width=86mm]{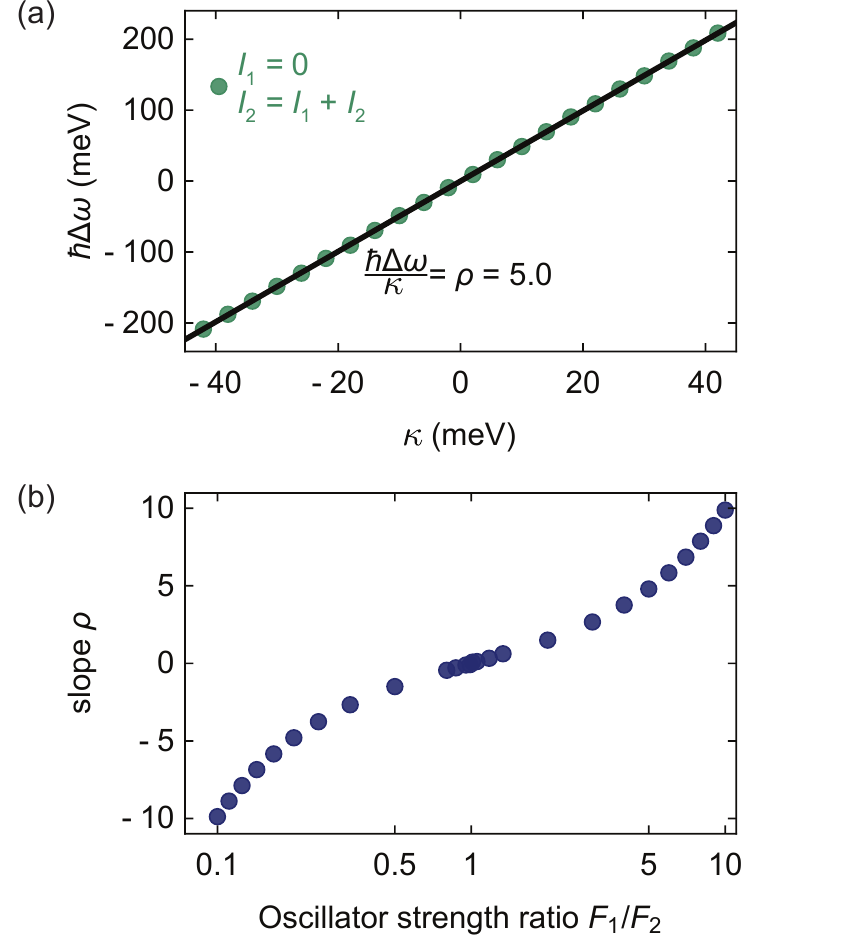}
\caption{(a) Plot of the energy detuning $\hbar \Delta \omega$ of the maximum/minimum absorption strength due to constructive/destructive interference in each eigenmode as a function of the coupling strength $\kappa$. For the calculation of the absorption strengths according to equation~\ref{int}, the parameters for the IE-A interaction shown in SI~Table~\ref{parameters} are used. The electric field offset $F_{z,0}$ and the polarisability $\beta_z$ are set to zero. The black line is a linear fit to the data and yields a slope of $\rho= 5.0$ for $F_1/F_2=5.159$. (b) Plot of the slope $\rho$ as fitted in (a) for varying oscillator strength ratios $F_1/F_2$. For a negative dipole moment $\mu$ (IE-B), the sign of the slopes is flipped.}
\label{minmax}
\end{figure}

Next, we look at the influence of the coupling strength and the oscillator strength ratio on the constructive/destructive interference. SI~Fig.~\ref{minmax}a shows the energy detuning $\hbar\Delta \omega$ of the maximum/minimum intensity due to constructive/destructive interference in each eigenmode for various coupling strength values $\kappa$. The sign of $\kappa$ is the same as the sign of the detuning $\hbar\Delta \omega$ needed to reach the extrema. A larger absolute value for $\kappa$ shifts the absorption strength extrema towards larger detuning. Additionally, the constructively/destructively interfered absorption strengths are spread over a larger detuning range for larger coupling strengths. Let's turn to the large positive coupling strength of the IE-B interaction with $\kappa =35.8$~meV. The effect of the constructive/destructive interference would only be visible at a high detuning of roughly $-175$~meV $\widehat{=}-5$~MV/cm (see SI~Fig.~\ref{minmax}a and SI~Fig.~\ref{cases} top row). Reaching such high electric fields is experimentally not feasible. \\

The dependence of the detuning position of the extreme points on $\kappa$ is fitted with a linear function yielding a slope of $\rho=5.0$ for $F_1/F_2=5.159$. SI~Fig.~\ref{minmax}b shows the slope values as a function of the oscillator strength ratio $F_1/F_2$. For a large imbalance between $F_1$ and $F_2$, the slopes increase. This means that even for a small $\kappa$ the detuning needed to reach the extrema is quite large (compare vertical dashed green lines in SI~Fig.~\ref{cases} top row). Towards $F_1/F_2 = 1$ the slopes approach zero. A smaller imbalance of $F_1$ and $F_2$ pushes the extrema towards zero detuning (compare vertical dashed green lines in SI~Fig.~\ref{cases} top and bottom row). From an experimental point of view, the interplay between coupling strength and oscillator strengths is important for the constructive/destructive interferences to be visible for reasonable experimental parameters (compare SI~Fig.~\ref{cases}).

\begin{figure}[t!]
\centering
\includegraphics[width=86mm]{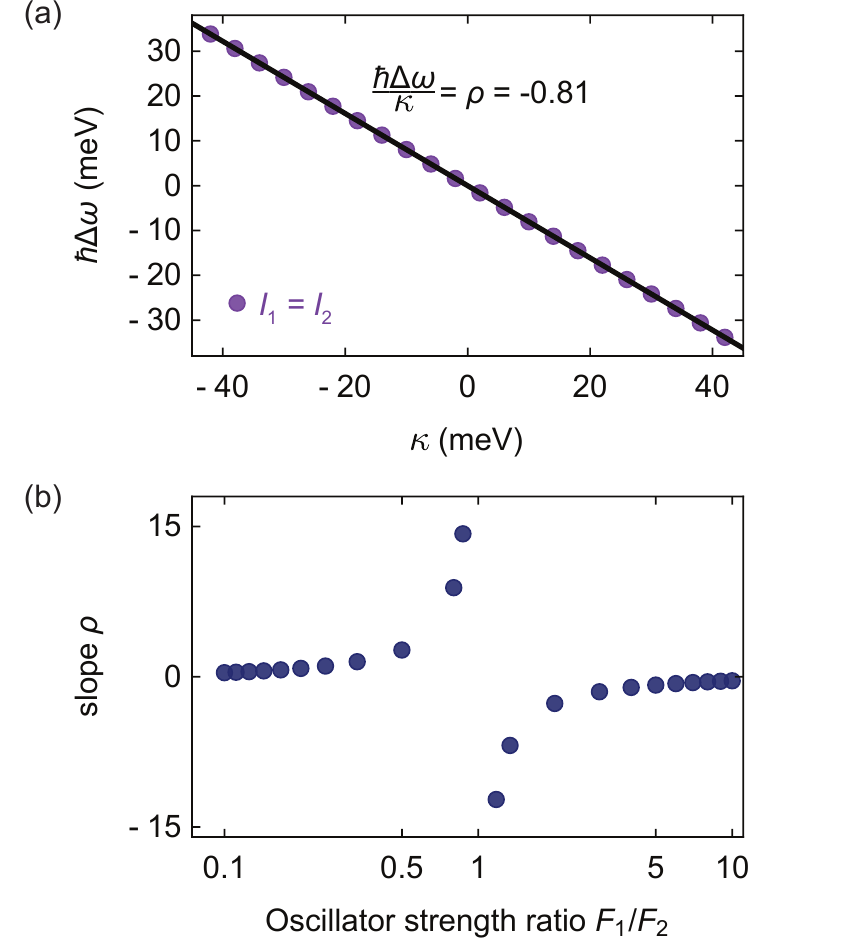}
\caption{(a) Plot of the energy detuning $\hbar \Delta \omega$ of the equal absorption strength point, $I_1=I_2$, as a function of the coupling strength $\kappa$. For the calculation of the absorption strengths according to equation~\ref{int}, the parameters for the IE-A interaction shown in SI~Table~\ref{parameters} are used. The electric field offset $F_{z,0}$ and the polarisability $\beta_z$ are set to zero. The black line is a linear fit to the data and yields a slope of $\rho=-0.81$ for $F_1/F_2=5.159$. (b) Plot of the slope $\rho$ where $I_1=I_2$ as a function of the oscillator strength ratio $F_1/F_2$. For a negative dipole moment $\mu$ (IE-B), the sign of the slopes is flipped. }
\label{equal}
\end{figure}

\subsection{Equal absorption strength point}

\begin{figure*}[ht]
\centering
\includegraphics[width=172mm]{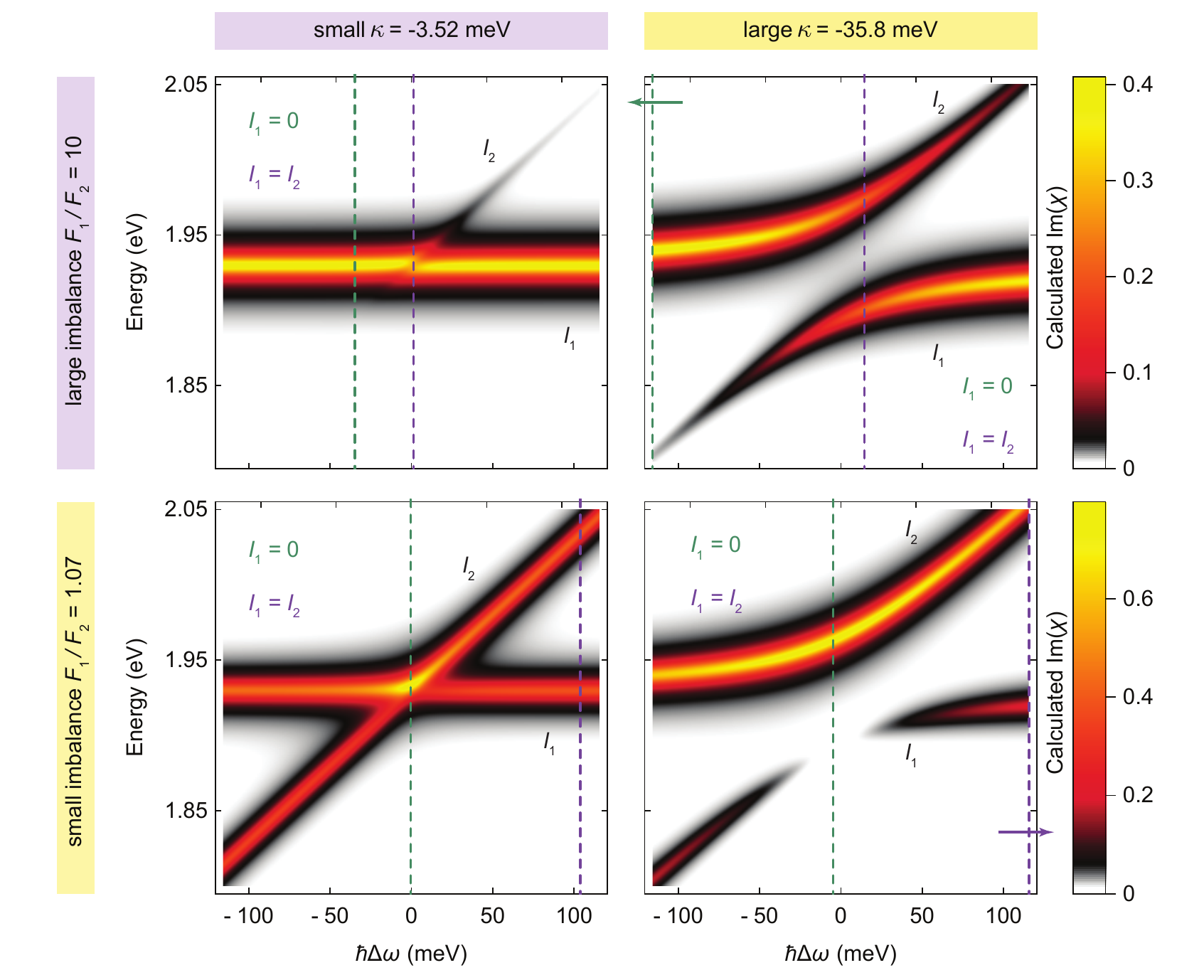}
\caption{Colour-map of the calculated absorption Im($\chi$) of an excitonic interaction as a function of energy detuning $\hbar\Delta\omega$ with negative coupling according to equation~\ref{abs2}. The left (right) column calculation uses a small (large) coupling $\kappa=-3.18$~meV ($\kappa=-35.8$~meV). The top (bottom) row calculation uses a large (small) imbalance of the oscillator strengths $F_1/F_2=10$ ($F_1/F_2=1.07$). $F_1$ is set to the A-exciton oscillator strength in table~\ref{parameters}. The electric field offset $F_{z,0}$ and the polarisability $\beta_z$ are set to zero. The other parameters used are from the IE-A interaction shown in SI~Table~\ref{parameters}. The vertical dashed green line indicates the detuning where the absorption strength $I_1$ vanishes due to the destructive interference in eigenmode 1; the absorption strength $I_2$ is maximised due to the constructive interference (see SI~Fig.~\ref{minmax}). The vertical dashed purple line indicates the detuning where the absorption strengths $I_1$ and $I_2$ are identical (see SI~Fig.~\ref{equal}). For a large $\kappa$ (right column), the extrema (top) and the equal absorption strength point (bottom) lie outside of the depicted detuning range. The extrema (right column, top) are shifted to $\hbar\Delta\omega\ll-1000$~meV; the equal absorption strength point (right column, bottom) is shifted to $\hbar\Delta\omega=1000$~meV. All colour-maps are interpolated to first order.}
\label{cases}
\end{figure*}

According to equation~\ref{int} we calculate numerically the equal absorption strength point of the two coupled resonances ($I_1=I_2$). The calculations based on the model predict that the equal absorption strength point is detuned from the crossing point of the uncoupled energy levels. SI~Fig.~\ref{equal}a plots the detuning $\hbar\Delta \omega$ needed to reach the equal intensity point for a certain coupling constant $\kappa$: the smaller $\left|\kappa\right|$, the lower the detuning of the equal absorption strength point. A linear fit to the calculated data yields a slope of $\rho= -0.81$ for $F_1/F_2=5.159$ where the condition $I_1=I_2$ is fulfilled. The influence of the oscillator strength ratio on the ratio $\rho$ is shown in SI~Fig.~\ref{equal}b. Increasing (decreasing) the oscillator strength imbalance decreases (increases) the ratio $\rho$. This means that a larger imbalance of the oscillator strengths pushes the equal intensity point towards the uncoupled energy crossing point (compare vertical dashed purple lines in SI~Fig.~\ref{cases} top and bottom row). \\

\section{Quantum-confined Stark effect of the A-exciton}

In an external electric field perpendicular to the MoS$_2$ layers $F_z$, excitons experience an energy shift $\Delta E$ given by
\begin{equation}
\Delta E =-\mu_z F_z- \beta_z F_z^2 \ .
\label{stark}
\end{equation}
where $\mu_z$ is the excitonic dipole moment and $\beta_z$ is the excitonic polarisability. This effect is called the quantum-confined Stark effect (QCSE) \cite{miller1984}. While the interlayer excitons have a significant out-of-plane dipole moment, the intralayer excitons have a near-zero out-of-plane dipole moment ($\mu_z \simeq 0$) leading to a quadratic QSCE \cite{roch2018}.  \\

The quadratic shift of the A-exciton is included in the model when fitting the IE-A interaction in the main text. This is carried out by adding the term $-\beta_z F_z^2$ to the uncoupled energy of the optical dipole 1 representing the A-exciton. Then, the eigenenergies have an additional term depending on the quadratic energy shift of the optical dipole 1 (A-exciton). The eigenenergies are
\begin{widetext}
\begin{equation}
\hbar\Omega_{\pm}=\hbar\omega_0+\kappa- \frac{1}{2}\beta_z F_z^2+\frac{1}{2}\hbar\Delta \omega \mp \frac{1}{2} \sqrt{\left(\hbar\Delta\omega\right)^2+4\kappa^2+2\left(\beta_z F_z^2\right)^2+2\beta_z F_z^2\hbar\Delta\omega} \ .
\label{eigenenergiesBeta}
\end{equation}
\end{widetext}

The matrix elements of $\textbf{U}$ that are not equal to one become
\begin{widetext}
\begin{equation}
U_{12,21}=\mp\frac{1}{2}\frac{\beta_z F_z^2}{\kappa}\mp\frac{1}{2}\rho \pm\sgn \left(\rho\right) \sqrt{1+\frac{1}{4}\rho^2+\frac{1}{4}\left(\frac{\beta_z F_z^2}{\kappa}\right)^2+\frac{1}{2}\frac{\beta_z F_z^2}{\kappa}\rho} \ .
\label{eigenvectorsBeta}
\end{equation}
\end{widetext}

The fits of the IE-A interaction shown in Main~Fig.~2 and in SI~Fig.~\ref{energyFits} use the equations modified by the quadradic Stark shift of the A-exciton. \\

The fit yields a polarisability of the A-exciton of $\beta_z=6.4 \times 10^{-9}$~D~m~V$^{-1}$. The A-exciton polarisability in the bilayer is roughly a factor of 10 larger than the polarisability of A-excitons in a monolayer MoS$_2$ \cite{roch2018}. In homobilayer MoS$_2$ the valence band of both layers are mixed \cite{gerber2019}. This leads to an increased effective quantum well width as compared to the monolayer which in turn also leads to an increased polarisability \cite{pedersen2016}. \\

The B-exciton can't be described nicely with a simple quadratic energy shift. Due to the large IE-B coupling the B-exciton is influenced by the IE even at zero electric field. This influence alters the energy evolution of the B-exciton beyond the quadratic Stark shift.

\begin{table*}[hbt!]
\begin{tabular}{@{}l|c|c|c|c|c|c|c|c|c|c|@{}}

\cline{2-11}
                                      & $\hbar\omega_0$ (eV) & $\kappa$ (meV) & $\mu$ ($e$ nm) & $F_{z,0}$ (MV/cm) & $F_1$   & $F_2$    & $F_1/F_2$ & $\hbar\gamma_1$ (meV) & $\hbar\gamma_2$ (meV) & $\beta_z$ (D$~$m$~$V$^{-1}$) \\ \midrule
\hline

\multicolumn{1}{|l|}{IE-A} & 1.930     & -3.517         & 0.4634            & -1.119            & 0.05008  & 0.009706 & 5.159     & 12                & 12              & $6.412\times 10^{-9}$                  \\ \midrule
\hline

\multicolumn{1}{|l|}{IE-B} & 2.099     & 35.76          & -0.5000            & -1.898            & 0.04645 & 0.009116 & 5.096     & 20              & 12          &                                    \\ \bottomrule
\hline

\end{tabular}
\caption{\label{parameters} Parameters extracted from the measured data shown in Main~Fig.~2 and used for the calculation of the excitonic interaction model. For IE-B, $\hbar\omega_0$ is the mean value of the fitted non-interacting B$_\text{L1}$ energies for all electric fields. $\hbar\gamma_{1,2}$ are set to make the calculated resonances have a comparable lindewidth as in the experiments. All other parameters are extracted from the measured data using the interaction model equations \ref{eigenenergies} and \ref{int}. For the IE-A interaction, the equations are adjusted to accommodate the polarisability $\beta_z$ of the A-exciton (see supplement section IV). We estimate the random errors in all parameters deduced with the coupled dipole model to be $\pm 10\%$.}
\end{table*}

\begin{figure*}[ht!]
\centering
\includegraphics[width=172mm]{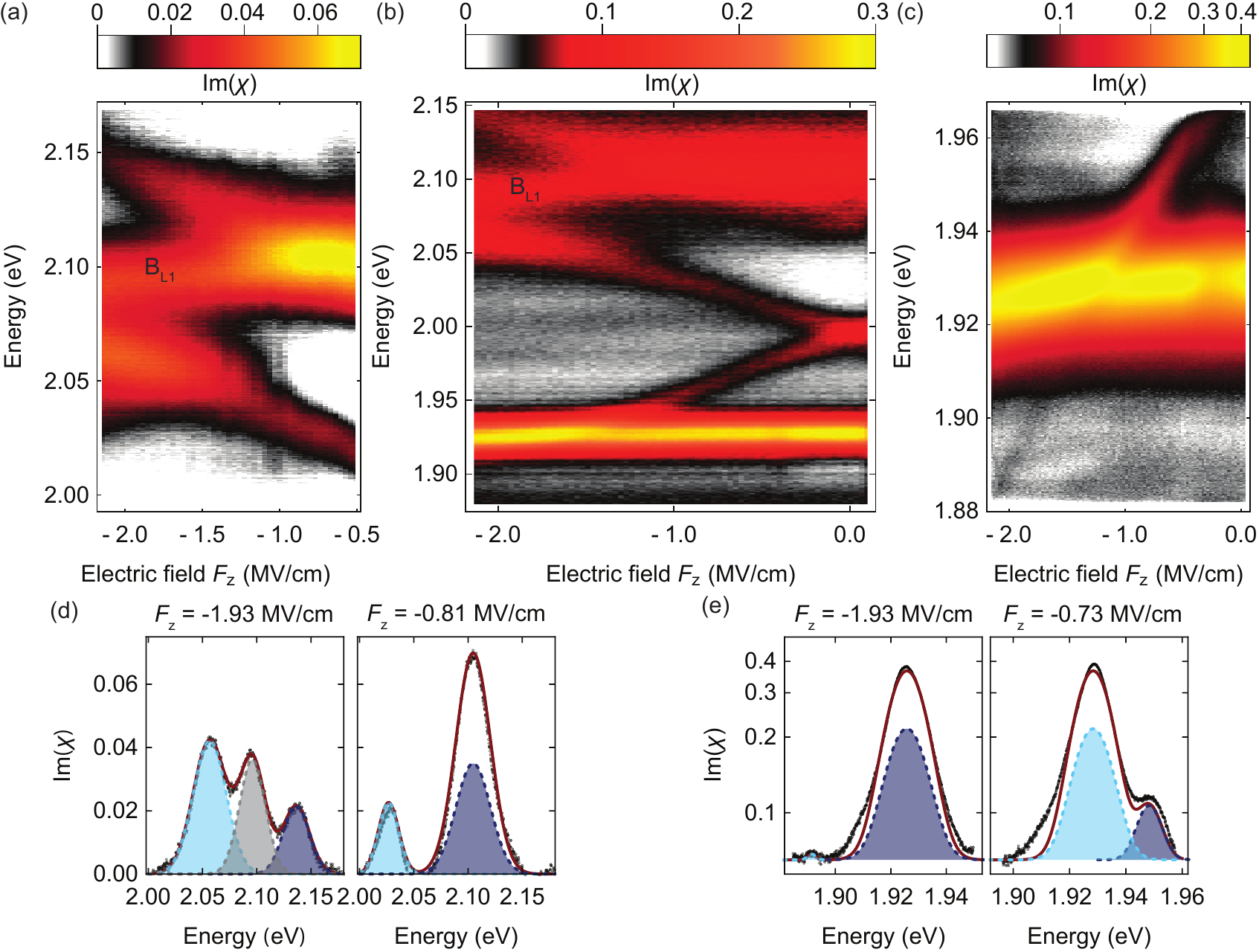}
\caption{Measured absorption ($\sigma^+$ polarisation) of homobilayer MoS$_2$ as a function of applied electric field $F_z$ (b) over the whole studied excitonic region, (a) centred around the B-exciton, and (c) centred around the A-exciton. The measurements were carried out at $T=4.2$~K and $B_z=9$~T. These are the same measurements as shown in Main~Fig.~2a-c without the subtraction of the non-interacting B$_\text{L1}$ and A$_\text{L2}$ excitons. (d,e) Two example spectra of (d) the IE-B interaction shown in (c) and (e) the IE-A interaction shown in (a). Each spectrum is fitted by a sum of three Gaussians. The grey Gaussian in the left spectrum in (d) corresponds to the non-interacting B$_\text{L1}$. The absorption in (c) and (e) is scaled logarithmically. }
\label{bl1}
\end{figure*}

\section{Measurement data processing}

The measured absorption spectra in Main~Fig.~2a-c are shown without the non-interacting B$_\text{L1}$ and A$_\text{L2}$ exciton peaks. The raw data of those three measurements are shown in SI~Fig.~\ref{bl1}. Each spectrum is fitted by a sum of three Gaussians. First, the removal procedure of the B$_\text{L1}$ in SI~Fig.~\ref{bl1}a and SI~Fig.~\ref{bl1}b is described. The linewidth and amplitude of the substracted B$_\text{L1}$ exciton at all electric fields is set to the fitted values at the highest electric field. The energy of the B$_\text{L1}$ exciton for the scan in SI~Fig.~\ref{bl1}a (SI~Fig.~\ref{bl1}b) is taken as the fitted values until $-0.90$~MV/cm ($-1.39$~MV/cm) and from then on taken as constant. The B$_\text{L1}$ exciton is then removed with these fitting parameters.\\

\begin{figure*}[ht!]
\centering
\includegraphics[width=172mm]{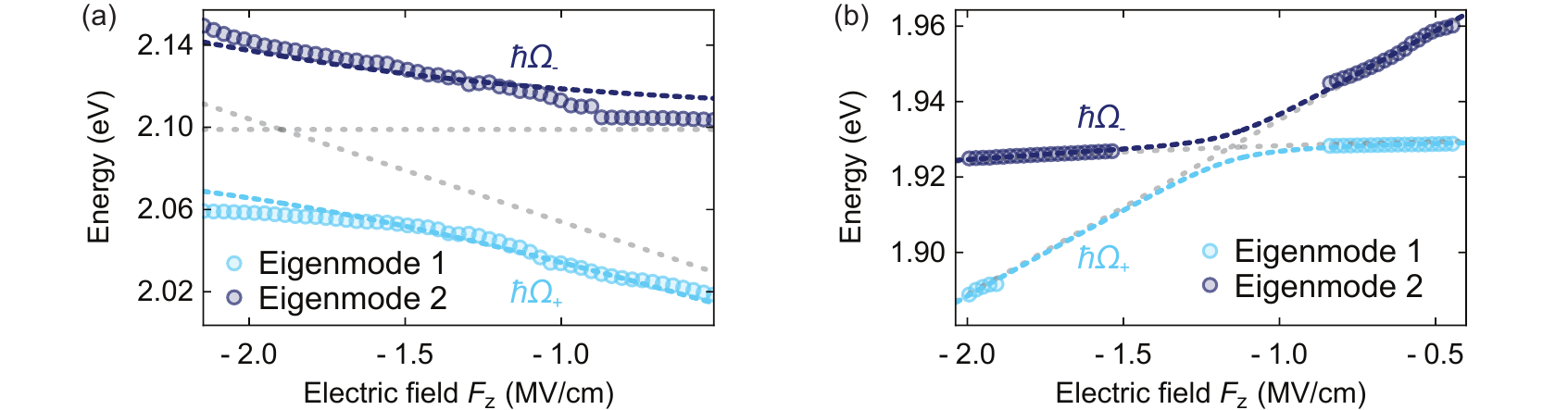}
\caption{(a) Peak energies extracted from the absorption spectra shown in SI~Fig.~\ref{bl1}a. The coloured dashed lines are fits to the energy according to equation~\ref{eigenenergies} for the eigenenergies $\hbar\Omega_\pm$. (b) Peak energies extracted from the absorption spectra shown in SI~Fig.~\ref{bl1}c. As the coupling strength is much smaller than the linewdith of the A-exciton, only the spectra far from the crossing point can be fitted unambiguously. The coloured dashed lines are fits to the energy according to equation~\ref{eigenenergiesBeta} which includes the quadratic Stark shift of the A-exciton. The coloured dashed lines in (a) and (b) are the same lines as shown overlaid on the measured absorption maps in Main~Fig.~2a and Main~Fig.~2c, respectively. All parameters extracted from the measured data are summarised in SI~Table~\ref{parameters}. The light grey dashed lines in (a) and (b) show the energy evolution with zero coupling.}
\label{energyFits}
\end{figure*}

Second, the substraction of the A$_\text{L2}$ exciton in SI~Fig.~\ref{bl1}b and SI~Fig.~\ref{bl1}c is discussed. The linewidth and amplitude of the substracted A$_\text{L2}$ exciton at all electric fields is set to the fitted values at zero electric field. The energy is taken to be the fitted zero-field energy modified by the quadratic Stark shift for finite fields. The polarisability is set to the value extracted from the coupled dipole model shown in SI~Table~\ref{parameters}.\\

The exciton energies and peak areas shown in Main~Fig.~2 are extracted from the unmodified measured absorption spectra shown in SI~Fig.~\ref{bl1}a and SI~Fig.~\ref{bl1}c. The fitting routines yielding the energies and peak areas are described. The IE-A interaction shown in SI~Fig.~\ref{bl1}c  is fitted as the sum of three Gaussians where the two Gaussians describing the A-excitons in each layer are set to the same energy and intensity (see SI~Fig.~\ref{bl1}e). This is a reasonable assumption if the spectra are energtically far away from the energy crossing point compared to the coupling strength. For the IE-B measurement shown in SI~Fig.~\ref{bl1}a, spectra at electric fields higher than $-0.81$~MV/cm are fitted by a sum of three independent Gaussians while for lower electric fields the two B-excitons in each layer are set to the same energy and intensity for reasonable fitting results (see SI~Fig.~\ref{bl1}d). \\

The absorption maps shown in Main~Fig.~2a and Main~Fig.~2c are overlaid by the eigenenergy evolution as determined by the coupled dipole model. SI~Fig.~\ref{energyFits} shows the extracted peak energies from the absorption spectra shown in SI~Fig.~\ref{bl1}a and SI~Fig.~\ref{bl1}c. The coloured dashed lines in SI~Fig.~\ref{energyFits}a are fits to the IE- and B-energies according to equation~\ref{eigenenergies}. The IE- and A-energies shown in SI~Fig.~\ref{energyFits}b are fitted by equation~\ref{eigenenergiesBeta}.

\section{Beyond DFT calculations}

\begin{figure}[ht!]
\centering
\includegraphics[width=86mm]{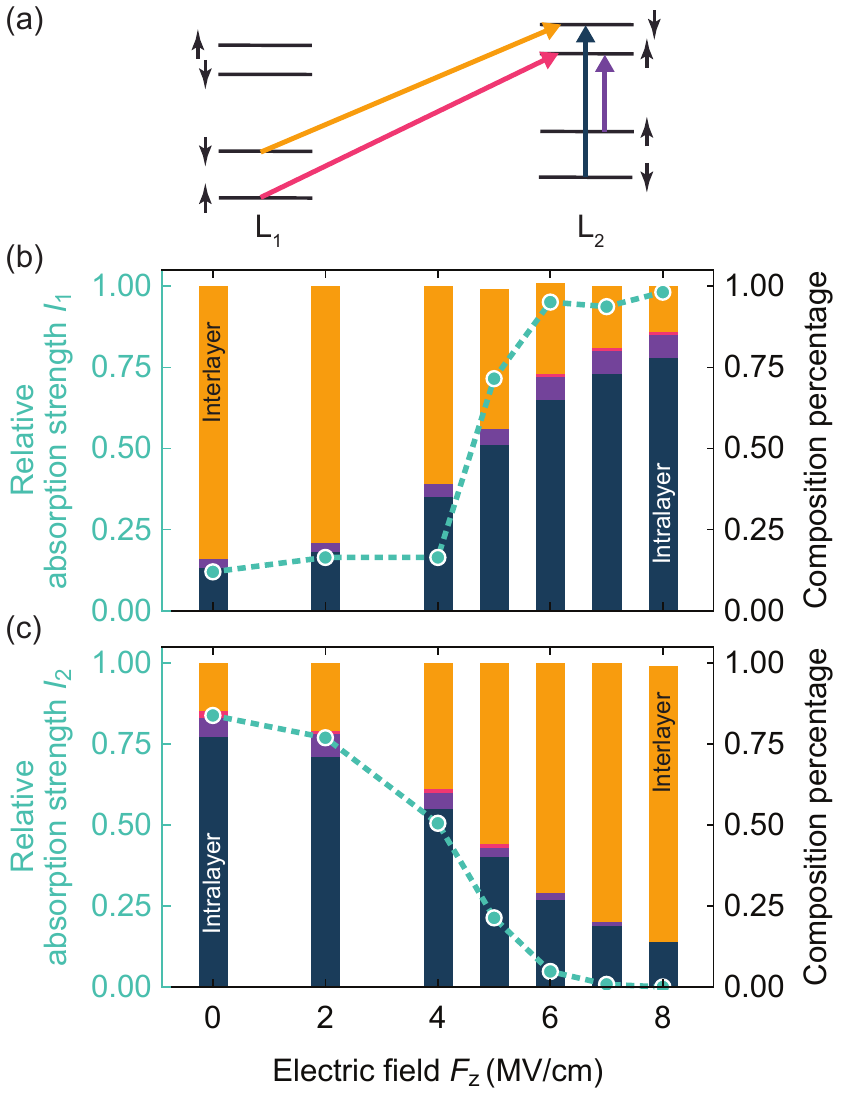}\\
\caption{(a)  Main excitonic transitions contributing to the IE$_2$-B$_\text{L2}$ dressed state. They are sketched as the coloured arrows in the schematic band structure at the $K$-point of bilayer MoS$_2$ at a finite positive electric field $F_z$. (b,c) Relative absorption strength of eigenmode 1 $I_1$ (b) and eigenmode 2 $I_2$ (c) of the IE$_2$-B$_\text{L2}$ interaction as a function of electric field determined by GW+BSE calculations. The cyan dashed line is a guide to the eye. The coloured bars indicate the composition percentage at each electric field step; the colours match those used to denote the excitonic transition in (a).}
\label{DFT_IE-B}
\end{figure}

\begin{figure}[ht!]
\centering
\includegraphics[width=86mm]{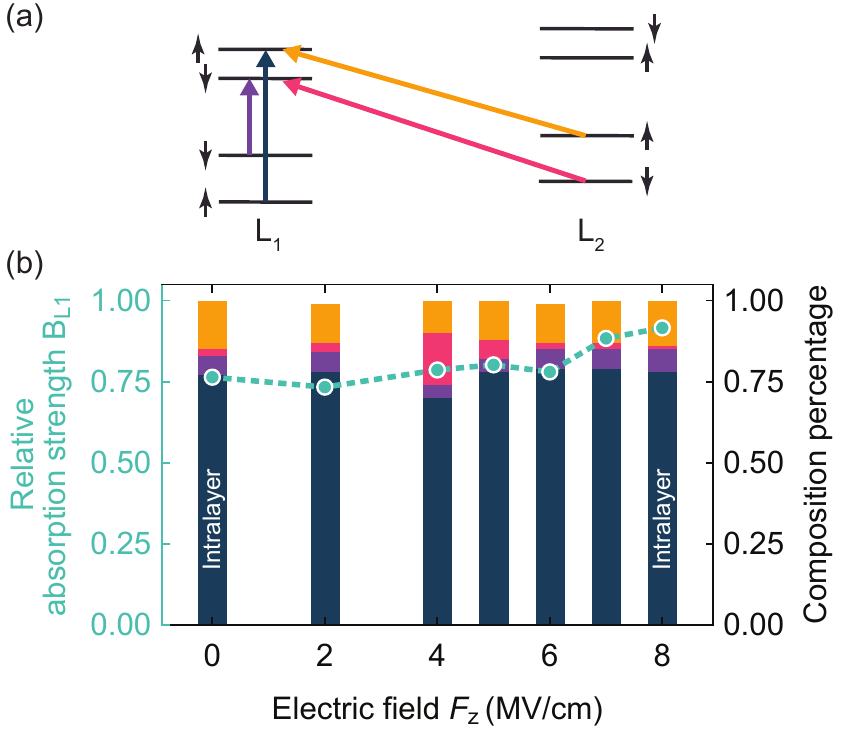}\\
\caption{(a) Main excitonic transitions contributing to the true B$_\text{L1}$ exciton. They are sketched as the coloured arrows in the schematic band structure at the $K$-point of bilayer MoS$_2$ at a finite positive electric field $F_z$. (b) Relative absorption strength of B$_\text{L1}$ as a function of electric field determined by GW+BSE calculations. The cyan dashed line is a guide to the eye. The coloured bars indicate the composition percentage at each electric field step; the colours match those used to denote the excitonic transition in (a).}
\label{DFT_BL1}
\end{figure}

\subsection{Computational details}

The atomic structures, the quasi-particle band structures, and the optical spectra were obtained using the VASP package \cite{Kresse:1993a,Kresse:1996a}. Core electrons were treated by the plane-augmented wave scheme \cite{blochl:prb:94,kresse:prb:99a}. A lattice parameter value of 3.22~\AA~was set for all calculation runs. For all the calculation cells, a grid of 15$\times$15$\times$1 k-points was used together with a vacuum height of  21.9~\AA. Van der Waals interactions between the layers were included by perfoming the geometry's optimisation process at the PBE-D3 level~\cite{Grimme:2010ij}. All atoms were allowed to relax with a force convergence criterion below $0.005$ eV/\AA. A Heyd-Scuseria-Ernzerhof (HSE) hybrid functional~\cite{heyd:jcp:04_a,heyd:jcp:05,paier:jcp:06} was used as an approximation of the exchange-correlation electronic term, including spin-orbit coupling (SOC). It was used to determine the eigenvalues and wave functions as an input for the full-frequency-dependent $GW$ calculations~\cite{Shishkin:2006a} performed at the $G_0W_0$ level. The electric field was applied at this step, just before $GW$ calculation process. Its application was only a small perturbation to the band structures, considering small/moderate electric field values. For partial occupancies, an energy cutoff of 400 eV and a gaussian smearing of 0.05 eV width were chosen. A tight electronic minimisation tolerance of $10^{-8}$~eV was set to determine the corresponding derivative of the orbitals with respect to $k$ needed in quasi-particle band structure calculations with a good precision. The direct band gap convergence was carefully checked (smaller than 0.1~eV as a function of k-points sampling). Afterwards, the total number of states included in the $GW$ procedure was set to 1280 together with an energy cutoff of 100~eV for the response function. A Wannier interpolation procedure performed by the WANNIER90 program~\cite{Mostofi:2008ff} was used to obtain the band structures.
All optical excitonic transitions were calculated by solving the Bethe-Salpeter Equation~\cite{Hanke:1979to,Rohlfing:1998vb}. The twelve highest valence bands and the sixteen lowest conduction bands were used to obtain eigenvalues and oscillator strengths on all systems. From these calculations, we use the imaginary part of the complex dielectric function to construct the absorbance with an artificial broadening of 12 meV.

\subsection{Oscillator strength evolution and interband transition composition of the true excitonic eigenstates}

Excitons in TMD monolayers do not consist of a single excitonic transition \cite{guo2019}. In fact, the true excitonic eigenstates are mixed states of multiple excitonic transitions. Let's start by looking at the composition of the true A- and B-excitons in the same valley in monolayer MoS$_2$ as an example. Our calculations for this case show that the A-$1s$ state contains 4.3\% of the B-state, while the B-$1s$ exciton contains 8.8\% of the A-transition. Those values are in line with the ones reported in Ref.~\cite{guo2019}. This mixing is due to a strong intravalley Coulomb exchange interaction and is also present in the bilayer case, as shown in Main~Fig.~4. In the following, we will further discuss the GW+BSE results on the MoS$_2$ bilayer.\\

\begin{figure}[ht!]
\centering
\includegraphics[width=86mm]{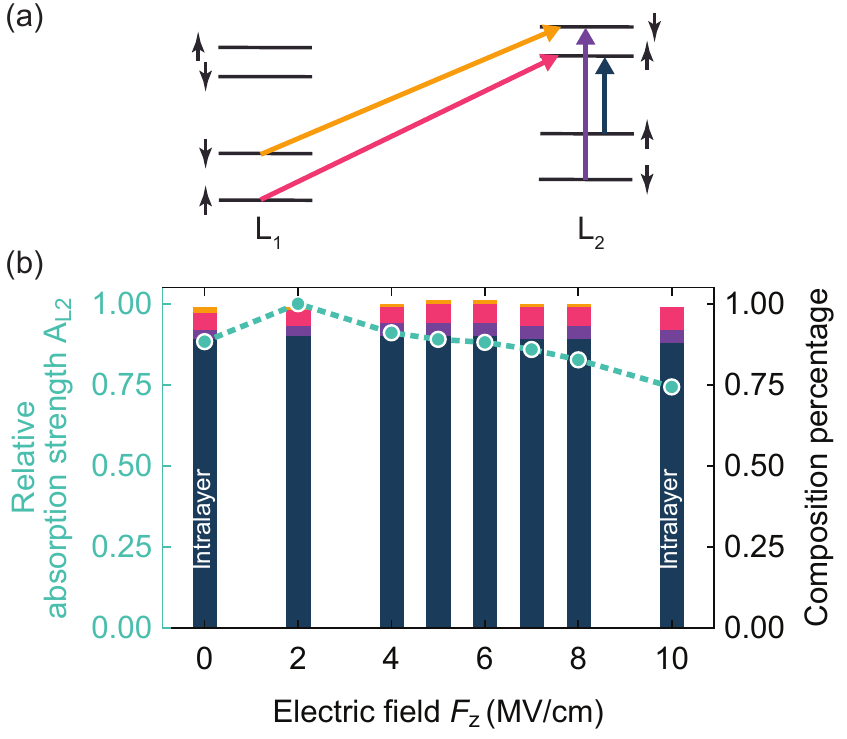}\\
\caption{(a) Main excitonic transitions contributing to the true A$_\text{L2}$ exciton. They are sketched as the coloured arrows in the schematic band structure at the $K$-point of bilayer MoS$_2$ at a finite positive electric field $F_z$. (b) Relative absorption strength of A$_\text{L2}$ as a function of electric field determined by GW+BSE calculations. The cyan dashed line is a guide to the eye. The coloured bars indicate the composition percentage at each electric field step; the colours match those used to denote the excitonic transition in (a).}
\label{DFT_AL2}
\end{figure}

\subsubsection{IE$_2$-B$_\text{L2}$ interaction}
SI~Fig.~\ref{DFT_IE-B} summarises the results from the GW+BSE calculations for eigenmode 1 and eigenmode 2 of the IE$_2$-B$_\text{L2}$ interaction. As already discussed in the main text, these two excitons interact strongly with each other. The evolution of the calculated relative absorption strength $I_1$ and $I_2$ as a function of electric field is shown as the cyan points in SI~Fig.~\ref{DFT_IE-B}b and SI~Fig.~\ref{DFT_IE-B}c, respectively. Upon tuning through the interaction region, a clear transfer of absorption strength between the two eigenmodes is visible. These beyond standard DFT results agree quite well with the measured absorption shown in Main~Fig.~2a and with the fitted absorption strength evolution shown in Main~Fig.~2d. \\

SI~Fig.~\ref{DFT_IE-B}b and SI~Fig.~\ref{DFT_IE-B}c also show the interband transition composition of the true excitonic eigenstates at each electric field step. The composition percentage indicates how much each interband transition contributes to the overall excitonic absorption strength. The involved interband transitions are colour-coded and shown in SI~Fig.~\ref{DFT_IE-B}a. The calculations reproduce nicely the layer character of each eigenmode upon tuning through the interaction region. Initially, eigenmode 1 (eigenmode 2) has a large interlayer IE$_2$, yellow (intralayer B$_\text{L2}$, blue) component. Upon increasing the electric field the IE$_2$ and B$_\text{L2}$ percentages in each eigenmode change: one increases at the cost of the other one. Finally, at high electric fields eigenmode 1 (eigenmode 2) has changed its layer character to an intralayer (interlayer) character. The swapping of the layer character of each eigenmode confirms the interaction of IE$_2$ and B$_\text{L2}$ \cite{leisgang2020}.\\

\subsubsection{B$_\text{L1}$ exciton}
To confirm that IE$_2$ only interacts with B$_\text{L2}$, we will now look at the calculation results for the B$_\text{L1}$ exciton. SI~Fig.~\ref{DFT_BL1} summarises the calculated relative absorption strength of B$_\text{L1}$ and its interband transition composition. The B$_\text{L1}$ absorption strength stays constant over the whole electric field range. The interband transition composition shows several interesting features. First, the intralayer B$_\text{L1}$ (blue) component dominates the oscillator strength and stays rather constant for all studied electric fields. These results are a confirmation that the B$_\text{L1}$ exciton does not interact with the IE$_2$ exciton. Second, at 4~MV/cm the B-interlayer exciton (BIE, pink) component is increased as compared to lower and higher electric fields. An explanation could be that B$_\text{L1}$ and BIE with opposite spin interact in a similar way as IE$_1$ and A$_\text{L1}$. From measurements we know that the BIE is energetically located just below the B-excitons at around $2.07$~meV \cite{leisgang2020}. An energetic crossing of B$_\text{L1}$ and BIE could therefore be expected around this electric field range. Experimentally, we can't resolve this interaction due to the very small oscillator strength of the BIE and the probably small coupling strength to B$_\text{L1}$. Last, the true B$_\text{L1}$ exciton retains a very large and rather constant IE$_1$ (yellow) component over the whole electric field range even though IE$_1$ is energetically detuned far away from the B-exciton. This indicates that the orbital hybridisation and consequently the large oscillator strength of the IE is not affected much in the studied electric field range. The IE oscillator strength is not diminished even after energetically tuning it below the A-excitons.\\

\subsubsection{$\text{A}_\text{L2}$ exciton}
Next, we will discuss the GW+BSE results for the true A$_\text{L2}$ exciton (see SI~Fig.~\ref{DFT_AL2}). Similarly to the B$_\text{L1}$ exciton discussed before, the absorption strength of A$_\text{L2}$ and its composition stay constant over the studied electric field range. This confirms that the A$_\text{L2}$ exciton does not interact with IE$_1$.\\

\subsubsection{Applying coupled dipole model to DFT results}
\begin{figure*}[t!]
\centering
\includegraphics[width=172mm]{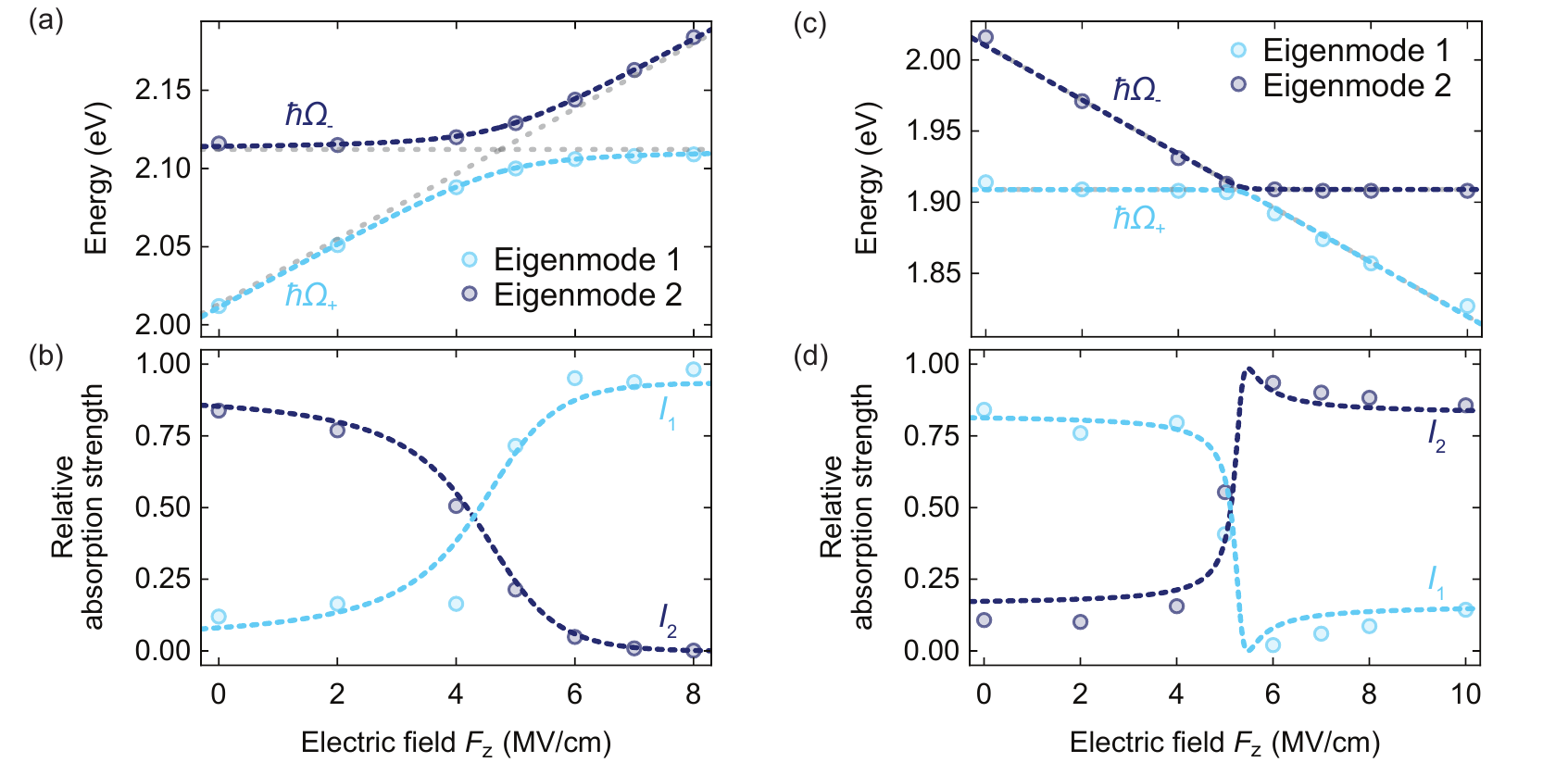}\\
\caption{(a) Peak energies and (b) relative absorption strengths of the IE-B interaction determined by GW+BSE calculations. The coloured dashed lines are fits of the energy according to equation~\ref{eigenenergies} and fits to the oscillator strength according to equation~\ref{int}. (c) Peak energies and (d) oscillator strengths of the IE-A interaction determined by GW+BSE calculations. The coloured dashed lines are fits of the energy and to the oscillator strength with the same functions as in (a) and (b), respectively. The light grey dashed lines in (a) and (c) show the energy evolution with zero coupling. All parameters extracted from the calculated data are summarised in the text. }
\label{dft}
\end{figure*}

The calculated GW+BSE excitonic peak energies and relative absorption strengths can also be fitted with the model presented in Section~III, similarly as the measured data shown in Main~Fig.~2. The energies are fitted with equation~\ref{eigenenergies} and the relative absorption strengths are fitted with equation~\ref{int}. The energy fit of the IE-B interaction shown in SI~Fig.~\ref{dft}a yields a coupling strength of $\kappa=+14.2$~meV, a B-exciton energy $\hbar\omega_0$ of $2.11$~eV, a crossing point $F_{z,0}$ of $4.74$~MV/cm, and a linear Stark shift $\mu$ of $0.21$~$e$~nm. The oscillator strength fit (SI~Fig~\ref{dft}b) yields an oscillator strength ratio of $F_1/F_2=6.26$ and confirms the positive sign of the IE-B coupling. \\

The energy fit of the IE-A interaction shown in SI~Fig.~\ref{dft}c yields an absolute coupling strength of $\kappa=-1.7$~meV,  an A-exciton energy $\hbar\omega_0$ of $1.91$~eV, a crossing point $F_{z,0}$ of $5.31$~MV/cm, and a linear Stark shift $\mu$ of $-0.19$~$e$~nm. The oscillator strength fit (SI~Fig.~\ref{dft}d) yields an oscillator strength ratio of $F_1/F_2=2.27$ and confirms the negative sign of the IE-A coupling. \\

In addition to the measurement results, the coupled dipole model is also able to parameterise the results from the post-DFT calculations quite well. 

\bibliographystyle{apsrev4-2}
\bibliography{supplement_bib}

\clearpage